\documentclass[acmsmall]{acmart}

\AtBeginDocument{%
  }

\setcopyright{cc}
\setcctype{by}
\acmJournal{PACMPL}
\acmYear{2025} \acmVolume{9} \acmNumber{OOPSLA1} \acmArticle{111} \acmMonth{4} \acmPrice{}\acmDOI{10.1145/3720452}

\usepackage[T1]{fontenc}
\usepackage{graphicx}
\usepackage{color}

\usepackage{booktabs}   
\usepackage{subcaption} 
\usepackage{algorithm}
\usepackage{algpseudocode}

\usepackage{mathpartir}
\usepackage{listings}
\usepackage{mathpartir}
\usepackage{amsmath,amsthm}
\usepackage{stmaryrd}
\usepackage{xspace}
\usepackage{tabularx}
\usepackage{wrapfig}
\usepackage{enumitem}
\usepackage{xcolor} 
\usepackage{longtable}
\usepackage{enumitem}
\usepackage{siunitx}
\setlistdepth{9}

\newenvironment{nop}{}{}

\definecolor{Bittersweet}{RGB}{255,102,102}
\definecolor{BrightBlue}{RGB}{0,0,255}
\definecolor{midnightblue}{rgb}{0.1, 0.1, 0.44}

\lstMakeShortInline[columns=fullflexible]|

\newcommand{\F}[1]{\mathsf{#1}}
\newcommand{\M}[1]{\mathcal{#1}}

\newcommand{\fstar}{F$^{\star}$\xspace}

\begin{document}

\title{Automatically Verifying Replication-aware Linearizability}
\author{Vimala Soundarapandian}
\affiliation{
	\institution{IIT Madras}
	\city{Chennai}
  \country{India}
}
\email{cs19d750@cse.iitm.ac.in}

\author{Kartik Nagar}
\affiliation{
	\institution{IIT Madras}
	\city{Chennai}
  \country{India}
}
\email{nagark@cse.iitm.ac.in}

\author{Aseem Rastogi}
\affiliation{
	\institution{Microsoft Research}
	\city{Bangalore}
  \country{India}
}
\email{aseemr@microsoft.com}

\author{KC Sivaramakrishnan}
\affiliation{
	\institution{IIT Madras and Tarides}
	\city{Chennai}
  \country{India}
}
\email{kcsrk@cse.iitm.ac.in}

\begin{abstract}
      Data replication is crucial for enabling fault tolerance and uniform low
      latency in modern decentralized applications. Replicated Data Types
      (RDTs) have emerged as a principled approach for developing replicated
      implementations of basic data structures such as counter, flag, set, map,
      etc. While the correctness of RDTs is generally specified using the notion of
      strong eventual consistency--which guarantees that replicas that have
      received the same set of updates would converge to the same state--a more
      expressive specification which relates the converged state to updates
      received at a replica would be more beneficial to RDT users.
      Replication-aware linearizability is one such specification, which
      requires all replicas to always be in a state which can be obtained by
      linearizing the updates received at the replica. In this work, we develop
      a novel fully automated technique for verifying replication-aware linearizability
      for Mergeable Replicated Data Types (MRDTs). We identify
      novel algebraic properties for MRDT operations and the merge function
      which are sufficient for proving an implementation to be linearizable and
      which go beyond the standard notions of commutativity, associativity, and
      idempotence. We also develop a novel inductive technique called bottom-up
      linearization to automatically verify the required algebraic properties.
      Our technique can be used to verify both MRDTs and state-based CRDTs. We
      have successfully applied our approach to a number of complex MRDT and
      CRDT implementations including a novel JSON MRDT.
\end{abstract}



\maketitle

\section{Introduction}
\label{sec:intro}
Modern decentralized applications often employ data replication across
geographically distributed locations to enhance fault tolerance, minimize data
access latency, and improve scalability. This practice is crucial for mitigating
the impact of network failures and reducing data transmission delays to end
users.  However, these systems encounter the challenge of concurrent
conflicting data updates across different replicas.

Recently, Mergeable Replicated Data Types (MRDTs) \cite{Kaki2019, Kaki2022,
Vimala} have emerged as a systematic approach to the problem of ensuring that
replicas remain eventually consistent despite concurrent conflicting updates.
MRDTs draw inspiration from the Git version control system, where each update
creates a new version, and any two versions can be merged explicitly through a
user-defined $\F{merge}$ function.  $\F{merge}$ is a ternary function that
takes as input the two versions to be merged and their Lowest Common Ancestor
(LCA), i.e., the most recent version from which the two versions diverged. As
opposed to Conflict-Free Replicated Data Types (CRDTs)\cite{Shapiro}, which may
have to carry around causal context metadata to ensure consistency, MRDTs can
rely on the underlying system model to provide the causal context through the
LCA.  This results in implementations that are comparatively simpler and also
more efficient. For example, if we consider state-based CRDTs, which are the
closest analogue to the MRDT model, then any counter CRDT implementation would
require $O(n)$ space, where $n$ is the number of replicas (a lower bound proved
by \cite{Burckhardt}), whereas a counter MRDT implementation only requires
$O(1)$ space. The states maintained by CRDT implementations need to form a join
semi-lattice, with all CRDT operations restricted to being monotonic functions
and merge restricted to the lattice join. While these restrictions simplify the
task of reasoning about correctness \cite{Verifx,Katara,EneaArxiv}, crafting
correct and efficient CRDT implementations itself becomes much harder.

MRDTs do not require any of the above restrictions, which helps in developing
implementations with better space and time complexity. However, reasoning about
correctness now becomes harder. Indeed, the MRDT system model allows arbitrary
replicas to merge their states at arbitrary points in time, and this can result
in subtle bugs requiring a very specific orchestration of merge actions. As
part of this work, we discovered such subtle bugs in MRDT implementations
claimed to be verified by previous works \cite{Vimala} (more details can be
found in \S\ref{subsec:bug}).  The MRDT state as well as the implementation of
data type operations and the $\F{merge}$ function have to be cleverly designed
to ensure strong eventual consistency. That is, despite concurrent conflicting
updates and arbitrary ordering of merges, all replicas will eventually converge
to the same state. Further, we would also like to show that an MRDT satisfies
the functional behavior of the data type, along with the user-defined conflict
resolution policy for concurrent conflicting updates (e.g., for a set data
type, an \emph{add-wins} policy that favors the $\F{add}$ operation over a
concurrent $\F{remove}$ of the same element at different replicas). There have
been a few works \cite{Kaki2019,Kaki2022,Vimala} that have looked at the problem
of specifying and verifying MRDTs. However, they either restrict the system
model by disallowing concurrent merges \cite{Kaki2019}, focus only on
convergence as the correctness specification \cite{Kaki2022,Kaki2019}, or do
not support automated verification \cite{Vimala}.

In this work, we couch the correctness of MRDTs using the notion of
\emph{Replication-Aware Linearizability} (RA-linearizability) \cite{Wang}, which says that the state
at any replica must be obtained by linearizing (i.e., constructing a sequence
of) update operations that have been applied at the replica. As a first contribution,
we adapt RA-linearizability to the MRDT system model (\S
\ref{sec:lin}), and develop a simple specification framework for MRDTs based on
conflict resolution policy for concurrent update operations. We show that an MRDT
implementation can be linearized only under certain technical constraints on
the conflict resolution policy and if the merge operation satisfies a weaker
notion of commutativity called \emph{conditional commutativity}. By ensuring
that the linearization order obeys the conflict resolution policy for
concurrent update operations and it remains the same across all replicas, we guarantee
both strong eventual consistency and adherence to the user-provided
specification.


Next, we propose a sound but not complete technique for proving
RA-linearizability for MRDT implementations. The main challenge
lies in showing that the $\F{merge}$ function generates a state which is a
linearization of its inputs. We develop a technique called \emph{bottom-up
linearization}, which relies on certain simple algebraic properties of the
$\F{merge}$ function to prove that it generates the correct linearization. We
then design an induction scheme to \emph{automatically} verify the required
algebraic properties of $\F{merge}$ for an arbitrary MRDT implementation. Our
main insight here is to leverage the fact that the merge inputs are themselves
linearizations, and hence, we can use induction over their operation sequences.
We extract a set of verification conditions (VCs) that are amenable to
automated reasoning, and prove that if an MRDT implementation satisfies the
VCs, it is linearizable  (\S \ref{sec:lemmas}). While our development is
focussed on MRDTs, our technique can be directly applied on state-based CRDTs.
State-based CRDTs also have a merge-based system model which is slightly
simpler than MRDTs as the merge function does not require any LCA. 



Finally, we develop a framework in the \fstar~\cite{fstar} programming language
that allows implementing MRDTs and automatically mechanically proving the VCs
required by our technique. The framework provides several advantages over
previous works. First, we require the programmer to specify only the MRDT
operations, the merge function, and the conflict resolution policy, in contrast
to the earlier work that also requires proof constructs such as abstract
simulation relations~\cite{Vimala}. Second, the VCs are simple enough that in
\emph{all} the case studies we have done, including data types such as counter,
set, map, boolean flag, and list, they are automatically discharged by \fstar.
Finally, we extract the verified implementations to OCaml using the \fstar
extraction pipeline and run them (\S \ref{sec:results}). We have also
implemented and verified a few state-based CRDTs using our framework. In the next
section, we present the main ideas of our work informally through a series of
examples.

\section{Overview}
\label{sec:overview}

\subsection{System Model}
\label{model}

The MRDT system model resembles a distributed version control system, such as
Git~\cite{Git}, with replication centred around versioned states in branches
and explicit merges.  A replicated data store handles multiple objects
independently \cite{Riak,Irmin}; in our presentation, we focus on modeling a
store with a single object. The state of the object is replicated across
multiple replicas $r_{1}, r_{2},\ldots \in \mathcal{R}$ in the store. 
Clients interact with the store by performing query or update operations 
on one of the replicas, with update operations modifying its state.
These replicas operate concurrently, allowing
independent modifications without synchronization. They periodically (and
non-deterministically) exchange updates with each other through a process
called \emph{merge}. Due to concurrent operations happening at multiple
replicas, conflicts may arise, which must be resolved by the merge operation in
an appropriate and consistent manner.
An object has a type $\tau \in Type$, whose type signature  $\langle O_\tau, Q_\tau, Val_\tau \rangle$
contains the set of supported update operations $O_{\tau}$, 
query operations $Q_{\tau}$ and 
their return values $Val_{\tau}$.

\begin{definition}
An MRDT implementation for a data type $\tau$ is a tuple $\mathcal{D}_{\tau}$ =
$\langle \Sigma, \sigma_0, \F{do}, \F{merge}, \F{query}, \\
\F{rc} \rangle$, where:
\begin{itemize}
	\item $\Sigma$ is the set of states, $\sigma_0 \in \Sigma$ is the initial state.
	\item $\F{do}$ : $\Sigma \times \mathcal{T} \times \mathcal{R} \times O_{\tau} \rightarrow \Sigma$
			implements all update operations in $O_\tau$, 
			where $\mathcal{T}$ is the set of timestamps.
	\item $\F{merge}$ : $\Sigma \times \Sigma \times \Sigma \rightarrow \Sigma$
				is a three-way merge function.
	\item $\F{query}$: $\Sigma \times Q_{\tau} \rightarrow Val_{\tau}$ implements all query operations in $Q_\tau$, returning a value in $Val_\tau$.
	\item $\F{rc} \subseteq O_{\tau} \times O_{\tau}$  is the conflict resolution policy to be followed for
			concurrent update operations.
\end{itemize}
\end{definition}

An MRDT $\mathcal{D}_\tau$ provides implementations of $\F{do}, \F{merge}$ and $\F{query}$ which will be invoked by the data store appropriately.
A client request to perform an update operation
$o \in O_{\tau}$ at a replica $r$ triggers the call $\F{do}(\sigma,t,r,o)$. 
This takes as input the current state $\sigma \in \Sigma$ of $r$, a unique timestamp $t \in \mathcal{T}$
and produces an updated
state which is then installed at $r$. The data store ensures that timestamps are unique across all operations
(which can be achieved through e.g. Lamport timestamps \cite{Lamport78}).

Replicas can also receive states from other replicas, which are merged with the
receiver's state using \emph{$\F{merge}$}. The \emph{$\F{merge}$}
function is called with the current states of both the sender and receiver
replicas and their lowest common ancestor (LCA), which represents the most
recent common state from which the two replicas diverged. 
Clients can query the state of the MRDT using the $\F{query}$ method.
This takes a MRDT state $\sigma \in \Sigma$ and 
a query operation as input and produces a return value. Note that a query operation cannot change the state at a replica.

\begin{wrapfigure}{r}{0.36\textwidth}
	\footnotesize
	\begin{algorithmic} [1]
		\State $\Sigma = \mathbb{N}$
		\State $O = \{\F{inc} \}$
		\State $Q = \{\F{rd} \}$
		\State $\sigma_0 = 0$
		\State $\F{do}(\sigma, \_, \_,\F{inc}) = \sigma+1$
		\State $\F{merge}(\sigma_\top,\sigma_1,\sigma_2) = \sigma_1 + \sigma_2 - \sigma_\top$
		\State $\F{query}(\sigma,rd) = \sigma$
		\State $\F{rc} = \emptyset$
	\end{algorithmic}
	\vspace{-0.5em}
	\caption{Counter MRDT implementation}
	\vspace{-1.5em}
	\label{fig:counter_impl}
\end{wrapfigure}

While merging, it may
happen that conflicting update operations may have been performed on the two states,
in which case, the implementation also provides a conflict resolution policy
$\F{rc}$.  The merge function should make sure that this policy is followed
while computing the merged state. To illustrate, we now present a couple of
MRDT implementations: an increment-only counter and an observed-remove set.

The counter MRDT implementation is given in Fig. \ref{fig:counter_impl}. The
state space of the counter MRDT is simply the set of natural numbers, and it
allows clients to perform only one update operation ($\F{inc}$) which increments the
value of the counter. For merging two counter states $\sigma_1$ and $\sigma_2$,
whose lowest common ancestor is $\sigma_\top$, intuitively, we want to find the
total number of increment operations across $\sigma_1$ and $\sigma_2$. Since
$\sigma_\top$ already accounts for the effect of the common increments in
$\sigma_1$ and $\sigma_2$, we need to count the newer increments and then add
them to $\sigma_\top$. This is achieved by adding $\sigma_1 - \sigma_\top$ and
$\sigma_2 - \sigma_\top$ to $\sigma_\top$, which simplifies to the merge definition
in Fig. \ref{fig:counter_impl}. For example, suppose we have replicas $r_1$ and
$r_2$ whose initial state was $\sigma_\top = 2$. Now, if there are 2 $\F{inc}$
operations at $r_1$ and 3 $\F{inc}$ operation at $r_2$, their states will be
$\sigma_1 = 4$ and $\sigma_2 = 5$. On merging $r_2$ at $r_1$,
$\F{merge}(\sigma_\top,\sigma_1,\sigma_2)$ will return 7, which reflects the total
number of increments. The $\F{query}$ method simply returns the 
current state of the counter. Finally, the increment operation commutes with itself, so
there is no need to define a conflict resolution policy.

\begin{wrapfigure}{l}{0.4\textwidth}
\footnotesize 
\begin{algorithmic} [1]
	\State $\Sigma = \mathcal{P}(\mathbb{E} \times \mathcal{T})$
	\State $O = \{\F{add}_a, \F{rem}_a \mid a \in \mathbb{E}\}$
	\State $Q = \{\F{rd} \}$
	\State $\sigma_0 = \{\}$
	\State ${\F{do}(\sigma,t,\_,\F{add}_{a}}) = \sigma\cup\{(a,t)\}$
	\State ${\F{do}(\sigma,\_,\_,\F{rem}_{a}}) = \sigma\textbackslash\{(a,i) \mid (a,i)\in \sigma\}$
	\State $\F{merge}(\sigma_\top,\sigma_1,\sigma_2) =$
	\Statex $ \quad (\sigma_\top \cap \sigma_1 \cap \sigma_2) \cup (\sigma_1 \textbackslash \sigma_\top) \cup (\sigma_2 \textbackslash \sigma_\top)$
	\State $\F{query}(\sigma,rd) = \{a \mid (a,\_) \in \sigma\}$
	\State $\F{rc} = \{(\F{rem}_a, \F{add}_a)\ \mid\ a \in \mathbb{E}\}$					
\end{algorithmic}
\caption{OR-set MRDT implementation}
\vspace{-1em}
\label{fig:orset_impl}
\end{wrapfigure}

An observed-remove set (OR-set) \cite{Shapiro} is an implementation of a set
data type that employs an add-wins conflict-resolution strategy, prioritizing
addition in cases of concurrent addition and removal of the same element.
Fig.~\ref{fig:orset_impl} shows the OR-set MRDT implementation. This
implementation is quite similar to the operation-based (op-based) CRDT
implementation of OR-set \cite{shapiro2011comprehensive}.  The state of the
OR-set is a set of element-timestamp pairs, with the initial state being an
empty set.  Clients can perform two operations for every element $a \in
\mathbb{E}$: $\F{add}_a$ and $\F{rem}_a$. The $\F{add}_a$ method adds the
element $a$ along with the (unique) timestamp at which the operation was
performed.  The $\F{rem}_a$ method removes all entries in the set corresponding
to the element $a$. An element $a$ is considered to be present in the set if
there is some $(a,t)$ in the state.

The \emph{$\F{merge}$} method takes as input the LCA set $\sigma_\top$ and the two
sets $\sigma_1$ and $\sigma_2$ to be merged, retains elements of $\sigma_\top$
that were not removed in both sets, and includes the newly added elements from
both sets. Since $\sigma_\top$ is the most recent state from which the two sets
diverged, the intersection of all three sets is the set of elements that were
not removed from $\sigma_\top$ in either branch, while the difference of either
set with the $\sigma_\top$ corresponds to the newly added elements.
The $\F{query}$ operation $rd$ returns all the elements in the set. The conflict
resolution relation $\F{rc}$ orders $\F{rem}_a$ before $\F{add}_a$ of the same
element in order to achieve the add-wins semantics. Note that all other pairs of
operations ($\F{add}\_$ and $\F{add}\_$, $\F{rem}\_$ and $\F{rem}\_$, 
and $\F{add}_x$ and $\F{rem}_y$ with $x \neq y$) commute with each other, 
hence $\F{rc}$ does not specify their ordering. We now consider whether 
the merge operation adheres to the conflict resolution policy.

\subsection{RA-linearizability for MRDTs}
\label{subsec:lin}
We would like to verify that an MRDT implementation is correct, in the sense
that in every execution, (a) replicas which have observed the same set of
update operations converge to the same state, and (b) this state reflects the
semantics of the implemented data type and the conflict resolution policy. 
Note that an update operation $o$ is considered to be visible to a replica
$r$ either if $o$ is directly applied by a client at $r$, or indirectly through
merge with another replica $r'$ on which $o$ was visible. To specify MRDT
correctness, we propose to use the notion of RA-linearizability
\cite{Wang}: the state at any replica during any execution must be achievable
by applying a sequence (or linearization) of the update operations visible to the
replica.  Further, this linearization should obey the user-specified conflict
resolution policy for concurrent operations, and the local replica order for
non-concurrent operations.

Our definition of RA-linearizability allows viewing the state of an MRDT replica as a sequence of update operations applied on the initial state, thus abstracting over the merge function and how it handles concurrent operations. Consequently, any formal reasoning (e.g. assertion checking, functional correctness, equivalence checking etc.) can now essentially forget about the presence of merges, and only focus on update operations, with the additional guarantee that operations would have been correctly linearized, taking into account the conflict resolution policy and local replica ordering.

Proving RA-linearizability for MRDTs is straightforward when there is only a
single replica on which all operations are performed, since there is no
interleaving among operations on a single replica. Complexity arises when update
operations happen concurrently across replicas, which are then merged. For a
merge operation, we need to show that the output can be obtained by applying a
linearization of update operations witnessed by both replicas being merged. However,
the states being merged would have been obtained after an arbitrary number of update
operations or even other merges. Further, the MRDT framework maintains only the
states, but not the update operations leading to those states, thus requiring the
verification technique to somehow infer the update operations leading to a state, and
then show that merge constructs the correct linearization.

We break down this difficult problem gradually with a series of observations.
We will start with an intuitively correct approach, show how it could be broken
through examples, and gradually refine it to make it work. As a starting point,
we first observe that we can leverage the following algebraic properties of the
MRDT update operations and the merge function: (i) commutativity of merge and
update operations, (ii) commutativity of merge, (iii) idempotence of merge, and (iv)
commutativity of update operations. To motivate this, we first introduce some
terminology. An event $e = \langle t,r,o \rangle$ is generated for every update
operation instance, where $t$ is the event's timestamp and $r$ is the replica
on which the update operation $o$ is applied.  Applying an event $e$ on a replica with
state $\sigma$ changes the replica state to $e(\sigma) = \F{do}(\sigma,t,r,o)$ using
the implementation of the operation $o$. Given a sequence of events $\pi = e_1
e_2 \ldots e_n$, we use the notation $\pi(\sigma)$ to denote
$e_n(\ldots(e_2(e_1(\sigma))))$. Now, the properties described above can be
formally defined as follows (forall $ \sigma_\top, \sigma_1, \sigma_2, e, e'$):

{\small
	\begin{enumerate}
	\item[(P1)] $\F{merge}(\sigma_\top,e(\sigma_1), \sigma_2) = e(\F{merge}(\sigma_\top, \sigma_1, \sigma_2 ))$
	\item[(P2)]$\F{merge}(\sigma_\top,\sigma_1, \sigma_2) = \F{merge}(\sigma_\top, \sigma_2, \sigma_1)$
	\item[(P3)] $\F{merge}(\sigma_\top, \sigma_\top, \sigma_\top) = \sigma_\top$
	\item[(P4)] $e(e'(\sigma)) = e'(e(\sigma))$
\end{enumerate}}

As per our proposed definition of RA-linearizability, we need to show that there exists a linearization of events visible at the replica such that the state of the replica can be obtained by applying this linearization. As mentioned earlier, an event can become visible at a replica either by a direct client application, or
by merging with another replica. To illustrate this, consider the scenario
\begin{wrapfigure}{r}{0.3\textwidth}
	\vspace{-1em}
	\begin{center}
		\includegraphics[angle=0, width=0.8\linewidth]{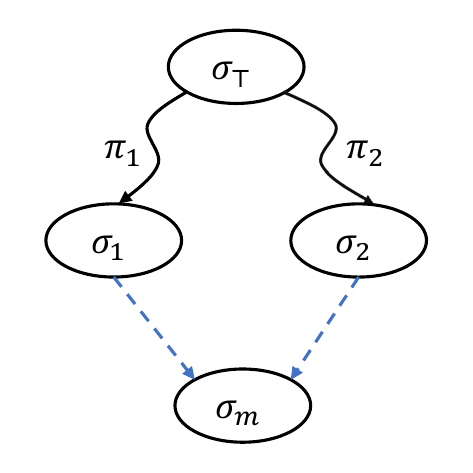}
	\end{center}
	\vspace{-1em}
	\caption{Linearizing a merge operation}
	\vspace{-1em}
	\label{fig:merge-lin}
\end{wrapfigure}
shown in Fig. \ref{fig:merge-lin} where two replicas with states $\sigma_1$ and
$\sigma_2$ are being merged. These states were obtained by applying a sequence
of events $\pi_1$ and $\pi_2$ respectively on the LCA state $\sigma_\top$. We call
the events in $\pi_1$ and $\pi_2$ as local to their respective replicas. Now,
when the two states are merged to create a new state $\sigma_m$ we would need
to show that the state $\sigma_m$ ($= \F{merge}(\sigma_\top, \sigma_1, \sigma_2)$)
can be obtained by linearizing all the events in $\pi_1$ and $\pi_2$, and
applying this linearization on the state $\sigma_\top$.

To show that the merge function constructs a linearization, we can take advantage of
properties (P1)-(P4). In particular, commutativity of merge and update operation
application (P1) allows us to move an event from the second argument of merge
to outside, and we can then repeatedly apply this property to peel off all the
events in $\pi_1$. More formally, by performing induction on the sequence
$\pi_1$ and using (P1), we can show that $\F{merge}(\sigma_\top,\pi_1(\sigma_\top),
\sigma_2 ) = \pi_1(\F{merge}(\sigma_\top, \sigma_\top, \sigma_2))$.  We can then use
commutativity of merge (P2) to swap the last two arguments of merge, and then
apply (P1) again to peel off all the events in $\pi_2$, thus establishing that
$\F{merge}(\sigma_\top, \sigma_\top, \pi_2(\sigma_\top)) =
\pi_2(\F{merge}(\sigma_\top,\sigma_\top, \sigma_\top))$. Finally, using merge
idempotence (P3), and combining all the previous results, we can infer that
$\F{merge}(\sigma_\top,\sigma_1, \sigma_2) = \pi_2(\pi_1(\sigma_\top))$.
Commutativity of update operations (P4) ensures that all linearizations of events in
$\pi_1$ and $\pi_2$ lead to the same state, thus ratifying the specific
linearization order $\pi_1 \pi_2$ that we constructed using properties P1-P3.
We call this process as bottom-up linearization, since we built the sequence
from end through property (P1), linearizing one event at a time.

\begin{wrapfigure}{r}{0.35\textwidth}
	\vspace{-1.8em}
	\begin{center}
		\includegraphics[angle=0, width=1\linewidth]{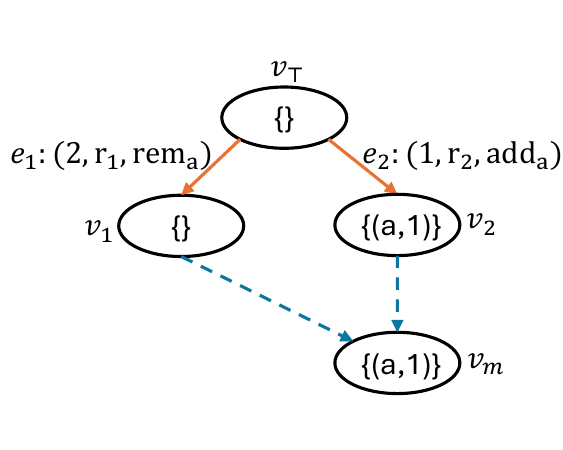}
	\end{center}
	\vspace{-1.5em}
	\caption{OR-set execution}
	\label{fig:orset_exec}
\end{wrapfigure}

It is also easy to see that the counter MRDT implementation in Fig.
\ref{fig:counter_impl} satisfies (P1)-(P4). In particular, commutativity of
integer addition and subtraction essentially gives us (P1)-(P4) for free. While
this strategy works for the counter MRDT, commutativity of all update operations is in
general a very strong requirement, and would fail for other
datatypes. For example, the OR-set MRDT of Fig. \ref{fig:orset_impl}
does not satisfy (P4), as the $\F{add}_a$ and $\F{rem}_a$ operations do not
commute.

In the presence of non-commutative update operations, the property (P1) now needs to
be altered, as we need to consider the conflict resolution policy
to decide the replica from which an event needs to be peeled off. To
illustrate this, consider an OR-set execution depicted in
Fig.~\ref{fig:orset_exec}. We show the version graph of the execution, where
each oval represents a version. The state of the version is depicted inside the
oval. The versions $v_1$ and $v_2$ are obtained by applying $\F{rem}_a$  and
$\F{add}_a$ operations to the version $v_\top$ on two different replicas ($r_1$
and $r_2$). Each edge is labeled with the event corresponding to the
application of an operation. Let $\sigma_\top = \{\}$ denote the state of the LCA
$v_\top$. The versions $v_1$ and $v_2$ are then merged at $r_2$ which gives rise
to a new version $v_m$ with state $\F{merge}(\sigma_\top,e_1(\sigma_\top),
e_2(\sigma_\top))$. Now, since $e_1$ and $e_2$ do not commute, the conflict
resolution policy of OR-set places $e_1$ (i.e. the remove operation)
before $e_2$ (i.e. the add operation). Hence, we want the merged version to
follow the linearization order $e_2(e_1(\sigma_\top))$. This requires us to
first peel off the event $e_2$ from the third argument of $\F{merge}$. To
achieve this, we can alter the property (P1) by making it aware of the conflict
resolution policy as follows:

\begin{itemize}
	\item[(P1$'$)]
		$(e_1,e_2) \in \F{rc} \implies
			\F{merge}(\sigma_\top, e_1(\sigma_1), e_2(\sigma_2)) =
			e_2(\F{merge}(\sigma_\top, e_1(\sigma_1), \sigma_2 ))$\footnote{Note that we are abusing the $\F{rc}$ notation slightly, since $\F{rc}$ is a
		relation over operations $O$, but we are considering it over operation
		instances (i.e. events)}
\end{itemize}

Property (P1$'$) would then allow us to establish the
required linearization order. Property (P4) also needs to be altered due to the presence of non-commutative update operations. 
We modify (P4) to enforce commutativity for non-$\F{rc}$ related events, which gives us flexibility to include such events in any order while constructing the linearization sequence:

\begin{itemize}
	\item[(P4$'$)] $(e_1,e_2) \notin \F{rc} \wedge (e_2,e_1) \notin \F{rc}  \implies e_1(e_2(\sigma)) = e_2(e_1(\sigma))$
\end{itemize}

However, we now face another major challenge: proving (P1$'$) for the OR-set
MRDT. For the counter MRDT, the operations and merge function used integer
addition and subtraction, which commute with each other. But for the OR-set,
$\F{add}_a$ uses set union, while $\F{merge}$ uses set difference and
intersection, which do not commute in general.  Hence, (P1$'$) does not hold
for arbitrary $\sigma_\top,\sigma_1,\sigma_2$.

To illustrate this concretely, consider the same execution of
Fig.~\ref{fig:orset_exec}, except assume that the state $\sigma_\top$ of the LCA
$v_\top$ is $\{(a,1)\}$. Let us try to establish (P1$'$) for the merge of versions
$v_1$ and $v_2$. First, note that as per the OR-set $\F{rc}$, the antecedent of
(P1$'$) is satisfied, as $(e_1,e_2) \in \F{rc}$. Now, the RHS in the consequent
must contain the tuple $(a,1)$, since the event $e_2$ adds $(a,1)$ to the
result of the merge. Does the LHS also contain $(a,1)$? Expanding the
definition of merge in the LHS, $(a,1)$ will not be present in $(\sigma_\top
\cap e_1(\sigma_\top) \cap e_2(\sigma_\top))$ (because $(a,1) \notin e_1(\sigma_\top)$,
as $e_1$ removes $a$). Similarly, since $(a,1)$ is in $\sigma_\top$, it will not be
present in $e_2(\sigma_\top) \setminus \sigma_\top$. It will not be in $e_1(\sigma_\top)
\setminus \sigma_\top$, as $e_1$ removes $a$. To conclude, $(a,1)$ will not be
present in the LHS, thus invalidating the consequent of (P1$'$).

However, we note that this particular execution is actually spurious, because
the tuple $(a,1)$ in the LCA could only have been added by another $\F{add}_a$
operation whose timestamp is the same as $e_2$. But this is not possible as the
data store ensures that timestamps are unique across all events. In the general
case, we would not be able to show (P1$'$) for OR-set because the tuple $(a,t)$
being added by the $\F{add}_a$ operation (event $e_2$) could also be present in
the LCA state. However, this situation cannot occur.

Thus, it is possible to show (P1$'$) for all \emph{feasible} states $\sigma_\top,
\sigma_1, \sigma_2$ that may occur during an actual execution. In
the case of OR-set, there are two arguments which are required to infer this:
(i) timestamps are unique across all events and (ii) if a tuple $(a,t)$ is
present in the state $\sigma$, then there must have been an $\F{add}_a$
operation with timestamp $t$ in the history of events leading to $\sigma$.
While the first argument is a property of the data store, the second argument
is an invariant linking a state with the history of events leading to that
state. Such arguments are in general hard to infer, and would also change
across different MRDTs. We now present our second major observation which
allows us to automatically verify (P1$'$) for feasible states without requiring
invariants like argument (ii) linking MRDT states and events.

\subsection{Verification using Induction on Event Sequences}
\label{subsec:induction}
In order to show property (P1$'$) for an MRDT implementation, we need to consider the feasible states which would be given as input to the merge function during an actual execution. We observe that we can leverage the RA-linearizability of the MRDT implementation, and hence characterize these feasible states by sequences of MRDT update operations (more precisely, events corresponding to update operation instances). We can now use induction over these sequences to establish property (P1$'$). Note that the input states to merge may themselves have been obtained through prior merges, but we can inductively assume that these prior merges resulted in correct linearizations. Since merge takes as input three states ($\sigma_\top, \sigma_1, \sigma_2$), we need to consider three sequences that led to these states and induct on all the three separately.

\begin{wrapfigure}{r}{0.28\textwidth}
	\vspace{-1.5em}
	\begin{center}
		\includegraphics[angle=0, width=0.9\linewidth]{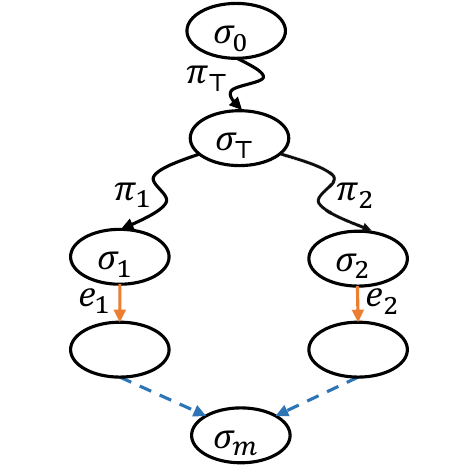}
	\end{center}
	\vspace{-1em}
	\caption{Induction on event sequences}
	\vspace{-1em}
	\label{fig:induction}
\end{wrapfigure}

Concretely, let $\pi_\top$ be a sequence of events which when applied on the
initial MRDT state $\sigma_0$ results in the state $\sigma_\top$. Since the LCA
state always contains events which are common to the states $\sigma_1$ and
$\sigma_2$, $\pi_\top$ will be the common prefix of the sequences leading to both
$\sigma_1$ and $\sigma_2$. We consider the
sequences $\pi_1$ and $\pi_2$ that consist of the local events which when
applied on $\sigma_\top$ led to $\sigma_1$ and $\sigma_2$ respectively. Fig.
\ref{fig:induction} depicts the situation. Notice that the last two events on
each replica before the merge are fixed to be $e_1$ and $e_2$, which would be
related by the $\F{rc}$ relation, as per the requirement of property (P1$'$).

\vspace{-2em}
\begin{align}
	& \F{merge} (\sigma_0, e_1(\sigma_0), e_2(\sigma_0)) = e_2 (\F{merge} (\sigma_0, e_1(\sigma_0), \sigma_0)) \label{eq:1}\\
	&\F{merge} (\sigma_\top, e_1(\sigma_\top), e_2(\sigma_\top)) = e_2 (\F{merge} (\sigma_\top, e_1(\sigma_\top), \sigma_\top))\nonumber \\
	&\implies \F{merge} (e(\sigma_\top), e_1(e(\sigma_\top)), e_2(e(\sigma_\top))) = e_2 (\F{merge} (e(\sigma_\top), e_1(e(\sigma_\top)), e(\sigma_\top)))  \label{eq:2}
\end{align}
\vspace{-1.5em}

We first induct on the sequence $\pi_\top$ which leads to the state $\sigma_\top$.
For this, we assume that $\pi_1 = \pi_2 = \epsilon$, and hence $\sigma_\top =
\sigma_1 = \sigma_2 = \pi_\top(\sigma_0)$. We also assume the antecedent of
property (P1$'$), i.e. $(e_1,e_2) \in \F{rc}$, and hence our goal is to show
its consequent. For the OR-set, $e_1$ will be a $\F{rem}_a$ event, while $e_2$
will be an $\F{add}_a$ event (say with timestamp $t$).

Eqn. \eqref{eq:1} is the base-case of the induction (where $\pi_\top = \epsilon$),
and this can be now directly discharged since $\sigma_0$ is an empty set, and
hence clearly won$'$t contain $(a,t)$. Eqn. \eqref{eq:2} is the inductive case,
which assumes that (P1$'$) is true for some LCA state $\sigma_\top$, and tries to
prove the property when one more update operation (signified by the event $e$) is
applied on the LCA (and also on both $\sigma_1$ and $\sigma_2$, since LCA
operations are common to both states to be merged). This can also be
automatically discharged with the property that events $e,e_1,e_2$ have
different timestamps. Intuitively, the inductive hypothesis establishes that
$(a,t) \notin \sigma_\top$, and since the timestamp of event $e$ is different from
$e_1$ and $e_2$, it cannot add $(a,t)$ to the LCA, thus preserving the property
that $(a,t) \notin e(\sigma_\top)$, thereby implying the consequent. This
completes the proof for property (P1$'$) for any arbitrary LCA state $\sigma_\top$
that may be feasible in an actual execution. A similar inductive strategy is used for proving property (P1$'$) for feasible states $\sigma_1$ and $\sigma_2$ (more details in \S \ref{sec:lemmas}).

\vspace{-0.5em}
\subsection{Intermediate Merges}
\label{subsec:inter}
In our linearization strategy for merges (given by properties (P1$'$-P4$'$)), we first considered the local update operations of each branch,
linearized them according to the conflict-resolution policy, and then applied
this sequence on the LCA. This effectively orders the update operations that led
to the LCA before the update operations local to each branch.

\begin{wrapfigure}{r}{0.35\textwidth}
	\vspace{-3em}
	\begin{center}
		\includegraphics[angle=0, width=1\linewidth]{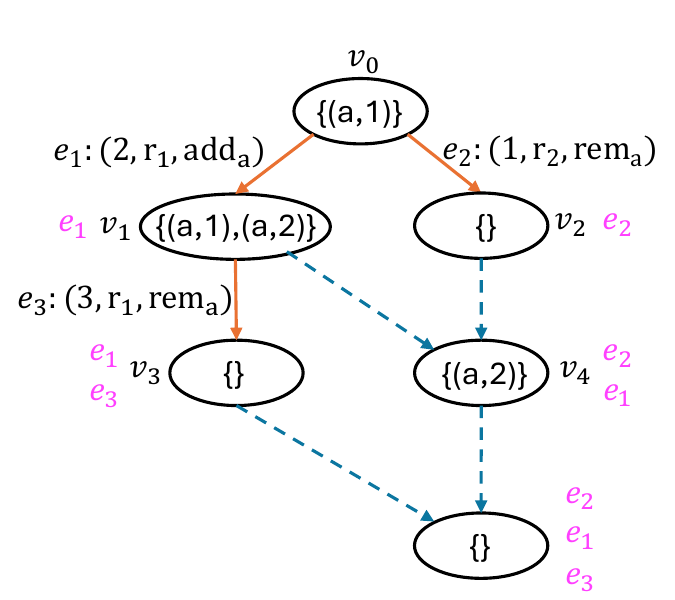}
	\end{center}
	\caption{Intermediate merge}
	\vspace{-1em}
	\label{fig:inter_merge}
\end{wrapfigure}

However, in a Git-based execution model, due to a phenomenon known as
intermediate merges, it may happen that update operations of the LCA may need to be
linearized after update operations local to a branch. To
illustrate this, consider an execution of the OR-set MRDT as shown in Fig.
~\ref{fig:inter_merge}. There are 3 operations and 2 merges being performed in
this execution, with the events $e_1,e_3$ at replica $r_1$ and event $e_2$ at
replica $r_2$.

Instead of merging with the latest version $v_3$ at replica $r_1$, replica
$r_2$ first merges with an intermediate version $v_1$ to generate the version
$v_4$. Next, this version $v_4$ is merged with the latest version $v_3$ of
replica $r_1$. However, note that for this merge, the LCA will be version
$v_1$. This is because the set of events associated with version $v_3$ is
$\{e_1,e_3\}$, while for version $v_4$, it is $\{e_1,e_2\}$. Hence, the set of
common events among both versions would be $\{e_1\}$, which corresponds to the
version $v_1$. Indeed, in the version graph, both $v_1$ and $v_0$ are ancestors
of $v_3$ and $v_4$, but $v_1$ is the lowest common ancestor\footnote{in \S \ref{sec:lin},
we will formally prove that the LCA of two versions according to the version
graph contains the intersection of events in both versions.}.

In Fig. \ref{fig:inter_merge}, we have also provided the linearization of
events associated with each version. Notice that for version $v_4$, which is
obtained through a merge of $v_1$ and $v_2$, the conflict resolution policy of
the OR-set linearizes $e_2$ before $e_1$. Now, for the merge of $v_3$ and
$v_4$, we have a situation where a local event ($e_2$ in $v_4$) needs to be
linearized before an event of the LCA ($e_1$ in $v_1$). This does not fit our
linearization strategy. Let us see why. If we were to try to apply (P1$'$), it would linearize $e_1$ after $e_3$,
since these are the last operations in the two states to be merged and the
conflict resolution policy orders $\F{add}_a (e_1)$ after $\F{rem}_a (e_3)$.
However, in the execution, $e_1$ and $e_3$ are causally related, i.e. $e_1$
occurs before $e_3$ on the same replica, and hence they should be linearized in
that order. Intuitively, property (P1$'$) does not work because it does not consider the possibility that the last event in one replica could be visible to the last event in another replica, and hence the linearization must obey the visibility relation.

In order to handle this situation, we consider another algebraic property (P1-1), which explicitly forces visibility relation among the last events by making one of them part of the LCA:

\begin{itemize}
	\item[(P1-1)] $\F{merge}(e_1(\sigma_0),e_3(\sigma_1), e_1(\sigma_2)) = e_3(\F{merge}(e_1(\sigma_0), \sigma_1, e_1(\sigma_2 )))$
\end{itemize}

Note that events in the LCA are visible to events on both replicas being merged. Hence, by having the same event $e_1$ in both the first and third argument to $\F{merge}$ in the LHS, $e_3$ would have to be linearized after $e_1$ to respect the visibility order, thus over-riding the $\F{rc}$ ordering among them. Property (P1-1) can be directly applied to the execution in Fig.
\ref{fig:inter_merge} for the merge of $v_3$ and $v_4$ (with $\sigma_0$ as the
state of $v_0$, $\sigma_1$ as the state of $v_1$ and $\sigma_2$ as the state of
$v_2$), constructing the correct linearization.

We will revisit the example in Fig. \ref{fig:inter_merge} and properties (P1$'$) and (P1-1) in a more formal setting in \S \ref{sec:lemmas}, renaming them as $\textrm{\sc{BottomUp-2-OP}}$ and $\textrm{\sc{BottomUp-1-OP}}$. We will also identify the conditions under which these properties can guarantee the existence of a correct linearization.
\section{Problem Definition}
\label{sec:lin}

In this section, we formally define the semantics of the replicated data store
on top of which the MRDT implementations operate (\S \ref{subsec:os}),
the notion of RA-linearizability for MRDTs (\S \ref{subsec:lin_def}), and the process of bottom-up linearization (\S \ref{subsec:bottom-up}).

\subsection{Semantics of the Replicated Data Store}
\label{subsec:os}

\begin{figure}[ht]
	\scriptsize
	\raggedright$\textrm{\sc{[CreateBranch]}}$
	\[
	\inferrule{r\in dom (H) \\ r'\notin dom (H) \\  v\notin dom (N) \\
		N' = N [v \mapsto N(H(r))] \\ H' = H[r' \mapsto v] \\
		L' = L [v \mapsto L(H(r))] \\ G' = (dom(N) \cup \{v\}, {E} \cup \{(H(r), v)\})}
	{(N, H, L, G, vis) \xrightarrow{\F{createBranch(r',r)}} (N', H', L', G', vis)\\}
	\]
		\raggedright$\textrm{\sc{[Apply]}}$
	\[
	\inferrule{e = (t,r,o) \\  o \in O_\tau \\  \forall e' \in \bigcup range(L).\ time (e') \neq t  \\ r\in dom (H) \\ v\notin dom (N) \\
		N' = N [v \mapsto \F{do} (N(H(r)), e)] \\
		H' = H [r \mapsto v] \\ L' = L [v \mapsto L(H(r)) \cup \{e\}] \\
		G' = (dom(N'), {E} \cup \{(H(r), v)\}) \\
		vis' = vis \cup (L (H(r)) \times \{e\})}
	{(N, H, L, G, vis) \xrightarrow{\F{apply(t,r,o)}} (N', H', L', G', vis')\\}
	\]

		\raggedright$\textrm{\sc{[Merge]}}$
	\[
	\inferrule{r_{1}, r_{2} \in dom (H) \\ v\notin dom (N)  \\ v_{\top} = LCA (H(r_{1}), H(r_{2})) \\
		N' = N [v \mapsto \F{merge} (N(v_{\top}), N(H(r_{1})), N(H(r_{2}))] \\
		H' = H [r_{1} \mapsto v] \\ L' = L [v \mapsto L(H(r_{1})) \cup L(H(r_{2}))] \\
		G' = (dom(N'), {E} \cup \{(H(r_{1}), v), (H(r_{2}), v)\})}
	{(N, H, L, G, vis) \xrightarrow{\F{merge (r_1, r_2)}} (N', H', L', G', vis)\\}
	\]
	\raggedright$\textrm{\sc{[Query]}}$
	\[
	\inferrule{r\in dom (H) \\ q \in Q_\tau \\ a = \F{query}(N(H(r)),q)} 
	{(N, H, L, G, vis) \xrightarrow{\F{query(r,q,a)}} (N, H, L, G, vis)\\}
	\]
	\vspace{-2em}
	\caption{Semantics of the replicated datastore}
	\vspace{-3.25em}
	\label{fig:sem}
\end{figure}

The semantics of the replicated store defines all possible executions of an
MRDT implementation.  Formally, the semantics are parameterized by an MRDT
implementation $\M{D} = \langle \Sigma, \sigma_0, \F{do}, \F{merge}, \F{query},\\
\F{rc}\rangle$ of type $\tau = \langle O_\tau, Q_\tau, Val_\tau \rangle$ and are represented by a labeled transition system
$\M{S_\M{D}}$ = ($\Phi$, $\rightarrow$). Each configuration in $\Phi$ maintains a set of
versions, where each version is created either by applying an MRDT operation to
an existing version, or by merging two versions. Each replica is associated
with a head version, which is the most recent version seen at the replica.
Formally, each configuration $C$ in $\Phi$ is a tuple $\langle N, H, L, G, vis
\rangle$, where:
 
\begin{itemize}
	\item $N : \F{Version} \rightharpoonup \Sigma$ is a partial function
that maps versions to their states ($\F{Version}$ is the set of all possible
versions). 
	\item $H: \M{R} \rightharpoonup \F{Version}$ is also a partial function
that maps replicas to their head versions. A replica is considered active if it 
is in the domain of $H$ of the configuration.
	\item $L: \F{Version} \rightharpoonup \mathbb{P}(\M{E})$ maps a version to the set of events 
that led to this version. Each event $e \in \M{E}$ is an update
operation instance, uniquely identified by a timestamp value (we define $\M{E}
= \M{T} \times \M{R} \times O$).
	\item $G = (dom(N),E)$ is the version graph, whose
vertices represent the versions in the configuration (i.e. those in the domain of
$N$) and whose edges represent a relationship between different versions (we explain the different types of edges below).
	\item $vis \subseteq \M{E} \times \M{E}$ is a partial order over events.
\end{itemize}

Figure~\ref{fig:sem} gives a formal description of the transition rules. $\textrm{\sc{CreateBranch}}$ forks a new replica $r'$ from an existing replica
$r$, installing a new version $v$ at $r'$ with the same state as the head
version $H(r)$ of $r$, and adding an edge $(H(r),v')$ in the version graph.
$\textrm{\sc{Apply}}$ applies an update operation $o$ on some replica $r$,
generating a new event $e$ with a timestamp different than all events generated
so far. $\bigcup\F{range}(L)$ denotes the set of events witnessed across all versions.
A new version $v$ is also created whose state is obtained by applying
$o$ on the current state of the replica $r$. The version graph is updated by
adding the edge $(H(r),v)$. The $vis$ relation as well as the function
$L$, which tracks events applied at each version, are also updated. In
particular, each event $e'$ already applied at $r$, i.e. $e' \in L(H(r))$, is
made visible to $e$: $(e',e) \in vis$, while $L'(v)$ is obtained by adding $e$
to $L(H(r))$.

$\textrm{\sc{Merge}}$ takes two replicas $r_1$ and $r_2$, applies the $\F{merge}$
function on the states of their head versions to generate a new version $v$, which is
installed as the new head version at $r_1$. Edges are added in the version graph
from the previous head versions of $r_1$ and $r_2$ to $v$. $L(v)$ is obtained by
taking a union of $L(r_1)$ and $L(r_2)$, and there is no change in the visibility
relation. $\textrm{\sc{Query}}$ takes a replica $r$ and a query operation $q$ and applies $q$ to the state at the head version of 
$r$, returning an output value $a$. Note that the $\textrm{\sc{Query}}$ transition 
does not modify the configuration and the return value of the query is stored as part of the transition label. While our operational semantics is based on and inspired by previous works 
\cite{Kaki2022,Vimala}, we note that it is more general and precisely
captures the MRDT system model as opposed to previous
works. In particular, \citet{Kaki2022} place
significant restrictions on the $\textrm{\sc{Merge}}$ transition, disallowing
arbitrary replicas to be merged to ensure that there is a total order on the
merge transitions. While the semantics in the work by \citet{Vimala}
does allow arbitrary merges, it is more abstract and high-level, and does not
even keep track of versions and the version graph. 

\textbf{Notation:} We now introduce some notation that will be used throughout the paper. Given a configuration $C$, we use $X(C)$ to project the component $X$ of $C$. For a relation $R$, we use $x \xrightarrow{R}
y$ to signify that $(x,y) \in R$. We use $R_{\mid S}$ to indicate the relation
as given by $R$ but restricted to elements of the set $S$. Let $R^*$ denote the
reflexive-transitive closure of $R$, and let $R^+$ denote the
transitive closure of $R$. For an event $e$, we use the projection functions
$\F{op}, \F{time}, \F{rep}$ to obtain the update operation, timestamp and replica
resp. For a sequence of events $\pi$, $\pi_{\mid S}(\sigma)$ denotes
application of the sub-sequence of $\pi$ restricted to events in $S$. For a
configuration $C$, we use $e_1 \mid\mid_C e_2$ to denote that $e_1$ and $e_2$
are concurrent, that is $\neg (e_1 \xrightarrow{\F{vis}(C)} e_2 \vee e_2
\xrightarrow{\F{vis}(C)} e_1)$. Given a total order over a set of events $\M{E}$,
represented by a sequence $\pi$, and $\F{lo} \subseteq \M{E} \times \M{E}$, we say that
$\pi$ extends $\F{lo}$ if $\F{lo} \subseteq \pi$. The relation $\F{rc}$ orders
update operations, but for convenience we sometime use it for ordering events, with
the intention that it is actually being applied to the underlying update operations.
We use $e_1 \neq e_2$ to indicate that $ \F{time}(e_1) \neq  \F{time}(e_2)$.

We define the initial configuration of $\M{S_\M{D}}$ as $C_0 =
\langle N_0, H_0, L_0, G_0, \emptyset \rangle$, which consists of only one
replica $r_0$.  Here, $H_0 = [r_0 \mapsto v_0]$, $N_0 = [v_0 \mapsto
\sigma_0]$, where $\sigma_0$ is the initial state as given by
$\M{D_{\tau}}$, while $v_0$ denotes the initial version and $L_0 = [v_0
\mapsto \emptyset]$. The graph $G_0 = (\{v_0\}, \emptyset)$ is the initial
version graph. An execution of $\M{S_D}$ is defined to be a finite sequence of
transitions, $C_0 \xrightarrow{t_1} C_1 \xrightarrow{t_2} C_2 \ldots
\xrightarrow{t_n} C_n$. Note that the label of a transition corresponds to its
type. 
Let $\llbracket \M{S_\M{D}} \rrbracket$ denote the set of all possible
executions of $\M{S_D}$.

Finally, as mentioned earlier, $\F{merge}$ is a ternary function, taking as
input the states of two versions to be merged, and the state of the lowest
common ancestor (LCA) of the two versions.
Version $v_1 \in V$ is defined to be a causal ancestor of version $v_2 \in V$
if and only if  $(v_1, v_2) \in E^*$.

\begin{definition}[LCA]
	Given a version graph $G = (V,E)$ and versions $v_1, v_2 \in V$, $v_\top \in
	V$ is defined to be the lowest common ancestor of $v_1$ and $v_2$ (denoted by
	$LCA(v_1,v_2)$) if (i) $(v_\top,v_1) \in E^*$ and $(v_\top,v_2) \in E^*$,
	(ii) $\forall v \in V. (v,v_1) \in E^* \wedge (v,v_2) \in E^* \implies
	(v,v_\top) \in E^*$.
\end{definition}

Note that the version history graph at any point in any execution is guaranteed
to be acyclic (i.e. a DAG),  and hence the LCA (if it exists) is guaranteed to
be unique. We now present an important property linking the LCA of two versions
with events applied at each version.

\begin{lemma}\label{lem:LCA}
	Given a configuration $C = \langle N,H,L,G,vis \rangle$ reachable in some
	execution $\tau \in \llbracket \M{S_D} \rrbracket$ and two versions $v_1,v_2
	\in dom(N)$, if $v_\top$ is the LCA of $v_1$ and $v_2$ in $G$, then
	$L(v_\top) = L(v_1) \cap L(v_2)$\footnote{All proofs are in the Appendix \S \ref{sec:app}}.
\end{lemma}

\begin{wrapfigure}{r}{0.3\textwidth}
	\vspace{-1.5em}
	\begin{center}
		\includegraphics[angle=0, width=0.8\linewidth]{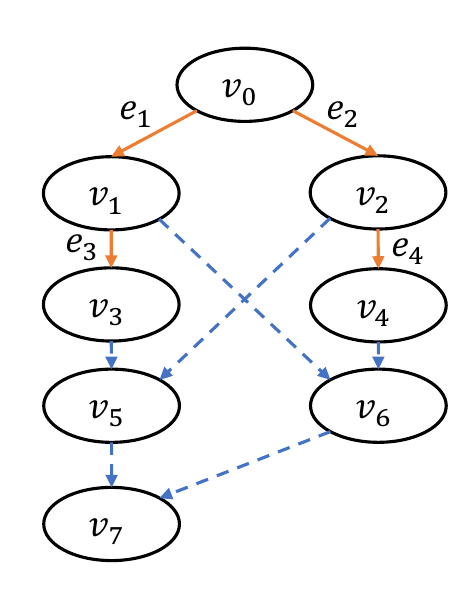}
	\end{center}
	\vspace{-1em}
	\caption{Version Graph with no LCA for $v_5$ and $v_6$}
	\label{fig:LCA}
	\vspace{-1em}
\end{wrapfigure}

Thus, the events of the LCA are exactly those applied at both the versions.
This intuitively corresponds to the fact that $LCA(v_1,v_2)$ is the most recent
version from which the two versions $v_1$ and $v_2$ diverged. Note that it is possible that the LCA may not exist for two versions. Fig.
\ref{fig:LCA} depicts the version graph of such an execution. Vertices with
in-degree 1 (i.e. $v_1,v_2,v_3,v_4$) have been generated by applying a new update
operation (with the orange edges labeled by the corresponding events $e_1,e_2,e_3,e_4$),
while vertices with in-degree 2 have been obtained by merging two other
versions (depicted by blue edges). The merge of $v_1$ and $v_4$ (leading to
$v_6$) has a unique LCA $v_0$, similarly, merge of $v_2$ and $v_3$ (leading to
$v_5$) also has a unique LCA $v_0$. However, if we now want to merge $v_5$ and
$v_6$, both $v_1$ and $v_2$ are ancestors, but there is no LCA. We note that
this execution will actually be prohibited by the semantics of
\citet{Kaki2022}, since the two merges leading to $v_5$ and $v_6$ are
concurrent.

Notice that $L(v_5) = \{e_1,e_2,e_3\}$, while $L(v_6) = \{e_1,e_2,e_4\}$.
Hence, by Lemma~\ref{lem:LCA}, $L(LCA(v_5,v_6)) = \{e_1,e_2\}$, but such a
version is not generated during the execution. To resolve this issue, we introduce the notion of \textit{potential} LCAs. 

\begin{definition}[Potential LCAs]
Given a version graph $G = (V,E)$ and versions $v_1, v_2 \in V$, $v_\top \in
	V$ is defined to be a potential LCA of $v_1$ and $v_2$  if 
	(i) $(v_\top,v_1) \in E^*$ and $(v_\top,v_2) \in E^*$,
	(ii) $\neg (\exists v. (v,v_1) \in E^* \wedge (v,v_2) \in E^* \wedge (v_\top,v) \in E^*)$.
\end{definition}

For merging $v_1$ and $v_2$, we first find all the potential LCAs, and recursively merge them to obtain the actual
LCA state. For the execution in Fig. \ref{fig:LCA}, the potential LCAs of $v_5$ and $v_6$ would be
$v_1$ and $v_2$ (with $L(v_1) = \{e_1\}$ and
$L(v_2) = \{e_2\}$); merging them would get us the actual LCA. 
  In \S \ref{subsec:lcaproof}, we prove that this recursive merge-based strategy is guaranteed to generate the
actual LCA.

\subsection{Replication-aware Linearizability for MRDTs}
\label{subsec:lin_def}

As mentioned in \S \ref{sec:overview}, our goal is to show that the state of every version $v$
generated during an execution is a linearization of the events in $L(v)$. We
use the notation $\F{lo}$ to indicate the linearization relation, which is a
binary relation over events. For an execution in $\M{S}_\M{D}$, we
want $\F{lo}$ between the events of the execution to satisfy certain desirable
properties: (i) $\F{lo}$ between two events should not change during an execution, (ii) $\F{lo}$ should obey the conflict resolution policy
for concurrent events and (iii) $\F{lo}$ should obey the replica-local
$\F{vis}$ ordering for non-concurrent events. This would ensure that two
versions which have observed the same set of events will have the same state (i.e. \textit{strong eventual consistency}), and this state would
also be a linearization of update operations of the data type satisfying the
conflict resolution policy.

While the $\F{lo}$ relation in classical linearizability literature is
typically a total order, in our context, we take advantage of commutativity
of update operations, and only define $\F{lo}$ over non-commutative events. As we
will see later, this flexibility allows us to have different sequences of
events which extend the same $\F{lo}$ relation between non-commutative events, and hence are guaranteed
to lead to the same state. We use the notation $e \rightleftarrows e'$ to
indicate that events $e$ and $e'$ commute with each other. Formally, this means
that $\forall \sigma.\;e(e'(\sigma)) = e'(e(\sigma))$. Two update operations
$o,o'$ commute if $\forall e,e'.\;\F{op}(e) = o \wedge \F{op}(e') = o'
\implies e \rightleftarrows e'$.  As mentioned earlier, the $\F{rc}$ relation
is also only defined between non-commutative update operations.

\begin{lemma}\label{lem:non-comm}
	Given a set of events $\M{E}$, if $\F{lo} \subseteq \M{E} \times \M{E}$ is defined over
	every pair of non-commutative events in $\M{E}$, then for any two sequences
	$\pi_1, \pi_2$ which extend $\F{lo}$, for any state $\sigma$, $\pi_1(\sigma)
	= \pi_2(\sigma)$.
\end{lemma}

Given a configuration $C = \langle N, H, L, G, vis \rangle$, let $\M{E}_C = \bigcup
\F{range}(L(C))$ denote the set of events witnessed across all versions in C.
Then, our goal is to define an appropriate linearization relation $\F{lo}_C
\subseteq \M{E}_C \times \M{E}_C$, which adheres to the $\F{rc}$ relation for concurrent
events, the $\F{vis}$ relation for non-concurrent events, and for every version $v
\in dom(N)$, $N(v)$ should be obtained by sequentializing the events in $L(v)$,
with the sequence extending $\F{lo}_C$. Note that this requires $\F{lo}^+$ to
be irreflexive\footnote{$\F{lo}$ need not be transitive, as we only want to
define $\F{lo}$ between non-commutative events, and non-commutativity is not a
transitive property}.

\begin{wrapfigure}{r}{0.3\textwidth}
	\vspace{-1em}
	\begin{center}
		\includegraphics[angle=0, width=0.8\linewidth]{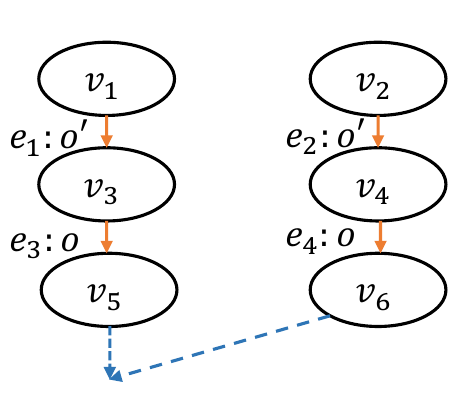}
	\end{center}
	\vspace{-1em}
	\caption{Example demonstrating cycle in $\F{lo}$}
	\vspace{-1.5em}
	\label{fig:conditional-commutativity}
\end{wrapfigure}

We now demonstrate that an $\F{lo}$ relation with all the desirable properties
may not exist for all executions. Suppose there are MRDT update operations $o,o'$ such
that $o \xrightarrow{\F{rc}} o'$. Fig. \ref{fig:conditional-commutativity}
contains a part of the version graph generated during some execution,
containing two instances of both $o$ and $o'$. We use $e_i:o_i$ to denote that
event $\F{op}(e_i) = o_i$. Notice that $e_1$ and $e_4$,
$e_2$ and $e_3$ are concurrent, while $e_1$ and $e_3$, $e_2$ and $e_4$ are
non-concurrent. Applying the $\F{rc}$ ordering on concurrent events, we would
want $e_3 \xrightarrow{\F{lo}} e_2$ and $e_4 \xrightarrow{\F{lo}} e_1$, while
applying $\F{vis}$ ordering, we would want $e_1 \xrightarrow{\F{lo}} e_3$ and
$e_2 \xrightarrow{\F{lo}} e_4$. However, this results in a $\F{lo}$-cycle, thus
making it impossible to construct a sequence of update operations for the merge of
$v_5$ and $v_6$, which adheres to the $\F{lo}$ ordering.

Notice that the above execution only requires the $\F{rc}$ relation to be
non-empty (i.e. there should exist some $(o,o') \in \F{rc}$). If the $\F{rc}$
relation is empty, then all update operations would commute with each other, and hence
the $\F{lo}$ relation would also be empty. If $\F{rc}$ is non-empty, $\F{rc}^+$
should be irreflexive to ensure irreflexivity of $\F{lo}^+$. Note that
$\F{rc}^+$ being irreflexive means that for any MRDT update operation $o$, $(o,o)
\notin \F{rc}$, and hence $o$ must commute with itself, since $\F{rc}$ relation
is defined for all pairs of non-commutative update operations. Furthermore, Fig.
\ref{fig:conditional-commutativity} shows that even if $\F{rc}^+$ is
irreflexive, it may still not be possible to construct an $\F{lo}$ relation
which can be extended to a total order and which adheres to the $\F{rc}$ relation
between all pairs of concurrent events. To ensure existence of an $\F{lo}$
relation such that $\F{lo}^+$ is irreflexive when $\F{rc}^+$ is irreflexive, we define it as follows:

\begin{definition}[Linearization relation]\label{def:lin-relation}
	Let $C$ be a configuration reachable in some execution in $\llbracket \M{S_D}
	\rrbracket$. Let $\M{E}_C$ be the set of events in $C$. Then, $\F{lo_C}$ is defined as:
	 \begin{align*}
		\forall e_1,e_2 \in \M{E}_C.\ e_1 \xrightarrow{\F{lo_C}} e_2  \Leftrightarrow &
		(e_1 \xrightarrow{\F{vis(C)}} e_2 \wedge \neg e_1 \rightleftarrows e_2) \\
		& \vee (e_1 \mid\mid_C e_2 \wedge e_1 \xrightarrow{\F{rc}} e_2 \wedge
		\neg(\exists e_3 \in \M{E}.\ e_2 \xrightarrow{\F{vis(C)}} e_3 \wedge \neg e_2
		\rightleftarrows e_3 ))
	\end{align*}
\end{definition}

$\F{lo}_C$ follows the visibility relation only between non-commutative
events. For concurrent non-commutative events $e_1$ and $e_2$ with $e_1
\xrightarrow{\F{rc}} e_2$, $\F{lo}_C$ follows the $\F{rc}$ relation only if
there is no event $e_3$ such that $e_2$ is visible to $e_3$ and $e_2$ does not commute with $e_3$. Applying this
definition to the execution in Fig. \ref{fig:conditional-commutativity}, for
the configuration obtained after merge, we would have neither $e_4
\xrightarrow{\F{lo}} e_1$, nor $e_3 \xrightarrow{\F{lo}} e_2$, thus avoiding
the cycle in $\F{lo}$.

\begin{lemma}\label{lem:irreflexive}
For an MRDT $\M{D}$ such that $\F{rc}^+$ is irreflexive, for any configuration $C$ reachable in
	$\M{S}_\M{D}$, $\F{lo}_C^+$ is irreflexive.
\end{lemma}

Going forward, we will assume that $\F{rc}^+$ is irreflexive for any MRDT $\M{D}$. 
We note that restricting $\F{lo}$ to not always obey the $\F{rc}$ relation by considering 
non-commutative update operations happening locally (and thus related by $\F{vis}$) is also
sensible from a practical perspective. For example, in the case of OR-set, even
though we have $\F{rem}_a \xrightarrow{\F{rc}} \F{add}_a$, if $\F{add}_a$ is
locally followed by another $\F{rem}_a$, it does not make sense to order a
concurrent $\F{rem}_a$ event before the $\F{add}_a$ event. More generally, if
an event $e_2$ is visible to another event $e_3$ with which it does not commute,
then $e_2$ is effectively "overwritten" by $e_3$, and hence there is no need to
linearize a concurrent event $e_1$ before $e_2$.

While $\F{lo}_C$ is now guaranteed to be irreflexive for any configuration $C$,
and hence can be extended to a sequence, it now no longer enforces an ordering
among all non-commutative pairs of events. Thus, there could exist sequences
$\pi_1,\pi_2$ extending an $\F{lo}_C$ relation which may contain a pair of
non-commutative events in different orders. For example, in Fig.
\ref{fig:conditional-commutativity}, for the configuration $C$ obtained after
the merge, $\F{lo}_C = \{(e_1,e_3), (e_2,e_4)\}$, resulting in sequences $\pi_1
= e_1 e_2 e_3 e_4$ and $\pi_2 = e_1 e_3 e_2 e_4$ which both extend $\F{lo}_C$,
but contain the non-commutative events $e_2$ and $e_3$ in different orders.
Thus, Lemma \ref{lem:non-comm} can no longer be applied, and it is not
guaranteed that $\pi_1$ and $\pi_2$ would lead to the same state. Notice that
in the sequences $\pi_1$ and $\pi_2$ above, even though $e_2$ and $e_3$ appear
in different orders, $e_4$ always appears after both. Indeed, $e_4$ must appear
after $e_2$ due to visibility relation, and since $e_3$ and $e_4$ commute with
each other (since both correspond to the same operation $o$), it is enough to
consider sequences where $e_4$ appears after $e_3$. Based on the above
observation, we now introduce a notion called conditional commutativity to
ensure that sequences such as $\pi_1,\pi_2$ would lead to the same state:

\begin{definition}[Conditional Commutativity]
	Events $e$ and $e'$ are said to conditionally commute with respect to event
	$e''$ (denoted by $e \overset{e''}{\rightleftarrows} e'$) if $\forall \sigma
	\in \Sigma.\ \forall \pi \in \M{E}^*.\ e''(\pi(e(e'(\sigma)))) =
	e''(\pi(e'(e(\sigma))))$.
\end{definition}

Update operations $o$ and $o'$ conditionally commute w.r.t. update operation $o''$ if
$\forall e,e',e''. \F{op}(e) = o \wedge \F{op}(e') = o' \wedge \F{op}(e'') =
o'' \Rightarrow e \overset{e''}{\rightleftarrows} e'$. For example, for the
OR-set MRDT of Fig. \ref{fig:orset_impl}, $\F{add}_a
\overset{\F{rem}_a}{\rightleftarrows} \F{rem}_a$. Even though \textit{add} and
\textit{remove} operations of the same element do not commute with each other,
if there is guaranteed to be a future \textit{remove} operation, then they do
commute. For the execution in Fig. \ref{fig:conditional-commutativity}, if
$e_2$ and $e_3$ conditionally commute w.r.t. $e_4$, then both the sequences
$\pi_1$ and $\pi_2$ will lead to the same state. For non-commutative update operations
that are not ordered by $\F{lo}$, we enforce their conditional commutativity
through the following property:
\begin{align*}
	\textrm{\sc{cond-comm}}(\M{D}) &  \triangleq  \forall o_1,o_2,o_3 \in O.\
	(o_1 \xrightarrow{\F{rc}} o_2 \wedge \neg o_2 \rightleftarrows o_3)
	\Rightarrow o_1 \overset{o_3}{\rightleftarrows} o_2
\end{align*}
$\textrm{\sc{cond-comm}}(\M{D})$ is a property of an MRDT $\M{D}$, enforcing
conditional commutativity of update operations $o_1$ and $o_2$ w.r.t. $o_3$ if $o_2$
does not commute with $o_3$. Connecting this with the definition of
linearization relation, if there are events $e_1,e_2,e_3$ performing operations
$o_1,o_2,o_3$ resp., and if $e_1 \xrightarrow{\F{rc}} e_2$, $e_2
\xrightarrow{\F{vis}} e_3$ and $\neg e_2 \rightleftarrows e_3$, then there
will not be a linearization relation between $e_1$ and $e_2$. However,
$\textrm{\sc{cond-comm}}(\M{D})$ would then ensure that the ordering of $e_1$
and $e_2$ will not matter, due to the presence of the event $e_3$. We also
formalize the requirement of an $\F{rc}$ relation between all pairs of
non-commutative update operations:
\begin{align*}
	\textrm{\sc{rc-non-comm}}(\M{D}) & \triangleq \forall o_1,o_2 \in O.\neg o_1
	\rightleftarrows o_2 \Leftrightarrow o_1 \xrightarrow{\F{rc}} o_2 \vee  o_2
	\xrightarrow{\F{rc}} o_1 \\
\end{align*}
\vspace{-3em}
\begin{lemma}\label{lem:convergence}
	For an MRDT $\M{D}$ which satisfies $\textrm{\sc{rc-non-comm}}(\M{D})$ and
	$\textrm{\sc{cond-comm}}(\M{D})$, for any reachable configuration $C$ in
	$\M{S}_\M{D}$, for any two sequences $\pi_1,\pi_2$ over $\M{E}_C$ which extend
	$\F{lo}_C$, for any state $\sigma$, $\pi_1(\sigma) = \pi_2(\sigma)$.
\end{lemma}

\begin{definition}[RA-linearizability of MRDT]
	\label{def:lin}
	Let $\M{D}$ be an MRDT which satisfies $\textrm{\sc{rc-non-comm}}(\M{D})$ and $\textrm{\sc{cond-comm}}(\M{D})$. Then, a configuration $C = \langle N, H, L, G, vis \rangle$ of $\M{S_D}$ is RA-linearizable if, for every active replica $r \in range(H)$, there exists a sequence $\pi$ consisting of all events in $L(H(r))$ such that  $\F{lo}(C)_{\mid L(H(r))} \subseteq \pi$ and $N(H(r)) = \pi(\sigma_0)$. 
An execution $\tau \in \llbracket \M{S_D} \rrbracket$ is RA-linearizable if all of its configurations are RA-linearizable. 
Finally, $\M{D}$ is RA-linearizable if all of its executions are RA-linearizable.
\end{definition}

For a configuration to be RA-linearizable, every active replica must have a state which can be obtained by applying a sequence of events witnessed at that replica, and that sequence must obey the linearization relation of the configuration. 
For an execution to be RA-linearizable, all of its configurations must be RA-linearizable.  Lemma \ref{lem:irreflexive} ensures the existence of a sequence extending the linearization relation, while Lemma \ref{lem:convergence} ensures that two versions which have witnessed the same set of events will have the same state (i.e. strong eventual consistency). Further, we also show that if an MRDT is RA-linearizable, then for any query operation in any execution, the query result is derived from the state obtained by applying the update events seen at the corresponding replica right before the query:

\begin{lemma}\label{lem:query}
	If MRDT $\mathcal{D}$ is RA-linearizable, then for all executions $\tau \in \llbracket \mathcal{S}_\mathcal{D} \rrbracket$, for all transitions $C \xrightarrow{query(r,q,a)} C'$ in $\tau$ where $C = \langle N, H, L, G, vis\rangle$, there exists a sequence $\pi$ consisting of all events in $L(H(r))$ such that $\F{lo}(C)_{\mid L(H(r))} \subseteq \pi$ and $a = \F{query}(\pi(\sigma_0),q)$.
\end{lemma}

Compared to the definition of RA-linearizability in the work by Wang et. al. \cite{Wang}, there is one major difference: Wang et. al. also consider a sequential specification in the form of a set of valid sequences of data-type operations, and requires the linearization sequence to belong to the specification. Our definition simply requires the state of a replica to be a linearization of the update operations applied to the replica, without appealing to a separate sequential specification.  Once this is done, we can separately show that a linearization of the MRDT operations obeys the sequential specification. For this, we can ignore the presence of the merge operation as well as the MRDT system model (which are taken care of by the RA-linearizability definition), thus boiling down to proving a specification over a sequential functional implementation, which is a well-studied problem.

\subsection{Bottom-up Linearization}
\label{subsec:bottom-up}

As demonstrated in \S \ref{sec:overview}, our approach to show RA-linearizability
of an MRDT implementation is based on using algebraic properties of merge
(specifically, commutativity of merge and update operation application) which allows
us to show that the result of a merge operation is a linearization of the
events in each of the versions being merged. We first describe a generic
template for the algebraic properties which can be used to prove
RA-linearizability:
	\[
\inferrule{\forall j.\ \pi_j \in \M{E} \cup \{\epsilon\} \quad l,a,b \in \Sigma \quad \pi \in \{\pi_0,\pi_1, \pi_2\}  \quad \forall j.\ \pi_j' = \pi_j - \pi }
{\F{merge}(\pi_0(l), \pi_1(a), \pi_2(b)) = \pi(\F{merge}(\pi_0'(l), \pi_1'(a)), \pi_2'(b)) ) )  }  \quad \quad \textrm{\sc{[BottomUpTemplate]}}
\]

The template for the algebraic property is given in the conclusion of the above
rule, while the premises describe certain conditions. Each $\pi_j$ for
$j \in \{0,1,2\}$ is a sequence of 0 or 1 event (i.e. either $\epsilon$ or a
single event $e_j$), while $l,a,b$ are arbitrary states of the MRDT. Note that applying the $\epsilon$ event on a state leaves it unchanged (i.e. $\epsilon(s) = s$). Then, we
can select one event $\pi$ which has been applied to the arguments of merge on
the LHS, and bring it outside, i.e. remove the event from each argument on
which it was applied, and instead apply the event to the result of merge. Note
that the notation $\pi_j^{'} = \pi_j - \pi$ means that if $\pi = \pi_j$, then
$\pi_j^{'} = \epsilon$, else $\pi_j^{'} = \pi_j - \pi$.

\begin{wrapfigure}{r}{0.3\textwidth}
	\vspace{-2em}
	\begin{center}
		\includegraphics[angle=0, width=0.8\linewidth]{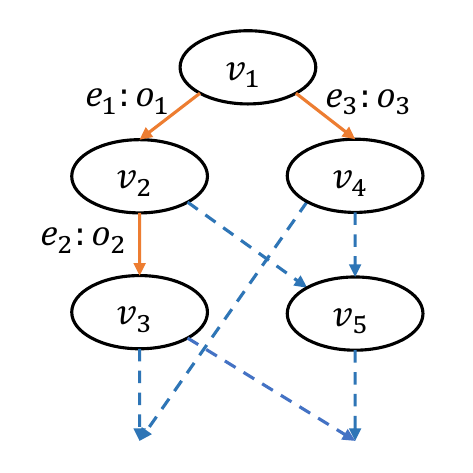}
	\end{center}
	\vspace{-1em}
	\caption{Example demonstrating the failure of bottom-up linearization in the presence of an $\F{rc}$-chain}
	\vspace{-1em}
	\label{fig:no-rc-chain}
\end{wrapfigure}

The rule (P1$'$) given in \S \ref{subsec:lin} can be seen as an instantiation of the above
template with $\pi_0 = \epsilon, \pi_1 = e_1, \pi_2 = e_2$ and $\pi = e_2$
where $e_1 \xrightarrow{\F{rc}} e_2$. Similarly, (P1-1) is another instantiation
with $\pi_0 = \pi_2 = e_1$, $\pi_1 = e_3$ and $\pi = e_3$ where $e_3 \neq e_1$.
Assuming that the input arguments to merge are obtained through sequences of events $\tau_0,
\tau_1, \tau_2$, the template rule builds the linearization sequence
$\tau = \tau' e$ where $e$ is the last event in one of the $\tau_i$s, and $\tau'$ is
recursively generated by applying the rule on $\tau^{'} = \tau - e$.
We call this procedure as \emph{bottom-up linearization}.  
The event $e$ should be chosen in such a way that the sequence $\tau$ is an
extension of the linearization relation (Def. \ref{def:lin-relation}).

However, bottom-up linearization might fail if the last event in the merge output 
is not the last event in any of the three arguments to merge.
For example, consider the execution shown in Fig. \ref{fig:no-rc-chain},
where there exists an $\F{rc}$-chain: $o_2 \xrightarrow{\F{rc}} 
o_3 \xrightarrow{\F{rc}} o_1$, and $o_1$ and $o_2$ are non-commutative.
$e_1$ is visible to $e_2$, while event $e_3$ is
concurrent to $e_1$ and $e_2$. Now, for the version obtained after merging
$v_3$ and $v_4$, the linearization relation would be $e_1 \xrightarrow[\F{vis}]{\F{lo}}
e_2$ and $e_2 \xrightarrow[\F{rc}]{\F{lo}} e_3$. Notably, even though
$e_1$ and $e_3$ are also concurrent, and $\F{rc}$ orders $o_3$ before $o_1$,
this will not result in a linearization relation from $e_3$ to $e_1$, due to
the presence of a non-commutative update operation $e_2$ to which $e_1$ is visible. 
The bottom-up linearization for the merge of $v_3$ and $v_4$, will result in
the sequence $e_1 e_2 e_3$, which is an extension of the linearization order.

However, suppose we first merge versions $v_2$ and $v_4$, to obtain the
version $v_5$, where the linearization relation is $e_3
\xrightarrow[\F{rc}]{\F{lo}} e_1$. Merging $v_3$ and $v_5$ (with LCA $v_2$) 
would have the same linearization relation as merging $v_3$ and $v_4$. 
However, the sequences
leading to $v_3$ and $v_5$ are $e_1 e_2$ and $e_3 e_1$ respectively, while the
only sequence which extends the linearization relation for their merge is $e_1
e_2 e_3$. Bottom-up linearization will then be constrained to pick either $e_1$
or $e_2$ to appear at the end, but such a sequence will not extend the linearization relation
resulting in the failure of bottom-up linearization. 
To avoid such cases, we place an additional constraint which prohibits the
presence of an $\F{rc}$-chain:
\begin{align*}
	\textrm{\sc{no-rc-chain}}(\M{D}) & \triangleq  \neg (\exists o_1,o_2,o_3 \in O.\ o_1 \xrightarrow{\F{rc}} o_2 \xrightarrow{\F{rc}} o_3)
\end{align*}
If there is an $\F{rc}$-chain, executions such as Fig. \ref{fig:no-rc-chain}
are possible, resulting in infeasibility of bottom-up linearization. However,
we will show that if an MRDT satisfies $\textrm{\sc{no-rc-chain}}(\M{D})$, then
we can use bottom-up linearization to prove that $\M{D}$ is linearizable. We
note that \textrm{\sc{no-rc-chain}} is a pragmatic restriction and consistent
with standard conflict-resolution strategies such as add/remove-wins,
enable/disable-wins, update/delete-wins, etc. which are typically used in MRDT
implementations.
\section{Verifying RA-linearizability of MRDTs}
\label{sec:lemmas}
In this section, we present our verification strategy for proving RA-linearizability of MRDTs using bottom-up linearization. According to Def. \ref{def:lin}, in order to prove that an MRDT $\M{D}$ is linearizable, we need to consider every configuration $C$ reachable in any execution, and show that all replicas in $C$ have states which can be obtained by linearizing the events applied to the replica, i.e. finding a sequence which obeys the linearization relation (Def. \ref{def:lin-relation}). We will assume that $\M{D}$ satisfies the three constraints ($\textrm{\sc{rc-non-comm}}$, $\textrm{\sc{cond-comm}}$ and $\textrm{\sc{no-rc-chain}}$) necessary for an MRDT to be linearizable, and for bottom-up linearization to succeed.

Our overall proof strategy is to use induction on the length of the execution and to extract generic verification conditions (VCs) which help us to discharge the inductive case. These VCs would essentially be instantiations of the $\textrm{\sc{BottomUpTemplate}}$ rule, proving that the merge operation results in a linearization of the events of the two versions being merged. Proving these VCs for arbitrary MRDTs is not straightforward (as discussed in \S \ref{subsec:induction}), and hence we propose another induction scheme over event sequences. We first discuss the instantiations of the $\textrm{\sc{BottomUpTemplate}}$ rule required for linearizing merges.

\subsection{Linearizing Merge Operations}

Consider an execution $\tau \in \llbracket \M{S_D} \rrbracket$ such that all
configurations in $\tau$ are linearizable. Suppose $\tau$ ends in the
configuration $C$. Now, we extend $\tau$ by one more transition, resulting in
the new configuration $C'$; we need to prove that $C'$ is also linearizable.
Let $C = \langle N, H, L, G, vis \rangle$, $C' = \langle N', H', L', G', vis'
\rangle$. It is easy to see if that this transition is caused due to
$\textrm{\sc{CreateBranch}}$ or $\textrm{\sc{Apply}}$ rules, then $C'$ will be
linearizable. For example, in the $\textrm{\sc{[Apply]}}$  transition, where a
new update operation $o$ is applied on a replica $r$ (generating a new event $e$),
only the state at $r$ changes, and this new state is obtained by directly
applying $e$ on the original state $\sigma$ at $r$. Since $\sigma$ was assumed
to be linearizable, there exists a sequence $\pi$ which extends
$\F{lo}(C)_{\mid L(H(r))}$, with $\sigma = \pi(\sigma_0)$ (recall that
$L(H(r))$ denotes the set of events applied at $r$). Then, the new state
$e(\sigma)$ is clearly linearizable through the sequence $\pi e$ which extends
$\F{lo}(C')_{\mid L'(H'(r))}$. 

We focus on the difficult case when there is a $\textrm{\sc{Merge}}$ transition
from $C$ to $C'$ which merges the replicas $r_1$ and $r_2$. Let $\sigma_1$ and
$\sigma_2$ be the states of the head versions $v_1$ and $v_2$ at $r_1$ and
$r_2$ respectively. Let $\sigma_\top$ be the state of the LCA version $v_\top$
of $v_1$ and $v_2$. Recall that $L(v_\top) = L(v_1) \cap L(v_2)$.  The
transition will install a new version with state $\sigma_m =
\F{merge}(\sigma_\top, \sigma_1, \sigma_2)$ at the replica $r_1$, leaving the
other replicas unchanged. Also, $L'(v_m) = L(v_1) \cup L(v_2)$. We need to show
that there exists a sequence $\pi$ of events in $L'(v_m)$ such that $\pi$
extends $\F{lo}(C')_{\mid L'(v_m)}$ and $\sigma_m = \pi(\sigma_0)$.

We first describe the structure of a sequence $\pi$ which extends
$\F{lo}(C')_{\mid L'(v_m)}$. For ease of readability, we use $L_1$ for
$L(v_1)$, $L_2$ for $L(v_2)$ and $L_\top$ for $L(v_\top)$, and $\F{lo_m}$ for
$\F{lo}(C')_{\mid L'(v_m)}$. We define the following sets of events:
\begin{align*}
	& L_1' = L_1 \setminus L_\top \quad \quad  L_2' = L_2 \setminus L_\top \\
	& L_1^b = \{e \in L_1^{'}\ \mid\ \exists e_\top \in L_\top.\ (e \xrightarrow{\F{lo_m}} e_\top \vee  \exists e' \in L_1^{'}.\ e \xrightarrow{\F{lo_m}} e^{'} \xrightarrow{\F{lo_m}} e_\top)\}\\
	& L_2^b = \{e \in L_2^{'}\ \mid\ \exists e_\top \in L_\top.\ (e \xrightarrow{\F{lo_m}} e_\top \vee \exists e' \in L_2^{'}.\ e \xrightarrow{\F{lo_m}} e^{'} \xrightarrow{\F{lo_m}} e_\top)\}\\
	& L_\top^{a} = \{e_\top \in L_\top \mid \exists e \in L_1^{b} \cup L_2^{b}. e  \xrightarrow{\F{lo_m}} e_\top\} \quad
	  L_1^a = L_1^{'} \setminus L_1^b \quad  \quad L_2^a = L_2^{'} \setminus L_2^b \quad \quad L_\top^{b} = L_\top \setminus L_\top^{a}
\end{align*}

\begin{wrapfigure}{r}{0.3\textwidth}
	\vspace{-1.5em}
	\begin{center}
		\includegraphics[angle=0, width=0.75\linewidth]{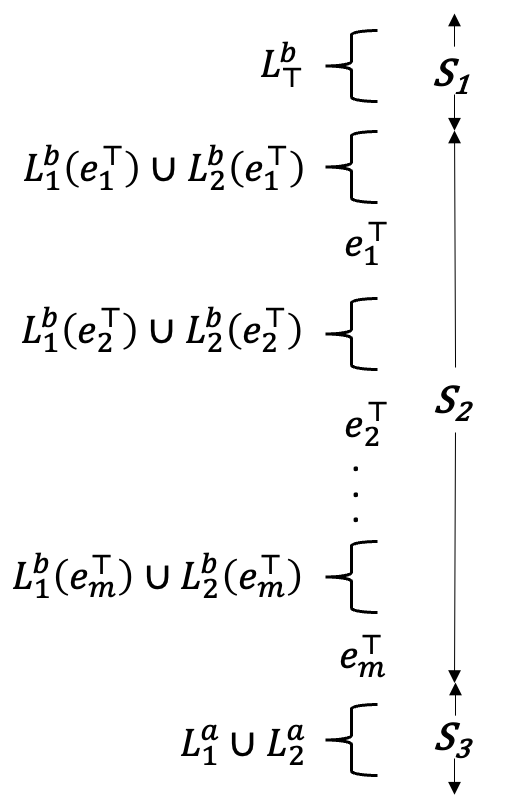}
	\end{center}
	\vspace{-1em}
	\caption{Structure of sequence $\pi$ extending $\F{lo}_m$}
	\vspace{-2em}
	\label{fig:order}
\end{wrapfigure}

$L_1^{'}$ and $L_2^{'}$ are the local events in each version. Note that any pair of events $e_1 \in L_1', e_2 \in L_2'$ will necessarily be concurrent. This is because, in any reachable configuration, any version $v$ is always \textbf{causally closed}, which means that if $e_1 \xrightarrow{\F{vis}} e_2$ and $e_2 \in L(v)$, then $e_1 \in L(v)$. Hence, for events $e_1 \in L_1', e_2 \in L_2'$, if $e_1 \xrightarrow{\F{vis}} e_2$ then $e_1 \in L_2'$, which would make $e_1$ a non-local event (i.e. part of the LCA). Bottom-up linearization first linearizes the local events across the two versions using the $\F{rc}$ relation for non-commutative events, and then linearizes events of the LCA. However, as demonstrated by the example in \S \ref{subsec:inter}, local events may also need to be linearized before events of the LCA (due to possible intermediate merges), and these events are collected in the sets $L_1^b$ and $L_2^b$. Specifically, $L_i^b (i = 1,2)$  contains those local events $e$ in $L_i'$ which either occur $\F{lo_m}$ before some event in the LCA, or which occur $\F{lo_m}$ before another local event $e'$ which occurs $\F{lo_m}$ before an LCA event. The events of the LCA which need to be linearized after local events are collected in $L_\top^a$. Finally, $L_1^a$ and $L_2^a$ contain local events which do not occur $\F{lo_m}$ before an LCA event.

\begin{example}\label{ex:1}
	Consider the execution in Fig. \ref{fig:inter_merge}, and the merge of versions $v_3$ and $v_4$, for which the LCA is $v_1$. For this merge, $L_1' = \{e_3\}$, $L_2' = \{e_2\}$, $L_1^b = \emptyset$, $L_2^b = \{e_2\}$, $L_\top^{a} = \{e_1\}$. For the merge of versions $v_1$ and $v_2$ (whose LCA is $v_0$), $L_1' = \{e_1\}$, $L_2' = \{e_2\}$, while $L_1^b,L_2^b,L_\top^a$ will all be empty (since no local event comes $\F{lo}$-before an LCA event).
\end{example}

We now show that there exists a sequence $\pi$ which extends $\F{lo_m}$ and which has events in $S_1 = L_\top^b$ followed by $S_2 = L_\top^a \cup L_1^b \cup L_2^b$ followed by $S_3 = L_1^a \cup L_2^a$ (later, we will discuss the ordering of events inside each set $S_i$). To prove this, we will demonstrate that there is no $\F{lo_m}$ from events in $S_i$ to events in $S_{i-1}$.  Based on the definitions of the $S_i$ sets, we can deduce some obvious facts: (i) there cannot be events $e \in S_3$, $e' \in L_\top$ such that $e \xrightarrow{\F{lo_m}} e'$, because otherwise, such an event $e$ would be in $L_1^b \cup L_2^b$ (and hence not in $S_3$), (ii) there cannot be events $e \in L_1^b \cup L_2^b$, $e' \in L_\top^b$ such that $e \xrightarrow{\F{lo_m}} e'$, because otherwise, such an event $e'$ would be in $L_\top^a$. In addition, using \textsc{no-rc-chain} and \textsc{rc-non-comm}, we also prove the following:

\begin{lemma}\label{lem:pi1}
	\begin{enumerate}
		\item For events $e \in L_1^a \cup L_2^a$, $e' \in L_1^b \cup L_2^b$, $\neg (e \xrightarrow{\F{lo_m}} e')$.
		\item For events $e \in L_\top^a$, $e' \in L_\top^b$, $\neg (e \xrightarrow{\F{lo_m}} e')$.
	\end{enumerate}
\end{lemma}

(1) from the above lemma ensures that there is no $\F{lo_m}$ relation from $S_3$ to $S_2$, while (2) ensures the same from $S_2$ to $S_1$. Hence a sequence with the structure $S_1\ S_2\ S_3$ would extend $\F{lo_m}$. Let us now consider the ordering among events in each set. First, for $S_3$, this set contains local events which are guaranteed to not come $\F{lo_m}$ before any event of the LCA. An event in $L_1^a$ will be concurrent with an event in $L_2^a$, and the linearization relation between them will depend upon the $\F{rc}$ relation between the underlying operations (if the events don't commute). We now instantiate $\textrm{\sc{BottomUpTemplate}}$ for the case where both $L_1^a$ and $L_2^a$ are non-empty in the rule $\textrm{\sc{BottomUp-2-OP}}$ in Fig. \ref{fig:bottom-up}, so that the linearization needs to consider the $\F{rc}$ relation between events in the two sets.

\begin{figure}[ht]
\vspace{-1em}
		\small
	\begin{align*}
		& \textrm{\sc{[BottomUp-2-OP]}} & &\textrm{\sc{[BottomUp-1-OP]}} & \\
		& \inferrule{e_1 \neq e_2 \quad e_1 \xrightarrow{\F{rc}} e_2 \vee e_1 \rightleftarrows e_2}{\F{merge}(l, e_1(a), e_2(b)) = e_2(\F{merge}(l, e_1(a), b))} & &\inferrule{(e_\top \neq \epsilon \wedge e_1 \neq e_\top) \vee (e_\top = \epsilon \wedge l = b) }{\F{merge}(e_\top(l), e_1(a), e_\top(b)) = e_1(\F{merge}(e_\top(l), a, e_\top(b)))}  
	\end{align*}
	\begin{align*}
		& \textrm{\sc{[BottomUp-0-OP]}} &    &\textrm{\sc{[MergeIdempotence]}} & 
		&\textrm{\sc{[MergeCommutativity]}} &\\
		& \inferrule{}{\F{merge}(e_\top(l), e_\top(a), e_\top(b)) = e_\top(\F{merge}(l, a, b))} & & \F{merge}(a,a,a) = a & & \F{merge}(l,a,b) = \F{merge}(l,b,a)
	\end{align*}
	\caption{Bottom-up Linearization}
	\vspace{-0.5em}
	\label{fig:bottom-up}
	\vspace{-1em}
\end{figure}

Note that $e_1,e_2$ and $l,a,b$ are all universally quantified. The $\textrm{\sc{BottomUp-2-OP}}$ rule is an algebraic property of $\F{merge}$ which needs to be separately shown for each MRDT implementation. For our case where we are trying to linearize $\F{merge}(\sigma_\top, \sigma_1, \sigma_2)$, we can apply $\textrm{\sc{BottomUp-2-OP}}$ with $l = \sigma_\top$, $e_1(a) = \sigma_1$ and $e_2(b) = \sigma_2$. Note that since $L_1^a$ and $L_2^a$ are both non-empty, $e_1\in L_1^a$, $e_2 \in L_2^b$ (in fact, $e_1$ and $e_2$ would be the maximal events in $L_1^a$ and $L_2^b$ according to $\F{lo_m}$). $\textrm{\sc{BottomUp-2-OP}}$ would then linearize $e_2$ at the end of the sequence. If $ e_1 \xrightarrow{\F{rc}} e_2$, then  $e_1 \xrightarrow{\F{lo_m}} e_2$, and thus linearizing $e_2$ at the end obeys the $\F{lo_m}$ ordering. Note that due to the $\textrm{\sc{no-rc-chain}}$ constraint, $e_2$ cannot come $\F{lo_m}$ before another concurrent event $e_3$. $\textrm{\sc{BottomUp-2-OP}}$  can now be recursively applied on $\F{merge}(l, e_1(a), b)$, by considering $e_1$ and the last event leading to the state $b$. By repeatedly applying $\textrm{\sc{BottomUp-2-OP}}$ all the remaining events in $L_1^a$ and $L_2^a$ can be linearized until one of the sets becomes empty.

Let us now consider the scenario where exactly one of  $L_1^a$ and $L_2^a$ is empty. WLOG, let $L_1^a$ be non-empty. We instantiate $\textrm{\sc{BottomUpTemplate}}$ for the case where $L_1^a$ is non-empty and $L_2^a$ is empty in the rule $\textrm{\sc{BottomUp-1-OP}}$ in Fig. \ref{fig:bottom-up}, so that the linearization orders all events of $L_1^a$ after events of $S_2$.

Let us consider the first clause in the premise where $e_\top \neq \epsilon$. To understand $\textrm{\sc{BottomUp-1-OP}}$, note that if $L_2^a$ is empty, then all local events in $L_2'$ are linearized before the LCA events. In this case, the last event which leads to the state $\sigma_2$ must be an LCA event. $\textrm{\sc{BottomUp-1-OP}}$ uses this observation, with $e_\top(l)=\sigma_\top$, $e_1(a)=\sigma_1$ and $e_\top(b)=\sigma_2 $. Notice that the last event in both the LCA and the second argument to merge are exactly the same. $e_\top$ will be the maximal event (according to $\F{lo_m}$ relation) in $L_\top^a$, while $e_1$ will be the maximal event in $L_1^a$. $\textrm{\sc{BottomUp-1-OP}}$ then linearizes $e_1$ at the end of the sequence, thus ensuring that all $L_1^a$ events are linearized after events in $S_1$ and $S_2$. It is possible that $L_\top^a$ is empty, in which case $L_2'$ will be empty, which is covered by the second clause where $e_\top = \epsilon$ and $l = b$ since there is no local event in the second state.

\begin{example}\label{ex:2}
Referring to Example \ref{ex:1} for the execution in Fig. \ref{fig:inter_merge}, recall that for the merge of $v_3$ and $v_4$, we have $L_1^a = \{e_3\}$, $L_2^a = \emptyset$ and $L_\top = \{e_1\}$. $\textrm{\sc{BottomUp-1-OP}}$ can be applied in this scenario to linearize $e_3$ at the end of the sequence.
\end{example}

$\textrm{\sc{BottomUp-2-OP}}$  and $\textrm{\sc{BottomUp-1-OP}}$ can thus be used to linearize all events in $S_3$. Let us now consider $S_2$, which contains both local events in $L_1^b \cup L_2^b$ and LCA events in $L_\top^a$. We first provide a more fine-grained structure of $\F{lo_m}$ among events in the set $S_2$.  Let $L_\top^a = \{e_1^{\top}, \ldots, e_m^{\top}\}$. For each $e_i^{\top}$, we collect all local events from $L_1^b$ and $L_2^b$ which need to be linearized before $e_i^{\top}$. For local events which need to be linearized before multiple $e_i^{\top}$s, we associate them with the smallest such $i$. We use $L_1^b(e_i^{\top})$ and $L_2^b(e_i^{\top})$ to denote these sets. Formally:
\begin{equation*}
	\begin{split}
		\forall e_i^{\top} \in L_\top^{a}.\ L_1^{b}(e_i^{\top}) = \{e \in L_1^{'} \mid (\forall j.\ j < i \implies e \notin L_1^{b}(e_j^{\top})) \wedge
		e \xrightarrow{\F{lo_m}} e_i^{\top} \vee \exists e^{'} \in L_1^{'}. e \xrightarrow{\F{lo_m}} e^{'} \xrightarrow{\F{lo_m}} e_i^{\top}\}
	\end{split}
\end{equation*}

$L_2^b(e_i^{\top})$ is defined in a similar manner. We now prove the following lemma using \textsc{no-rc-chain} and \textsc{rc-non-comm}:

\begin{lemma}\label{lem:pi2} \begin{enumerate}
		\item For all events $e_i^{\top}, e_j^{\top} \in L_\top^a$, where $L_\top^a = \{e_1^{\top}, \ldots, e_m^{\top}\}$, $\neg (e_i^{\top} \xrightarrow{\F{lo_m}} e_j^{\top})$
		\item 	For events $e \in L_1^b(e_i^{\top}) \cup L_2^b(e_i^{\top})$, $e' \in L_1^b(e_j^{\top}) \cup L_2^b(e_j^{\top})$ where $j<i$, $\neg (e \xrightarrow{\F{lo_m}} e')$.
	\end{enumerate}
\end{lemma}

From (1) in the above lemma, since there is no $\F{lo_m}$ relation among events in $L_\top^a$, consider the sequence $e_1^{\top} e_2^{\top} \ldots e_m^{\top}$ as a starting point for the sequence of events in $S_2$ which extends $\F{lo_m}$. We then inject $L_1^b(e_i^{\top}) \cup L_2^b(e_i^{\top})$ before each $e_i^{\top}$ in the sequence $e_1^{\top} e_2^{\top} \ldots e_m^{\top}$, as shown in Fig. \ref{fig:order}.  Note that in Fig.\ref{fig:order}, we have only presented various segments of the sequence, with the ordering within those segments determined by $\F{vis}$ and $\F{rc}$. By (2) in Lemma \ref{lem:pi2}, we can show that such a sequence will extend $\F{lo}_m$ among the events in $S_2$.

To show that $\F{merge}$ follows the sequence $\pi$ for $S_2$, we now instantiate  $\textrm{\sc{BottomUpTemplate}}$ for the case where $L_1^a$ and $L_2^a$ are empty (i.e. $S_3$ has already been linearized) in the rule $\textrm{\sc{Bottom-0-OP}}$ in Fig. \ref{fig:bottom-up}. Following the structure of $\pi$ in Fig. \ref{fig:order}, $e_\top$ would be the event $e_m^{\top} \in L_\top^a$. Note that since $e_m^{\top}$ is an LCA event, it will be present in both states being merged. $\textrm{\sc{BottomUp-0-OP}}$ then allows this event to be linearized first at the end.

\begin{example}\label{ex:3}
	Following on from Example \ref{ex:2} for the execution in Fig. \ref{fig:inter_merge} for the merge of $v_3$ and $v_4$, after $\textrm{\sc{BottomUp-1-OP}}$ is applied to linearize $e_3$, the states to be merged would be the versions $v_1$ and $v_4$ (with LCA $v_1$), both of whose last operation is $e_1$. Hence, $\textrm{\sc{BottomUp-0-OP}}$ would be applicable, which would linearize $e_1$.
\end{example}

After applying $\textrm{\sc{BottomUp-0-OP}}$ to linearize the LCA event $e_m^{\top}$, we then need to linearize events in $L_1^b(e_m^{\top}) \cup L_2^b(e_m^{\top})$. However, the event $e_m^{\top}$ has already been linearized, so none of the events in $L_1^b(e_m^{\top}) \cup L_2^b(e_m^{\top})$ appear $\F{lo_m}$ after an LCA event. This scenario can now be handled using $\textrm{\sc{BottomUp-2-OP}}$ (if both $L_1^b(e_m^{\top})$ and $L_2^b(e_m^{\top})$ are non-empty) or  $\textrm{\sc{BottomUp-1-OP}}$ (if one of 2 sets is empty). These rules will appropriately linearize the events in $L_1^b(e_m^{\top}) \cup L_2^b(e_m^{\top})$ taking into account the $\F{rc}$ relation for concurrent events and $\F{vis}$ relation for non-concurrent events. Once $L_1^b(e_m^{\top}) \cup L_2^b(e_m^{\top})$ becomes empty, we then encounter the next LCA event in $L_\top^a$, which can again be linearized using $\textrm{\sc{BottomUp-0-OP}}$.

The three instantiations of $\textrm{\sc{BottomUpTemplate}}$ can thus be repeatedly applied to linearize the rest of the events in $S_2$. Following this, all the local events would have been linearized, leaving only the LCA events in $S_1$. This would result in all three arguments to $\F{merge}$ being equal, in which case we can use the
$\textrm{\sc{MergeIdempotence}}$ rule in Fig. \ref{fig:bottom-up}. Using $\textrm{\sc{MergeIdempotence}}$, we can equate the output of $\F{merge}$ to it's argument, which has already been assumed to be appropriately linearized.

In order to avoid mirrored versions of $\textrm{\sc{BottomUp-2-OP}}$ and $\textrm{\sc{BottomUp-1-OP}}$ where the second and third arguments are swapped, we also require the $\textrm{\sc{MergeCommutativity}}$ property in Fig. \ref{fig:bottom-up}. We now state our soundness theorem linking the various properties with RA-linearizability of MRDT:

\begin{theorem}\label{thm:1}
	If an MRDT $\M{D}$ satisfies  $\textrm{\sc{BottomUp-2-OP}}$,  $\textrm{\sc{BottomUp-1-OP}}$,  $\textrm{\sc{BottomUp-0-OP}}$, $\textrm{\sc{MergeIdempotence}}$ and $\textrm{\sc{MergeCommutativity}}$, then $\M{D}$ is linearizable.
\end{theorem}

The proof closely follows the informal arguments that we have presented in this sub-section, using induction on the size of the various sets $L_1^a, L_2^a, L_1^b \cup L_2^b, L_\top^a$.

\subsection{Automated Verification}
While we have identified the sufficient conditions to show RA-linearizability of an MRDT using bottom-up linearization, proving these conditions for arbitrary MRDTs is not straightforward. Further, while the $\textrm{\sc{BottomUp-X-OP}}$ properties as shown in the previous sub-section had universal quantification over MRDT states $l,a,b$, in general, for proving RA-linearizability, we only need to show these properties for feasible states that may arise during an actual execution. 

We now leverage the fact that the feasible states would have been obtained through linearization of the visible events at the respective versions. In particular, we can characterize the states on which merge can be invoked through the various events sets $L_1^a, L_2^a, L_1^b, L_2^b, L_\top^a, L_\top^b$ that we had defined in the previous sub-section. We only need to prove the $\textrm{\sc{BottomUp-X-OP}}$ properties for states which have been obtained through linearizations of events in these event sets. For this purpose, we propose an induction scheme which establishes the required properties while traversing the event sets as depicted in Fig. \ref{fig:order} in a top-down fashion.
\newcommand{\comp}{\!\cdot\!}

\begin{table}[htpb]
	\caption{Induction scheme for  $\textrm{\sc{BottomUpTemplate}}$. For clarity, we use $\cdot$ for function composition, and $\mu$ for merge.}
	\scriptsize
	\vspace{-1em}
	\makebox[\textwidth][c]{
		\begin{tabular}{|>{\raggedright\arraybackslash}p{0.5cm}|>{\raggedright\arraybackslash}p{2.2cm}|>{\raggedright\arraybackslash}p{3.8cm}|>{\raggedright\arraybackslash}p{4.5cm}|}

			\hline
			\textbf{VC Name} & \multicolumn{2}{c|}{\textbf{Pre-condition}} & \textbf{Post-condition} \\
			\hline

			$\psi^{L_\top^b}_\F{base}$ &
			&
			$ $ &
			$\mu(\pi_0(\sigma_0), \pi_1(\sigma_0), \pi_2(\sigma_0)) = \pi \comp \mu(\pi_0'(\sigma_0),\pi_1'( \sigma_0), \pi_2'(\sigma_0))$\\
			\hline

			$\psi^{L_\top^b}_\F{ind}$&
			&
			$\mu(\pi_0(l), \pi_1(l), \pi_2(l)) = \pi \comp \mu(\pi_0'(l), \pi_1'(l), \pi_2'(l))$ &
			$\mu (\pi_0 \comp e_\top(l), \pi_1 \comp e_\top(l), \pi_2 \comp e_\top(l)) = \pi \comp \mu(\pi_0' \comp e_\top(l), \pi_1' \comp e_\top(l), \pi_2' \comp e_\top(l))$\\
			\hline

			$\psi^{L_\top^a}_\F{ind}$ &
			$\exists e.\ e \xrightarrow{\F{rc}} e_\top$ &
			$\mu(\pi_0(l), \pi_1(a), \pi_2(b)) = \pi \comp \mu(\pi_0'(l), \pi_1'(a), \pi_2'(b))$ &
			$\mu (\pi_0 \comp e_\top(l), \pi_1 \comp e_\top(a), \pi_2 \comp e_\top(b)) = \pi \comp \mu(\pi_0' \comp e_\top(l), \pi_1' \comp e_\top(a), \pi_2' \comp e_\top(b))$\\
			\hline

			$\psi^{L_1^b}_\F{ind1}$ &
			$e_b \xrightarrow{\F{rc}} e_\top$ &
			$\mu(\pi_0 \comp e_\top(l), \pi_1 \comp e_\top(a), \pi_2 \comp e_\top(b)) = \pi \comp \mu(\pi_0' \comp e_\top(l), \pi_1' \comp e_\top(a), \pi_2' \comp e_\top(b))$ &
			$\mu(\pi_0 \comp e_\top(l), \pi_1 \comp e_\top \comp e_b(a), \pi_2 \comp e_\top(b)) = \pi \comp \mu(\pi_0' \comp e_\top(l), \pi_1' \comp e_\top \comp e_b(a), \pi_2' \comp e_\top(b))$\\
			\hline

			$\psi_\F{ind2}^{L_1^b}$ &
			$e_b \xrightarrow{\F{rc}} e_\top \wedge \neg e \rightleftarrows e_b$ &
			$\mu(\pi_0 \comp e_\top(l), \pi_1 \comp e_\top \comp e_b(a), \pi_2 \comp e_\top(b)) = \pi \comp \mu(\pi_0' \comp e_\top(l), \pi_1' \comp e_\top \comp e_b(a), \pi_2' \comp e_\top(b))$ &
			$\mu(\pi_0 \comp e_\top(l), \pi_1 \comp e_\top \comp e_b \comp e(a), \pi_2 \comp e_\top(b)) = \pi \comp \mu(\pi_0' \comp e_\top(l), \pi_1' \comp e_\top \comp e_b \comp e(a), \pi_2' \comp e_\top(b))$ \\
			\hline

			$\psi^{L_2^b}_\F{ind1}$ &
			$ e_b \xrightarrow{\F{rc}} e_\top$ &
			$\mu(\pi_0 \comp e_\top(l), \pi_1 \comp e_\top(a), \pi_2 \comp e_\top(b)) = \pi \comp \mu(\pi_0' \comp e_\top(l), \pi_1' \comp e_\top(a), \pi_2' \comp e_\top(b))$ &
			$\mu(\pi_0 \comp e_\top(l), \pi_1 \comp e_\top(a), \pi_2 \comp e_\top \comp e_b(b)) = \pi \comp \mu(\pi_0' \comp e_\top(l), \pi_1' \comp e_\top(a), \pi_2' \comp e_\top \comp e_b(b))$\\
			\hline

			$\psi_\F{ind2}^{L_2^b}$ &
			$ e_b \xrightarrow{\F{rc}} e_\top \wedge \neg e \rightleftarrows e_b$ &
			$\mu(\pi_0 \comp e_\top(l), \pi_1 \comp e_\top \comp e_b(a), \pi_2 \comp e_\top(b)) = \pi \comp \mu(\pi_0' \comp e_\top(l), \pi_1' \comp e_\top \comp e_b(a), \pi_2' \comp e_\top(b))$ &
			$\mu(\pi_0 \comp e_\top(l), \pi_1 \comp e_\top \comp e_b(a), \pi_2 \comp e_\top \comp e_b \comp e(b)) = \pi \comp \mu(\pi_0' \comp e_\top(l), \pi_1' \comp e_\top \comp e_b(a), \pi_2' \comp e_\top \comp e_b \comp e(b))$ \\
			\hline

			$\psi_\F{ind}^\F{L_1^a}$ &
			&
			$\mu(\pi_0(l), \pi_1(a), \pi_2(b)) = \pi \comp \mu(\pi_0'(l), \pi_1'(a), \pi_2'(b))$ &
			$\mu(\pi_0(l), \pi_1 \comp e(a), \pi_2(b)) = \pi \comp \mu(\pi_0'(l), \pi_1' \comp e(a), \pi_2'(b))$\\
			\hline

			$\psi_\F{ind}^\F{L_2^a}$ &
			&
			$\mu(\pi_0(l), \pi_1(a), \pi_2(b)) = \pi \comp \mu(\pi_0'(l), \pi_1'(a), \pi_2'(b))$ &
			$\mu(\pi_0(l), \pi_1(a), \pi_2 \comp e(b)) = \pi \comp \mu(\pi_0'(l), \pi_1'(a), \pi_2' \comp e(b))$\\
			\hline

		\end{tabular}
	}
	\label{tbl:vc}
	\vspace{-1em}
\end{table}
Here, we present the induction scheme for the generic $\textrm{\sc{BottomUpTemplate}}$ rule. The scheme can then be instantiated for all the three $\textrm{\sc{BottomUp-X-OP}}$ rules. Table \ref{tbl:vc} contains the verification conditions corresponding to the base case and inductive case over the different event sets. Every VC has the form $(\text{pre-condition} \implies \text{post-condition})$, and all variables are universally quantified. Our goal is to show the $\textrm{\sc{BottomUpTemplate}}$ rule for all feasible MRDT states $l,a,b$, where $l$ is the state of the LCA of $a$ and $b$. Let $L_\top,L_1,L_2$ be the event sets corresponding to $l,a,b$ respectively. We define the event sets $L_1^a, L_2^a, L_1^b, L_2^b, L_\top^a, L_\top^b$ in exactly the same manner as the previous sub-section, based on the linearization relation of the configuration obtained by the $\F{merge}(l,a,b)$ transition. Note that the events in $\pi_0,\pi_1,\pi_2$ (used in the $\textrm{\sc{BottomUpTemplate}}$ rule) would also come from the above event sets, but in the following discussion, we freeze these events, i.e. all our assertions about the events sets will be modulo these events.

We start with the VC $\psi^{L_\top^b}_\F{base}$, which corresponds to the case where every event set is empty. There is no pre-condition, and the post-condition requires $\textrm{\sc{BottomUpTemplate}}$ to hold on the initial MRDT state $\sigma_0$. For example, for the $\textrm{\sc{BottomUp-2-OP}}$ rule, $\psi^{L_\top^b}_\F{base}$ VC would be $\F{merge}(\sigma_0, e_1(\sigma_0), e_2(\sigma_0)) = e_2(\F{merge}(\sigma_0, e_1(\sigma_0), \sigma_0))$, where $e_1 \xrightarrow{\F{rc}} e_2$ or $e_1$ and $e_2$ commute. Notice that $e_1$ and $e_2$ would be events in $L_1^a$ and $L_2^a$, and our assertion about all event sets being empty is modulo these events.

Next, the VC $\psi^{L_\top^b}_\F{ind}$ corresponds to the inductive case on $L_\top^b$, where we assume every event set except $L_\top^b$ to be empty. The pre-condition corresponds to the inductive hypothesis, where we assume the property to hold for some event set $L_\top^b$, and the post-condition asserts that the property holds while adding another event $e_\top$ to $L_\top^b$. Recall that $L_\top^b$ corresponds to the LCA events which come $\F{lo}$ before all local events. Since all the other event sets are empty, this translates to the same state $l$ for all the three arguments to merge in the pre-condition, and applying the LCA event $e_\top$ to all three arguments in the post-condition.

Next, we induct on the set $L_\top^a$, i.e. the set of LCA events which occur $\F{lo}$ after a local event. The base case, where $\mid L_\top^a \mid = \emptyset$  exactly corresponds to the result of the induction on $L_\top^b$. The inductive case is covered by the VC $\psi^{L_\top^a}_\F{ind}$, which adds an LCA event $e_\top$ to all three arguments of merge from pre-condition to post-condition. Notice that we also have another pre-condition which requires the existence of some event $e$ which should come $\F{rc}$-before $e_\top$, which is necessary for $e_\top$ to be in $L_\top^a$. The post-condition just adds a new LCA event $e_\top$. The events in $L_1^b(e_\top)$ and $L_2^b(e_\top)$ will be added by the next 4 VCs.

$\psi^{L_1^b}_\F{ind1}$ and $\psi^{L_1^b}_\F{ind2}$ add an event in $L_1^b$ from the pre-condition to the post-condition. $\psi^{L_1^b}_\F{ind1}$ considers an event $e_b$ which occurs $\F{rc}$-before the LCA event $e_\top$. Notice that the pre-condition of  $\psi^{L_1^b}_\F{ind1}$  is exactly the same as the post-condition of $\psi^{L_\top^a}_\F{ind}$. In the post-condition of $\psi^{L_1^b}_\F{ind1}$, the event $e_b$ is applied before $e_\top$ on the argument $a$ to merge, thus reflecting that this is an event in $L_1^b$. $\psi^{L_1^b}_\F{ind2}$ adds an event $e \in L_1^b$ which does not commute with an existing event $e_b \in L_1^b$ (see the definition of $L_1^b$).   $\psi^{L_2^b}_\F{ind1}$ and $\psi^{L_2^b}_\F{ind2}$ are analogous and do the same thing for the argument $b$ to merge.

Finally, $\psi_\F{ind}^\F{L_1^a}$ and $\psi_\F{ind}^\F{L_2^a}$ add events from $L_1^a$ and $L_2^a$. The base cases for the two sets would exactly correspond to the result of the induction carried out so far on the rest of the event sets. For the inductive case, in $\psi_\F{ind}^\F{L_1^a}$ (resp. $\psi_\F{ind}^\F{L_2^a}$), a new event $e$ is added on the second argument $a$ (resp. third argument $b$) from the pre-condition to the post-condition. This establishes the rule $\textrm{\sc{BottomUpTemplate}}$ for any feasible input arguments to merge during any execution. We denote the set of VCs in Table \ref{tbl:vc} by $\psi^*(\textrm{\sc{BottomUpTemplate}})$.

\begin{theorem}\label{thm:2}
	If an MRDT $\M{D}$ satisfies  the VCs $\psi^*(\textrm{\sc{BottomUp-2-OP}})$,  $\psi^*(\textrm{\sc{BottomUp-1-OP}})$,\\  $\psi^*(\textrm{\sc{BottomUp-0-OP}})$, $\textrm{\sc{MergeIdempotence}}$ and $\textrm{\sc{MergeCommutativity}}$, then $\M{D}$ is linearizable.
\end{theorem}
\vspace{-1em}
\section{Experimental Evaluation}
\label{sec:results}

We have implemented our verification technique in the \fstar programming
language and verified several MRDTs using it. We also extracted OCaml code from
the verified implementations and ran them as part of Irmin~\cite{Irmin}, a
Git-like distributed database which follows the MRDT system model described in
\S 3. This distinguishes our work from prior works in automated RDT
verification~\cite{Kartik} which focuses on verifying abstract models rather
than actual implementations.

Our framework in \fstar consists of an \fstar interface that
defines signatures for an MRDT implementation (Fig. \ref{fig:orset_impl}) and
the VCs described in Table \ref{tbl:vc}; these are encoded as \fstar lemmas.
This interface contains 200 lines of \fstar code.  An MRDT developer
instantiates the interface with their specific MRDT implementation and calls upon \fstar to prove
the lemmas (i.e., the VCs). Once this is done, our metatheory
(Theorem \ref{thm:2}) guarantees that the MRDT implementation is linearizable.

\begin{table}[htpb]
\centering
\footnotesize
	\vspace{-1em}
	\caption{Verified MRDTs. $^*$ denotes MRDT implementations not present in prior work.}
	\vspace{-0.5em}
	\begin{tabular}{@{}llcS[table-format=3.2]@{}}
		\toprule
		\textbf{MRDT} & \textbf{$\F{rc}$ Policy} & \textbf{\#LOC} & \textbf{Verification Time (s)} \\
		\midrule
		Increment-only counter~\cite{Kaki2019} & $\F{none}$ & 6 & 0.72 \\
		PN counter~\cite{Vimala} & $\F{none}$ & 10 & 1.64 \\
		Enable-wins flag$^*$ & $\F{disable} \xrightarrow{\F{rc}} \F{enable}$ & 30 & 29.80 \\
		Disable-wins flag$^*$ & $\F{enable} \xrightarrow{\F{rc}} \F{disable}$ & 30 & 37.91 \\
		Grows-only set~\cite{Kaki2019} & $\F{none}$ & 6 & 0.45 \\
		Grows-only map~\cite{Vimala} & $\F{none}$ & 11 & 4.65 \\
		OR-set~\cite{Vimala} & $\F{rem_a} \xrightarrow{\F{rc}} \F{add_a}$ & 20 & 4.53 \\
		OR-set (efficient)$^*$ & $\F{rem_a} \xrightarrow{\F{rc}} \F{add_a}$ & 34 & 660.00 \\
		Remove-wins set$^*$ & $\F{add_a} \xrightarrow{\F{rc}} \F{rem_a}$ & 22 & 9.60 \\
		Set-wins map$^*$ & $\F{del_k} \xrightarrow{\F{rc}} \F{set_k}$ & 20 & 5.06 \\
		Replicated Growable Array~\cite{Attiya} & $\F{none}$ & 13 & 1.51 \\
		Optional register$^*$ & $\F{unset} \xrightarrow{\F{rc}} \F{set}$ & 35 & 200.00 \\
		Multi-valued Register$^*$ & $\F{none}$ & 7 & 0.65 \\
		JSON-style MRDT$^*$ & $\F{Fig}.~\ref{fig:json_impl}$ & 26 & 148.84 \\
		\bottomrule
	\end{tabular}
	\label{tab:mrdts}
	\vspace{-1em}
\end{table}

We instantiate the interface with MRDT implementations of several datatypes
such as counter, flag, set, map, and list (Table~\ref{tab:mrdts}).  All the results were obtained on a
Intel\textregistered Xeon\textregistered Gold 5120 x86-64 machine running
Ubuntu 22.04 with 64GB of main memory.  While some of the MRDTs have been taken
from previous works \cite{Vimala,Kaki2019,Attiya} or translated from their CRDT
counterparts, we also develop some new implementations, denoted by $^*$ in
Table \ref{tab:mrdts}.  We also uncovered bugs in previous MRDT implementations
(Enable-wins flag and Efficient OR-set) from \cite{Vimala}, which we fixed (more
details in \S \ref{subsec:bug}). We note that in all our experiments, all the
VCs were automatically discharged by \fstar in a reasonable amount of time. 

While our approach ensures that the MRDT implementations are verified in the 
\fstar framework, it is important to note that the user is obligated to trust the \fstar
language implementation, the extraction mechanism, the OCaml language implementation, 
the OCaml runtime, and the hardware.

We now highlight several notable features about our verified MRDTs. We have
designed and developed the first correct implementations of both an enable-wins
and disable-wins flag MRDT. Our implementation of efficient
OR-set maintains a per-replica, per-element counter instead of adding different
versions of the same element (as done by the OR-set implementation of Fig.
\ref{fig:orset_impl}), thus matching the theoretical lower bound in terms of
space-efficiency for any OR-set CRDT implementation (as proved in
\cite{Burckhardt}). 
We have developed the first known MRDT implementation of a remove-wins set datatype.
Finally, as a demonstration of vertical compositionality, we have developed a
JSON MRDT which is composed of several component MRDTs, with its correctness
guarantee being directly derived from the correctness of the component MRDTs.

\subsection{Case study: A verified polymorphic JSON-style MRDT}
\label{subsec:json}

JSON is a notable example of a data type which is composed of several other
datatypes. JSON is widely used as a data interchange format in many databases
and web services ~\cite{Json}.  Our JSON MRDT is modeled as an unordered collection of key/value pairs, where the values
can be any primitive types, such as counter, list, etc., or they can be JSON
type themselves. We assume that keys are update-only; that is, key-value mappings
can be added and modified, but once a key is added, it cannot be deleted.
Previous works, such as Automerge ~\cite{Automerge}, have developed  similar
JSON-style CRDT models. However, these models are monomorphic, which means that
the data type of the values must be known in advance.  Our goal is to develop a
more generic JSON-style MRDT that supports polymorphic values, i.e. we leave
the value data type as an abstract type which can be instantiated with
any concrete MRDT.

\begin{figure}[ht]
	\footnotesize
	\vspace{-1.5em}
	\begin{algorithmic} [1]
		\State $\Sigma_\F{json} : (k:(\F{string }\times \Omega)) \rightarrow \Sigma_{\F{snd}(k)}$
		\State $O_\F{json} = \{\F{set}(k, o) \mid ~ o  \in O_{\F{snd}(k)} \}$
		\State $Q_\F{json} = \{\F{get}(k, q) \mid ~ q \in Q_{\F{snd}(k)} \}$
		\State $\sigma_{0_\F{json}} = \lambda (k: \F{string} \times \Omega).\ \sigma_{0_{\F{snd} (k)}}$
		\State $\F{do}(\sigma,t,r,\F{set}(k,o)) =  \sigma [k \mapsto o (\sigma(k), t, r)]$
		\State $\F{merge}_\F{json}(\sigma_{\top}, \sigma_1, \sigma_2) = \lambda (k: \F{string} \times \Omega).\ \F{merge}_{\F{snd}(k)}(\sigma_\top(k), \sigma_1(k), \sigma_2(k))$
		\State $\F{query}_\F{json} (\sigma, get(k, q)) = \F{query}_{\F{snd}(k)}(\sigma(k), q)$
		\State $\F{rc}_\F{json} = \{(\F{set}(k_1,o_1),\F{set}(k_2,o_2)) \in O_\F{json} \times O_\F{json}\ \mid\ k_1 = k_2 \wedge (o_1,o_2) \in \F{rc}_{\F{snd}(k_1)} \}$
	\end{algorithmic}
	\vspace{-1em}
	\caption{JSON-style MRDT implementation}
	\label{fig:json_impl}
	\vspace{-1.5em}
\end{figure}

Fig.~\ref{fig:json_impl} shows the implementation of the JSON MRDT. It uses a map to maintain the association between keys and values. Notice that the key is a tuple consisting of the identifier string and an MRDT type $\alpha \in \Omega$ which denotes the type of the value. The type $\alpha$ can be any arbitrary MRDT with implementation $\M{D}_\alpha = (\Sigma_\alpha, \sigma_{0_\alpha}, \F{merge}_\alpha, \F{query}_\alpha,\F{rc}_\alpha)$. Different key strings can now map to different value MRDT types. We also allow overloading: the same key string can be associated with multiple values of different types. The JSON MRDT allows update operations of the form $\F{set}(k,o)$ where $o$ is an operation of the underlying value MRDT associated with the key $k$. $\F{set}(k,o)$ simply applies the operation $o$ on the value associated with $k$, leaving the other key-value pairs unchanged. The JSON merge calls the underlying MRDT merge on the values associated with each key. The query operation of the form $\F{get}(k,q)$ retrieves the value associated with $k$ in $\sigma$ and applies the query operation $q$ of the underlying data type to it. The conflict resolution policy of JSON operations ($\F{rc}_\F{json}$) depends on the conflict resolution of the value types
when two operations update the same key (i.e. same identifier and value type). Every other pair of JSON operations commute with each other.

Notably, the proof of RA-linearizability of the JSON MRDT is directly derived from the proofs of the underlying value MRDT types. If all the MRDTs in $\Omega$ are linearizable, then the JSON MRDT is also linearizable. We have proved all the VCs for the JSON MRDT in \fstar by using the VCs of the underlying value MRDTs. We can now instantiate $\Omega$ with any set of verified MRDTs, thereby obtaining the verified JSON MRDT for free.

\vspace{-0.5em}
\subsection{Buggy MRDT Implementation in~\cite{Vimala}}
\label{subsec:bug}

We now present some details of one of the buggy MRDTs, Enable-wins flag, that we discovered using our framework in the work by \citet{Vimala}. The state of the enable-wins flag MRDT consists of a pair: a counter and a flag. The counter tracks the number of
the enable events, while the flag is set to true on an enable event. The desired specification for this flag is that it should be true when there is at least one enable event not visible to any disable event. In our framework, we can express this specification as $\F{disable} \xrightarrow{\F{rc}}\F{enable}$, linearizing the enable operation after a concurrent disable.
When we attempted to verify this implementation in our framework, we discovered that one of the
VCs, $\psi^{L_2^b}_\F{ind2-1op}$, was failing. Our investigation revealed that the implementation
violated the specification. The bug appeared in an execution with intermediate merges.

\begin{wrapfigure}{r}{0.35\textwidth}
	\vspace{-1em}
	\begin{center}
		\includegraphics[width=0.36\textwidth]{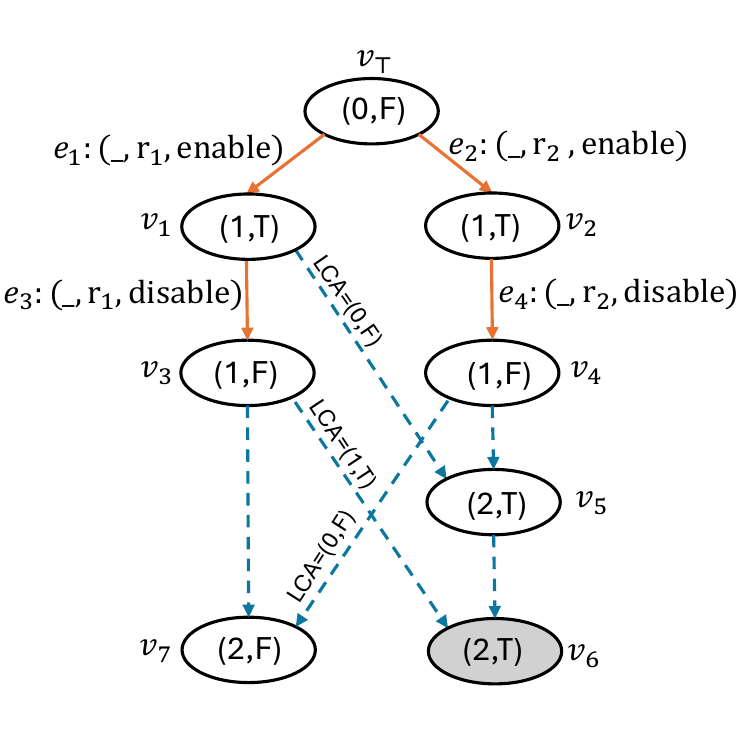}
	\end{center}
	\vspace{-2em}
	\caption{An enable-wins flag execution}
	\label{fig:ewflag-counter}
\end{wrapfigure}

Consider the execution depicted in Fig.~\ref{fig:ewflag-counter}.  When merging
versions $v_3$ and $v_5$ (with LCA $v_1$), since the counter value of $v_5$ is greater than $v_1$,
the flag in the merged version $v_6$ is set to true.  However, this contradicts the Enable-wins flag
specification, which states that the flag should be true only when there is an
enable event that is not visible to any disable event.  All enable
events in the execution are disabled by subsequent disable events on their individual replicas, yet the flag is true at
$v_6$. Notice that the version $v_5$ is obtained due to an intermediate merge. 
We discovered that \citet{Vimala} had an implementation bug in the framework. 
The framework expects a simulation relation from the MRDT developer, in addition to the specification and the implementation. 
This simulation relation serves as a proof artefact. \citet{Vimala} check whether the developer-provided simulation relation is valid
and the bug occurred during the validity-checking procedure. Due to this, \citet{Vimala} admitted the buggy enable-wins flag implementation\footnote{Buggy implementation can be found in \S \ref{subsec:bug_impl}}. 

We further note that this buggy implementation does not even satisfy strong eventual consistency.
In Fig.~\ref{fig:ewflag-counter}, merging $v_3$ and $v_4$ results in $v_7$, where the flag
is false. Note that both versions $v_6$ and $v_7$ have observed the same set of updates on both replicas,
yet they lead to divergent states. This violates strong eventual consistency.
We fixed this implementation by maintaining a counter-flag pair for every replica, 
i.e. changing the state to a map from replica-IDs to counter-flag pair.

\subsection{Verifying state-based CRDTs}
\label{subsec:crdts}
Although the developement in the paper so far has focused on  verifying MRDTs, we note that our framework can also directly verify state-based CRDTs. The only difference between the two is that state-based CRDTs do not maintain the LCA, and merge is a binary function. Our VCs (Table \ref{tbl:vc}) can be directly applied on state-based CRDTs, by simply ignoring the LCA argument for all merges. Note that while the merge function in state-based CRDTs does not use the LCA, our VCs still use the LCA to determine whether an event is local or common to both replicas, and appropriately linearize events taking into account both $\F{rc}$ and $\F{vis}$ relations. The entire set of VCs retrofitted for state-based CRDTs can be found in Table \ref{tab:full_crdts}. We have also successfully implemented and verified 7 state-based CRDTs in our framework: Increment-only counter, PN counter, Observed-Remove set, Two-Phase set, Grows-only set, Grows-only map and Multi-valued register.

\vspace{-0.5em}
\subsection{Limitations}

Our framework is currently unable to verify some MRDT implementations such as
Queue from previous works \cite{Vimala,Kaki2019}. The Queue MRDT follows at-least-once semantics for dequeues, which allows concurrent dequeue operations to return the same element from the queue, thereby having the effect of a single dequeue. Such an implementation is clearly not linearizable as per our definition, since we cannot omit any event while constructing the linearization. It would be possible to modify our notion of linearization to also allow events to be omitted; we leave this investigation as part of future work. Our verification technique is also not complete, but in practice we have been able to successfully verify all MRDT implementations (except Queue) from earlier works. 

\section{Related Work and Conclusion}
\label{sec:related}

Reconciling concurrent updates is a challenging problem in distributed systems.
CRDTs~\cite{Annette,Roh,Shapiro} (and more recently MRDTs) have emerged as a
principled approach for building correct and efficient replicated
implementations. Numerous works have focused on specifying and verifying
CRDTs~\cite{Attiya, Burckhardt, Gotsman, Gomes, Liu, Kartik, Wang, Nair,
Zeller, Katara, Verifx}. Op-based CRDTs have a considerably different system
model than MRDTs, where every operation instance at a replica is individually
sent to other replicas. Hence, verification efforts targeting them
\cite{Gomes,Liu,Kartik,Wang,Nair} are mostly orthogonal to our work.  

The system model of state-based CRDTs is similar to MRDTs, as it also requires 
a merge function to be implemented for reconciling concurrent updates. However, state-based CRDTs have 
stricter requirements for convergence and consistency: CRDT states must form a join-semilattice, 
updates must be monotonic, and the merge function must be the lattice join operation. 
The three algebraic properties of a semilattice: idempotence, commutativity, and 
associativity guarantee convergence.

Some CRDT works focus solely on ensuring convergence 
without addressing functional correctness. For instance, ~\citet{Verifx} do
not fully capture the user intent when verifying state-based CRDTs. Consider a
Counter CRDT with only an increment operation and an \emph{incorrect} merge
function that ignores its input states and always returns 0. Such an
implementation is still convergent. 
However, it clearly does not capture the developer intent, which is
that the value of the counter should be equal to the number of increment
operations. Functional correctness is as important as convergence for 
replicated data types. Our framework addresses both
by couching both in terms of RA-linearizability. We will flag the above
implementation as incorrect, since the state after merge cannot be obtained by
linearizing the operations performed on both the replicas. 

In the context of CRDTs, \citet{Wang} proposed the notion of replication-aware
linearizability, which requires all replicas to have a state which can be
obtained by linearizing the update operations visible to the replica according
to the sequential specification.  However, they do not propose any automated
verification methodology for RA-linearizability. Further, though the main paper
\citet{Wang} focuses on op-based CRDTs, the extended version \citet{EneaArxiv}
does address state-based CRDTs, but they also require a semi-lattice-based
formulation of the CRDT states for proving RA-linearizability.

Few works \cite{Vimala,Kaki2022} have explored the problem of verifying MRDT
implementations. ~\citet{Kaki2022} only focus on verifying convergence, but not
functional correctness. Moreover, they significantly restrict the underlying
system model by synchronizing all merge operations, which as mentioned in the
paper itself could lead to longer convergence times. ~\citet{Vimala} verify both
convergence and functional correctness, without requiring synchronized merges.
However, their approach is not fully automated, and
requires developers to provide a simulation relation linking concrete MRDT
states with an abstract state which is based on a event-based declarative
model. Their specification language is also based on an event-based model and
is not very intuitive or developer-friendly. A few MRDT implementations from
\cite{Vimala} were found to be buggy, and these errors were due to faulty
simulation relations.

To conclude, in this work, we present the first, fully-automated verification
methodology for MRDTs. We introduce the notion of replication-aware
linearizability for MRDTs, as well as a simple specification framework based on
ordering non-commutative update operations. We identify certain restrictions on the
specification to ensure existence of a consistent linearization. We then
leverage the definition of replication-aware linearizability to propose an automated verification
methodology based on induction on operation sequences. We have successfully
applied the technique on a number of complex MRDTs. While the foundations have
been laid in this work, we believe there is a lot of scope for enriching the
technique even further by considering more complex linearization strategies.
\vspace{-2em}
\vspace{-0.4em}
\section{Data-Availability Statement}
\label{sec:data}
The artifact supporting this paper is available on Zenodo \cite{vimalaartifact} 
as well as in our GitHub repository \cite{vimalagit}.
It provides our framework in the F$^\star$ programming language that allows implementing 
MRDTs/CRDTs and automatically proving the verification conditions (VCs) required by our technique.

\bibliographystyle{ACM-Reference-Format}
\bibliography{main}

\appendix
\section{Appendix}
\label{sec:app}

\subsection{Proofs of \S \ref{sec:lin}}
\label{subsec:lcaproof}
\noindent \textbf{Lemma~\ref{lem:LCA}} Given a configuration $C = \langle N,H,L,G,vis \rangle$ reachable in some execution $\tau \in \llbracket \M{S_D} \rrbracket$ and two versions $v_1,v_2 \in dom(N)$, if $v_\top$ is the LCA of $v_1$ and $v_2$ in $G$, then $L(v_\top) = L(v_1) \cap L(v_2)$.
\begin{proof}
If $(v,v') \in E$, then $L(v) \subseteq L(v')$. This is because either $L(v') = L(v) \cup \{e\}$ for some event $e$ due to the $\F{apply}$ transition, or $L(v') = L(v) \cup L(v'')$ due to the $\F{merge}$ transition. 

Hence, if $(v,v') \in E^*$, then $L(v) \subseteq L(v')$.

Since $(v_\top,v_1) \in E^*$ and $(v_\top,v_2) \in E^*$, hence $L(v_\top) \subseteq L(v_1)$ and $L(v_\top) \subseteq L(v_2)$. Hence, $L(v_\top) \subseteq L(v_1) \cap L(v_2)$.

Consider vertices $u,w$ and event $e$ such that $(u,w) \in E$, $e \notin L(u)$, $e \in L(w)$ and in-degree of $w$ is 1. Then $w$ is called the $generator$ vertex of event $e$. Note that there will always be a unique generator vertex for each event.

\begin{proposition}\label{prop:lca}
	For all versions $v$, events $e$, if $e \in L(v)$, and $w$ is the generator version of $e$, then $(w,v) \in E^*$.
\end{proposition}
 
Consider $e \in L(v_1) \cap L(v_2)$. Then if $w$ is the generator version of $e$, by Proposition \ref{prop:lca} $(w,v_1) \in E^*$ and $(w,v_2) \in E^*$. Then, by definition of LCA, $(w,v_\top) \in E^*$. Hence, $L(w) \subseteq L(v_\top)$. This implies that $e \in L(v_\top)$. Thus, $ L(v_1) \cap L(v_2)  \subseteq L(v_\top)$.

We now prove Proposition \ref{prop:lca}. If $v$ has in-degree 2, then suppose $(w_1,v) \in E$, $(w_2,v) \in E$ and $L(v) = L(w_1) \cup L(w_2)$. Then either $e \in L(w_1)$ or $e \in L(w_2)$. WLOG, suppose $e \in L(w_1)$. We now recursively apply Proposition \ref{prop:lca} on $w_1$. Then, $(w,w_1) \in E^*$, which implies $(w,v) \in E^*$. 

If $v$ has in-degree 1, then suppose $(u,v) \in E$. If $e \in L(u)$, we recursively apply Proposition \ref{prop:lca} on $u$. If $e \notin L(u)$, then $v$ itself is the generator version of $e$, and clearly, $(v,v) \in E^*$.

Note that everytime we move backwards along an edge by recursively applying Proposition \ref{prop:lca}, we are either decreasing the number of events in the source vertex, or the number of unvisited vertices in the graph while still retaining $e$. Since the graph is acyclic and finite, and the number of events are also finite, eventually, we will hit the generator version. \qed

\end{proof}

\noindent \textbf{Recursive Merge Strategy}: For a given version graph $G = (V,E)$, for versions $v_1,v_2$, if the LCA does not exist, then our strategy is to find potential LCAs. For each potential LCA $v_p$, $(v_p,v_1) \in E^*$, $(v_p,v_2) \in E^*$ and $\nexists v.\ (v,v_1) \in E^* \wedge (v,v_2) \in E^* \wedge (v_p,v) \in E^*$. Note that since the version graph is rooted at the initial version $v_0$, a common ancestor of any two versions $v_1$ and $v_2$ always exist. Let $V_c$ be the set of all common ancestors of $v_1$ and $v_2$.
$$
V_c = \{v \in V\ \mid\ (v,v_1) \in E^* \wedge (v,v_2) \in E^*\}
$$

For two common ancestors $v,v' \in V_c$, either there is a path between them or there isn't. If there is a path, say $(v,v') \in E^*$, then $v$ can neither be a potential or actual LCA. In this way, we eliminate all common ancestors which cannot be potential or actual LCAs. Finally, we are left with the set of potential LCAs $V_p$. Hence, for any $v,v' \in V_p$, $(v,v') \notin E^*$ and $(v',v) \notin E^*$. It is then clear to see that if $V_p = \{v_\top\}$, i.e. $V_p$ is singleton, then $v_\top$ must be the actual LCA, because every other common ancestor $v$ must have been eliminated due to $(v,v_\top) \in E^*$.

Otherwise, if $V_p$ is not singleton, we pairwise invoke $\F{merge}$ on every pair of versions in $V_p$. Note that we would have to repeat the same merge strategy while merging any two versions in $V_p$. We now show that if $v_m$ is the version obtained by merging all the versions in $V_p$, then $L(v_m) = L(v_1) \cap L(v_2)$. Since every version $v \in V_p$ is a common ancestor of $v_1$ and $v_2$, $L(v) \subseteq L(v_1) \cap L(v_2)$, and hence $L(v_m) \subseteq  L(v_1) \cap L(v_2)$. Consider $e \in  L(v_1) \cap L(v_2)$. Now, consider the generator version $w$ of $e$. By Proposition \ref{prop:lca}, $w$ is a common ancestor of $v_1$ and $v_2$. Either $w \in V_p$, in which case by merging $w$ to get $v_m$, we would have $e \in L(v_m)$. Or else, $w$ would have been eliminated, in which case there will exist some version $v \in V_p$ such that $(w,v) \in E^*$. Hence, $e \in L(v)$, which implies $e \in L(v_m)$.  \\

\noindent \textbf{Lemma~\ref{lem:non-comm}}
	Given a set of events $\M{E}$, if $\F{lo} \subseteq \M{E} \times \M{E}$ is defined over
	every pair of non-commutative events in $\M{E}$, then for any two sequences
	$\pi_1, \pi_2$ which extend $\F{lo}$, for any state $\sigma$, $\pi_1(\sigma)
	= \pi_2(\sigma)$.
\begin{proof}
	If $\pi_1=\pi_2$, then the result trivially holds. Consider the first point of difference between $\pi_1$ and $\pi_2$.\\
	$\pi_1 = \tau.e_1.\tau_1, \pi_2 = \tau.e_2.\tau_2$.\\
	Then $e_1$ must appear somewhere in $\tau_2$.\\
	$\pi_2 = \tau.e_2.\tau_3.e_1.\tau_4$.\\
	We consider two cases here:\\
	\textbf{Case 1: $(\tau_3 = \phi)$}\\
	Since $e_1$ and $e_2$ are in different orders in $\pi_1$ and $\pi_2$, neither $e_1 \xrightarrow{\F{lo}} e_2$ nor
	$e_2 \xrightarrow{\F{lo}} e_1$. Since $\F{lo}$ is defined over every pair of non-commutative events, but is not defined between $e_1$ and $e_2$, they must commute, Hence, we can flip $e_2$ and $e_1$ in $\pi_2$, leading to the same state.\\
\textbf{Case 2: $(\tau_3 \neq \phi)$}\\
	$\pi_2 = \tau.e_2.\tau_5.e_3.e_1.\tau_4$. Then in $\pi_1$, $e_3$ is not present in $\tau$, hence it must be present after $e_1$.
	Now $e_1$ and $e_3$ are in different orders in $\pi_1$ and $\pi_2$, hence neither $e_1 \xrightarrow{\F{lo}} e_3$ nor
	$e_3 \xrightarrow{\F{lo}} e_1$. 
	
	By the same argument as above applied on $e_1$ and $e_2$, we can flip $e_1$ and $e_3$ in $\pi_2$.
	We keep doing this for all events in $\tau_5$ until $e_2$ is adjacent to $e_1$ after which we can flip them.
	Thus we can change $\pi_2$ such that $e_1$ will appear in the same position in $\pi_1$. We can keep 
	doing this until $\pi_1$ and $\pi_2$ are identical.
\end{proof}

\noindent \textbf{Lemma~\ref{lem:irreflexive}} For an MRDT $\mathcal{D}$ such that $\F{rc}^+$ is irreflexive, for any configuration $C$ reachable in
	$\mathcal{S}_\mathcal{D}$, $\F{lo}_C^+$ is irreflexive.\\

To prove that $\F{lo}_C^+$ is irreflexive, we need to prove that there cannot be cycles formed out of $\F{lo}_C$ edges.
\begin{proof}
A cycle cannot be formed using only $\F{vis}$ edges, as $\F{vis}^+$ is irreflexive. Similarly, a cycle cannot be formed using only $\F{rc}$ edges, as $\F{rc}^+$ is irreflexive. Therefore, any potential cycle must consist of adjacent $\xrightarrow{\F{rc}}$ and $\xrightarrow{\F{vis}}$ edges.
Consider three events $e_1, e_2, e_3$ such that $e_1 \xrightarrow[\F{rc}]{\F{lo}} e_2 \xrightarrow[\F{vis}]{\F{lo}} e_3$.
Since $e_1 \xrightarrow[\F{rc}]{\F{lo}} e_2$, this implies $e_1 \xrightarrow{\F{rc}} e_2 \wedge e_1 \mid\mid_C e_2$. Given that $e_2 \xrightarrow{\F{vis}} e_3$, the relation $e_1 \xrightarrow[\F{rc}]{\F{lo}} e_2$ is not possible. Thus, this case is also not feasible.
Hence, there cannot be cycles formed out of $\F{lo}_C$ edges. 
\end{proof}

\noindent \textbf{Lemma~\ref{lem:convergence}}  For an MRDT $\M{D}$ which satisfies $\textrm{\sc{rc-non-comm}}(\M{D})$ and $\textrm{\sc{cond-comm}}(\M{D})$, for any reachable configuration $C$ in $\M{S}_\M{D}$, for any two sequences $\pi_1,\pi_2$ over $E_C$ which extend $\F{lo}_C$, for any state $\sigma$, $\pi_1(\sigma) = \pi_2(\sigma)$.
\begin{proof}
Consider the first point of difference between $\pi_1$ and $\pi_2$.\\
$\pi_1 = \tau.e_1.\tau_1, \pi_2 = \tau.e_2.\tau_2$.\\
Then $e_1$ must appear somewhere in $\tau_2$.\\
$\pi_2 = \tau.e_2.\tau_3.e_1.\tau_4$.\\
We consider two cases here:\\
\textbf{Case 1: $(\tau_3 = \phi)$}\\
Since $e_1$ and $e_2$ are in different orders in $\pi_1$ and $\pi_2$, neither $e_1 \xrightarrow{\F{lo}} e_2$ nor
$e_2 \xrightarrow{\F{lo}} e_1$. If either $e_1 \xrightarrow{\F{vis}} e_2$  or $e_2 \xrightarrow{\F{vis}} e_1$, it 
must be the case that $e_1 \rightleftarrows e_2$. In this case, we can flip the order of $e_1$ and $e_2$ in $\pi_2$
leading to the same state. 
Suppose $e_1 \mid\mid_C e_2$, if neither $e_1 \xrightarrow{\F{rc}} e_2$ nor $e_2 \xrightarrow{\F{rc}} e_1$, $e_1 \rightleftarrows e_2$.
In this case, we can again flip them in $\pi_2$.
Suppose $e_1 \xrightarrow{\F{rc}} e_2$, since $\neg (e_1 \xrightarrow{\F{lo}} e_2)$, by definition of $\F{lo}$,
$\exists e_3. e_2 \xrightarrow{\F{lo}} e_3$. Then $\neg (e_2 \rightleftarrows e_3)$. By \textrm{\sc{cond-comm}},
it must be the case that $e_1  \overset{e_3}{\rightleftarrows}  e_2$. Since $e_2 \xrightarrow{\F{lo}} e_3$, $e_3$ must be 
present in $\tau_4$. By definition of \textrm{\sc{cond-comm}}, we can flip $e_2$ and $e_1$ in $\pi_2$, leading to the same state.
Similar argument can be applied to $e_2 \xrightarrow{\F{rc}} e_1$.\\
\textbf{Case 2: $(\tau_3 \neq \phi)$}\\
$\pi_2 = \tau.e_2.\tau_5.e_3.e_1.\tau_4$. Then in $\pi_1$, $e_3$ is not present in $\tau$, hence it must be present after $e_1$.
Now $e_1$ and $e_2$ are in different orders in $\pi_1$ and $\pi_2$, hence neither $e_1 \xrightarrow{\F{lo}} e_3$ nor
$e_3 \xrightarrow{\F{lo}} e_1$. 

By the same argument as above applied on $e_1$ and $e_2$, we can flip $e_1$ and $e_3$ in $\pi_2$.
We keep doing this for all events in $\tau_5$ until $e_2$ is adjacent to $e_1$ after which we can flip them.
Thus we can change $\pi_2$ such that $e_1$ will appear in the same position in $\pi_1$. We can keep 
doing this until $\pi_1$ and $\pi_2$ are identical.
\end{proof}

\noindent \textbf{Lemma~\ref{lem:query}} If MRDT $\mathcal{D}$ is RA-linearizable, then for all executions $\tau \in \llbracket \mathcal{S}_\mathcal{D} \rrbracket$, and for all transitions $C \xrightarrow{query(r,q,a)} C'$ in $\tau$, where $C = \langle N, H, L, G, vis\rangle$, there exists a sequence $\pi$ consisting of all events in $L(H(r))$ such that $\F{lo}(C)_{\mid L(H(r))} \subseteq \pi$ and $a = \F{query}(\pi(\sigma_0), q)$.

\begin{proof}
Consider an MRDT $\mathcal{D}$ that is RA-linearizable. Let $\tau = C_0 \xrightarrow{t_1} C_1 \xrightarrow{t_2} C_2 \ldots \xrightarrow{t_n} C$ 
be an execution of $\mathcal{S}_\mathcal{D}$, where $\{t_1, \dots, t_n\}$ are the labels of the transition system. For a transition $C \xrightarrow{query(r, q, a)} C'$ in $\tau$, where $C = \langle N, H, L, G, vis\rangle$, we know that $C$ is RA-linearizable from Def.~\ref{def:lin}. That is, for every active replica $r \in \mathrm{range}(H)$, there exists a sequence $\pi$ such that $\F{lo}(C)_{\mid L(H(r))} \subseteq \pi$ and $N(H(r)) = \pi(\sigma_0)$.
According to the semantics, we have $a = \F{query}(N(H(r)), q)$. Thus $a = \F{query}(\pi(\sigma_0), q)$.
\end{proof}

\subsection{Proofs of \S \ref{sec:lemmas}}

\noindent \textbf{Lemma \ref{lem:pi1}}
(1) For events $e \in L_1^a \cup L_2^a$, $e' \in L_1^b \cup L_2^b$, $\neg (e \xrightarrow{\F{lo_m}} e')$.
\begin{proof}
Suppose $e \xrightarrow{\F{lo_m}} e'$ is true. There are 2 possibilities:
\begin{enumerate}
	\item $e \xrightarrow[\F{vis}]{\F{lo}} e':$  
	By definition of $L_i^b$, there are 2 cases:
	\begin{enumerate}
		\item $\exists e_\top \in L_\top. e' \xrightarrow{\F{lo_m}} e_\top:$ But this would require $e$ to be in $L_1^b \cup L_2^b$.
		\item $\exists e_\top \in L_\top, e'' \in L_1' \cup L_2'. e' \xrightarrow{\F{lo_m}} e'' \xrightarrow{\F{lo_m}} e_\top:$
		\begin{enumerate}
			\item $e' \xrightarrow[\F{vis}]{\F{lo}} e'':$ Due to transitivity of $\F{vis}$, $e \xrightarrow{\F{vis}} e''$. This would require $e \in L_1^b \cup L_2^b$.
			\item $e'' \xrightarrow[\F{vis}]{\F{lo}} e_\top$ is not possible as $L_\top^a$ is causally closed.
			\item $e' \xrightarrow[\F{rc}]{\F{lo}} e'' \xrightarrow[\F{rc}]{\F{lo}} e_\top$ is not possible due to \textrm{\sc{no-rc-chain}} restriction.
		\end{enumerate}
	\end{enumerate}
	
	\item $e \xrightarrow[\F{rc}]{\F{lo}} e':$ 
	By definition of $L_i^b$, there are 2 cases:
	\begin{enumerate}
		\item $\exists e_\top \in L_\top. e' \xrightarrow{\F{lo_m}} e_\top:$ 
			\begin{enumerate}
				\item $e' \xrightarrow[\F{vis}]{\F{lo}} e_\top$ is not possible as $L_\top^a$ is causally closed. 
				\item $e' \xrightarrow[\F{rc}]{\F{lo}} e_\top$ is not possible due to \textrm{\sc{no-rc-chain}} restriction.
				Since $e \mid\mid_C e_\top$, we have $e \xrightarrow[\F{rc}]{\F{lo}} e_\top$ which requires $e \in L_1^b \cup L_2^b$.
			\end{enumerate}
		\item $\exists e_\top \in L_\top, e'' \in L_1' \cup L_2'. e' \xrightarrow{\F{lo_m}} e'' \xrightarrow{\F{lo_m}} e_\top:$
		\begin{enumerate}
			\item $e'' \xrightarrow[\F{vis}]{\F{lo}} e_\top$ is not possible as $L_\top^a$ is causally closed.
			\item $e' \xrightarrow[\F{rc}]{\F{lo}} e''$ is not possible due to \textrm{\sc{no-rc-chain}} restriction.
			\item $e' \xrightarrow[\F{vis}]{\F{lo}} e'' \xrightarrow[\F{rc}]{\F{lo}} e_\top:$ 
				$e' \xrightarrow{\F{rc}} e''$ creates RC-chain.
				Since $e' \mid\mid_C e_\top$, we have $e' \xrightarrow[\F{rc}]{\F{lo}} e_\top$
				which violates the \textrm{\sc{no-rc-chain}} restriction.
				$e'' \xrightarrow{\F{rc}} e'$ would requires $e$ and $e'$ to conditionally commute with each other.
				So $e \xrightarrow[\F{rc}]{\F{lo}} e'$ does not hold true.
		\end{enumerate}
	\end{enumerate}
\end{enumerate}
\end{proof}

(2) For events $e \in L_\top^a$, $e' \in L_\top^b$, $\neg (e \xrightarrow{\F{lo_m}} e')$.
\begin{proof}
By definition of $L_\top^a$, $\exists e'' \in L_1^b \cup L_2^b. e'' \xrightarrow{\F{lo_m}} e$.
	$e'' \xrightarrow{\F{vis}} e$ is not possible as $L_\top^a$ is causally closed.
Suppose $e \xrightarrow{\F{lo_m}} e'$ is true. There are 3 possibilities:
\begin{enumerate}
	\item $e \xrightarrow[\F{vis}]{\F{lo}} e':$  
	\begin{enumerate}
			\item $e'' \xrightarrow[\F{rc}]{\F{lo}} e:$ $e \xrightarrow{\F{rc}} e'$ causes RC-chain.
			Since $e'' \mid\mid_C e'$, we have
			 $e'' \xrightarrow[\F{rc}]{\F{lo}} e'$ which requires $e' \in L_\top^a$.
			 $e' \xrightarrow{\F{rc}} e$ cause $e''$ and $e$ to conditionally commute with each other. So this case does not hold true.
	\end{enumerate}
	
	\item $e \xrightarrow[\F{rc}]{\F{lo}} e':$  
	$e'' \xrightarrow{\F{vis}} e$ is not possible as $L_\top^a$ is causally closed.
	\begin{enumerate}
			\item $e'' \xrightarrow[\F{rc}]{\F{lo}} e:$ causes RC-chain.		 
	\end{enumerate}
\end{enumerate}
\end{proof}

\noindent \textbf{Lemma \ref{lem:pi2}}
(1) For events $e_i^{\top}, e_j^{\top} \in L_\top^a$, where $L_\top^a = \{e_1^{\top}, \ldots, e_m^{\top}\}$, $\neg (e_i^{\top} \xrightarrow{\F{lo_m}} e_j^{\top})$.
\begin{proof}
By definition of $L_\top^a$, $\exists e \in L_1^b(e_i^{\top}) \cup L_2^b(e_i^{\top}). e \xrightarrow{\F{lo_m}} e_i^{\top}$. 
$e \xrightarrow{\F{vis}} e_i^{\top}$ is not possible as $L_\top^a$ is causally closed.
Suppose $e_i^{\top} \xrightarrow{\F{lo_m}} e_j^{\top}$. There are 3 possibilities.
\begin{enumerate}
	\item $e_i^{\top} \xrightarrow[\F{vis}]{\F{lo}} e_j^{\top}:$  
	\begin{enumerate}
		\item $e \xrightarrow[\F{rc}]{\F{lo}} e_i^{\top}:$ $e_i^{\top} \xrightarrow{\F{rc}} e_j^{\top}$ causes RC-chain. Since $e \mid\mid_C e_j^{\top}$, we have
			 $e \xrightarrow[\F{rc}]{\F{lo}} e_j^{\top}$ which requires $e \in L_1^b(e_j^{\top}) \cup L_2^b(e_j^{\top})$.
			 But $e$ belongs to $L_1^b(e_i^{\top}) \cup L_2^b(e_i^{\top})$.
			 $e_j^{\top} \xrightarrow{\F{rc}} e_i^{\top}$ cause $e$ and $e_i^{\top}$ to conditionally commute with each other. So this case does not hold true.
		
	\end{enumerate}

	\item $e_i^{\top} \xrightarrow[\F{rc}]{\F{lo}} e_j^{\top}:$
	\begin{enumerate}
		\item $e \xrightarrow[\F{rc}]{\F{lo}} e_i^{\top}:$ By \textrm{\sc{no-rc-chain}} restriction, this case cannot happen.
	\end{enumerate}
\end{enumerate}
\end{proof}

(2) For events $e \in L_1^b(e_i^{\top}) \cup L_2^b(e_i^{\top})$, $e' \in L_1^b(e_j^{\top}) \cup L_2^b(e_j^{\top})$ where $j<i$, $\neg (e \xrightarrow{\F{lo_m}} e')$.
\begin{proof}
Suppose $e \xrightarrow{\F{lo_m}} e'$, $\neg (e \rightleftarrows e')$.
By definition of $L_1^b(e_i^{\top})$ and $L_2^b(e_i^{\top})$, we know that $e \xrightarrow{\F{lo}} e_i^{\top}$ and $e' \xrightarrow{\F{lo}} e_j^{\top}$.
We consider several possibilities based on this:
\begin{enumerate}
	\item Neither $e \xrightarrow[\F{vis}]{\F{lo}} e_i^{\top}$ nor $e' \xrightarrow[\F{vis}]{\F{lo}} e_j^{\top}$ is true because
		$L_\top^a$ is causally closed.
	\item $e \xrightarrow[\F{rc}]{\F{lo}} e_i^{\top} \wedge e' \xrightarrow[\F{rc}]{\F{lo}} e_j^{\top}:$
	\begin{enumerate}
		\item $e \xrightarrow{\F{rc}} e' \vee e' \xrightarrow{\F{rc}} e$ creates RC chain.
	\end{enumerate}	
\end{enumerate}
\end{proof}

\noindent \textbf{Theorem~\ref{thm:1}}
If an MRDT $\M{D}$ satisfies  $\textrm{\sc{BottomUp-2-OP}}$,  $\textrm{\sc{BottomUp-1-OP}}$,  $\textrm{\sc{BottomUp-0-OP}}$, 
\\$\textrm{\sc{MergeIdempotence}}$ and $\textrm{\sc{MergeCommutativity}}$, then $\M{D}$ is linearizable.
\begin{proof}
To prove that $\M{D}$ is linearizable, we will prove that any execution $\tau \in \llbracket \M{S_D} \rrbracket$ is linearizable, for which we will show that all of its configurations are linearizable.
Let $\tau = C_0 \xrightarrow{t_1} C_1 \xrightarrow{t_2} C_2 \ldots \xrightarrow{t_n} C$
be an execution of $\M{S_D}$, where $\{t_1,\dots, t_n\}$ are labels of the transition system. 
We prove by induction on the length of $\tau$. Base case of $C_0$ which consists of only 
one replica $r_0$ is trivially satisfied, as no operations are applied on the head version $v_0$ at $r_0$. 
Assuming the required result holds in the execution $C_0 \rightarrow^{*} C$, 
and suppose there is a new transition $C \rightarrow C^{'}$, we need to prove that $C$ is linearizable.
There are four cases corresponding to the four transition rules given in Fig. 8.

\subsubsection{Case $({\textrm{\sc{CreateBranch}})}$:}
Assume that a new replica $r^{'}$ is forked off from the origin replica $r$. Let $C = \langle N, H, L, G, vis \rangle$ 
and $C^{'} = \langle N^{'}, H^{'}, L^{'}, G^{'}, vis \rangle$ be the configurations of the replica before and after the branch creation.
According to the semantics, we have $L(H(r)) = L^{'}(H^{'}(r^{'}))$ and $N(H(r)) = N^{'}(H^{'}(r^{'}))$. We need to prove that Def.~\ref{def:lin} 
holds for $C^{'}$. This is obvious since Def.~\ref{def:lin} holds for C by the induction assumption.

\subsubsection{Case $({\textrm{\sc{Apply}}})$:}
Assume that an event $e$ is applied on a replica $r$. Let $C = \langle N, H, L, G, vis \rangle$ 
and $C^{'} = \langle N^{'}, H^{'}, L^{'}, G^{'}, vis^{'} \rangle$ be the configurations of the replica before and after the apply operation.
By semantics we have $L^{'}(H^{'}(r)) = L(H(r)) \cup \{e\}$. We need to prove that Def.~\ref{def:lin} holds for $C^{'}$. 
By induction assumption, $\exists \pi.\ \F{lo}(C)_{\mid L(H(r))} \subseteq \pi \wedge N(H(r)) = \pi(\sigma_0)$. 
Here $\F{lo}(C^{'})_{\mid L^{'}(H^{'}(r))}$ is the linearization order $\F{lo}(C)_{\mid L(H(r))}.e$ and $\pi^{'} = \pi.e$.
We need to show that $\pi^{'}$ extends  $\F{lo}(C^{'})_{\mid L^{'}(H^{'}(r))}$. 
We have $N^{'}(H^{'}(r)) = e(\pi(\sigma_0))$. Event $e$ is visible to all events in $\pi$ according to the semantics of $\F{apply}$. 
Since $\forall e^{'} \in \pi. e^{'} \xrightarrow[\F{vis}]{\F{lo}} e$,  $e \xrightarrow[\F{vis}]{\F{lo}} e^{'}$ is not possible due to
anti-symmetry of $\F{vis}$. $e \xrightarrow[\F{rc}]{\F{lo}} e^{'}$ is also
not possible as it would require $e$ and $e^{'}$ to be concurrent events. 
Hence, $\pi^{'}$ is a total order which extends $\F{lo}(C^{'})_{\mid L^{'}(H^{'}(r))}$. This proves the required result.

\subsubsection{Case $({\textrm{\sc{Merge}}})$:}
Consider there is a $\F{merge}(r_1,r_2)$ transition to $C^{'}$ where $r_2$ merges with $r_1$.
Let $C = \langle N, H, L, G, vis \rangle$, $C' = \langle N', H', L', G', vis \rangle$ , and let $H(r_1) = v_1, H(r_2) = v_2$. Let $v_\top$ be the 
LCA of $v_1$ and $v_2$ in $G$. Let $N(v_1) = a$, $N(v_2) = b$, $N(v_\top) = l$. 
The transition will install a version $v_m$ with state $m = \F{merge}(l, a, b)$ at 
the replica $r_1$, leaving the other replicas unchanged. Also, $L'(v_m) = L(v_1) \cup L(v_2)$. We need to show that 
there exists a sequence $\pi_m$ of events in $L'(v_m)$ such that $\pi_m$ extends $\F{lo}(C')_{\mid L'(v_m)}$ and $m = \pi(\sigma_0)$.
For ease of readability, we use $L_1$ for $L(v_1)$, $L_2$ for $L(v_2)$ and $L_\top$ for $L(v_\top)$, and $\F{lo_m}$ for $\F{lo}(C')_{\mid L'(v_m)}$. 

We repeat the definitions of various event sets below:
\begin{align*}
	& L_1' = L_1 \setminus L_\top \quad \quad  L_2' = L_2 \setminus L_\top \\
	& L_1^b = \{e \in L_1^{'}\ \mid\ \exists e_\top \in L_\top.\ (e \xrightarrow{\F{lo_m}} e_\top \vee  \exists e' \in L_1^{'}.\ e \xrightarrow{\F{lo_m}} e^{'} \xrightarrow{\F{lo_m}} e_\top)\}\\
	& L_2^b = \{e \in L_2^{'}\ \mid\ \exists e_\top \in L_\top.\ (e \xrightarrow{\F{lo_m}} e_\top \vee \exists e' \in L_2^{'}.\ e \xrightarrow{\F{lo_m}} e^{'} \xrightarrow{\F{lo_m}} e_\top)\}\\
	& L_\top^{a} = \{e_\top \in L_\top \mid \exists e \in L_1^{b} \cup L_2^{b}. e  \xrightarrow{\F{lo_m}} e_\top\}\\
	&  L_1^a = L_1^{'} \setminus L_1^b \quad  \quad L_2^a = L_2^{'} \setminus L_2^b \quad \quad L_\top^{b} = L_\top \setminus L_\top^{a}
\end{align*}

Let $\F{lo_i} = \F{lo}(C')_{\mid L_i}$ for $i=1,2$.

First we will prove that $\F{lo}$ between two events should remain the same in all versions.
$\forall e, e^{'} \in L_i. e \xrightarrow{\F{lo_i}} e^{'} \Leftrightarrow e \xrightarrow{\F{lo_m}} e^{'}$. Note that $\F{vis}$ and $\F{rc}$ ordering between events remains same in
$L_i$ and $L'(v_m)$. 
\begin{itemize}
	\item If $e \xrightarrow{\F{rc}} e^{'}, e \mid\mid_C e^{'}$ and $\neg(\exists e^{''} \in L(v_i). e^{'} \xrightarrow{\F{vis}} e^{''} \wedge \neg e^{'} \rightleftarrows e^{''})$, 
	then these constraints will continue to hold in $L_m$. Because it is not possible that $e^{'} \in L_1^{'} , e^{''} \in L_2^{'}$
	such that $e^{'} \xrightarrow{\F{vis}} e^{''}$. Because otherwise $e^{'} \in L_2^{'} \Rightarrow e^{'} \in L_\top$. 
	\item If $e\xrightarrow{\F{vis}} e^{'} \wedge \neg e \rightleftarrows e^{'}$ in $L_i$, then it continues to hold in $L_m$.
\end{itemize}

By induction assumption, we know that\\
$\exists \pi_a.\ \F{lo}(C)_{\mid L(v_1)} \subseteq \pi_a \wedge a = \pi_a(\sigma_0)$\\
$\exists \pi_b.\ \F{lo}(C)_{\mid L(v_2)} \subseteq \pi_b \wedge b = \pi_b(\sigma_0)$\\
$\exists \pi_\top.\ \F{lo}(C)_{\mid L(v_\top)} \subseteq \pi_\top \wedge l = \pi_\top(\sigma_0)$\\

To start off, let's consider the set $L_1^a \cup L_2^a$. These are all local events of $v_1$ and $v_2$, which are not linearized before events of the LCA. 
We consider different cases depending on the size if this set.\\

\noindent\textrm{\sc{Case 1}}: $(\mid L_1^{a} \cup L_2^{a} \mid = 0)$\\
We note that in this case, $a,b$ can be defined as follows:
$a= {\pi_a}_{\mid (L_\top^{b} \cup L_1^{b} \cup L_\top^{a})}(\sigma_0)$,
$b = {\pi_b}_{\mid (L_\top^{b} \cup L_2^{b} \cup L_\top^{a})}(\sigma_0)$.\\
We need to show that there exists a sequence $\pi_m$ that extends $\F{lo_m}$ such that $\F{merge}(l,a,b) = \pi_m(\sigma_0)$.
Here, we induct on the size of the set $L_\top^a$.\\

\noindent\textrm{\sc{Base Case 1}}: $(\mid L_\top^a \mid  = 0)$\\
Then $L_1^b \cup L_2^b = \phi$. So $l = a = b$. $\F{merge}(l, l, l) = l$ is inferred by $\textrm{\sc{MergeIdempotence}}$. We know that $l$ is correctly linearized, hence the required result follows.\\

\noindent\textrm{\sc{Inductive Case 1}}: $(\mid L_\top^a  \mid > 0)$\\
Let $L_\top^a = \{e_1^{\top}, \dots, e_{m-1}^{\top}, e_m^{\top}\}$. Let $S = \{e_1^{\top}, \dots, e_{m-1}^{\top}\}$.
By IH, for the set $S$, we have the required result. We define $l',a',b'$ based on the above set $S$: $l'=\pi_{l_{\mid L_\top^b \cup S}}(\sigma_0)$, $a' = \pi_{a_{\mid L_\top^b \cup \bigcup_{e \in S} L_1^b(e) \cup S}}(\sigma_0)$, $b' = \pi_{b_{\mid L_\top^b \cup \bigcup_{e \in S} L_2^b(e) \cup S}(\sigma_0)}$. Note that in this case, all the LCA events which are linearized after local events are already taken as part of the states $l',a',b'$. Now, suppose we add one more LCA event $e_m^{\top}$ to all states. We define $a'',b''$ such that
$a'' = {\pi_a}_{\mid L_1^b(e_m^{\top})} (a')$,
$b'' = {\pi_b}_{\mid L_2^b(e_m^{\top})} (b')$. 

Then, $l = e_m^{\top}(l'), a = e_m^{\top}(a''), b = e_m^{\top}(b'')$.  $e_m^{\top}$ is not linearized before any of the events in $L_\top^{b} \cup L_1^{b} \cup L_2^b \cup S$  based on the definition of $L_\top^a$.

Now, by $\textrm{\sc{BottomUp-0-OP}}$ rule,

\begin{equation}\label{eq:0op}
	\F{merge}(e_m^{\top}(l'), e_m^{\top}(a''), e_m^{\top}(b'')) = e_m^{\top}(\F{merge}(l',a'',b''))
\end{equation}

Now that we have linearized $e_m^{\top}$, we need to linearize the events that led to $\F{merge}(l',a'',b'')$. 
Let's denote $L_1^b(e_m^{\top})$ as $M_1^a$ and $L_2^b(e_m^{\top})$ as $M_2^a$. Now we induct on the size of the set $M_1^{a} \cup M_2^{a}$.\\

\noindent\textrm{\sc{Base Case 1.1}}:$(\mid M_1^{a} \cup M_2^{a} \mid = 0)$\\
$a'' = a', b'' = b'$. By induction assumption, $\exists \pi.\ \F{lo}(C)_{\mid (L_\top^b \cup \bigcup_{e \in S} L_1^b(e) \cup \bigcup_{e \in S} L_2^b(e) \cup S)} \subseteq \pi$\\ 
and $\F{merge}(l',a',b') = \pi(\sigma_0)$. Hence, $\pi_m = \pi. e_m^{\top}$.\\

\noindent\textrm{\sc{Inductive Case 1.1}}:$(\mid M_1^{a} \cup M_2^{a} \mid > 0)$\\
We have 2 cases here: \\(1.1.1) Either of $M_1^{a}$ or $M_2^{a}$ is $\phi$ \\(1.1.2) Both $M_1^{a}$ and $M_2^{a}$ are not $\phi$.\\

\noindent\textrm{\sc{Case 1.1.1}}:$(M_1^a \neq \phi \wedge M_2^a = \phi)$\\
Consider $e_1 \in M_1^a$ such that there does not exist $e \in M_1^a$ and $e_1 \xrightarrow{\F{lo_m}} e$, i.e. $e_1$ is the maximal event according to $\F{lo_m}$. Since $\F{lo}$ ordering between events remains the same in all versions, and since versions $v_1$ and $v_2$ (which are being merged) were already linearizable, there would exist sequences leading to the states $a$ such that $e_1$ would appear at the end. Hence, there exists $a'''$ such that $a'' = e_1(a''')$. Since $M_2^a$ is empty, all local events in $L_2$ are linearized before the rest of the  LCA events.
Suppose $L_\top^a \setminus {e_m^{\top}} \neq \phi$ or $L_\top^b \neq \phi$, the last event which leads to the state $l',b''$ must be an LCA event.  Let's consider $e_\top$ to be the maximal event in $L_\top$ according to $\F{lo_m}$.
Hence there exists states $l'', b'''$ such that $l' = e_\top(l''), b'' = e_\top(b''')$.
By $\textrm{\sc{BottomUp-1-OP}}$ rule
\begin{equation}\label{eq:1op9}
	\F{merge}(e_\top(l''), e_1(a'''), e_\top(b''')) = e_1(\F{merge}(e_\top(l''),a''',e_\top(b''')))
\end{equation}

If both  $L_\top^a \setminus {e_m^{\top}} = \phi$ and $L_\top^b = \phi$, then $l' = b'' = \sigma_0$. By $\textrm{\sc{BottomUp-1-OP}}$

\begin{equation*}
	\F{merge}(\sigma_0, e_1(a'''), \sigma_0) = e_1(\F{merge}(\sigma_0,a''', \sigma_0))
\end{equation*}

From the induction assumption, we get that $\F{merge}(e_\top(l''),a''',e_\top(b'''))$ is already obtained by 
the linearization of events applied on the initial state $\sigma_0$. That is, there exists a sequence $\pi'$ over events in $L_\top^b \cup \bigcup_{e \in S} L_1^b(e) \cup \bigcup_{e \in S} L_2^b(e) \cup S \cup M_1^a \setminus e_1$ which extends $\F{lo_m}$ relation such that $\F{merge}(e_\top(l''),a''',e_\top(b''')) = \pi'(\sigma_0)$. Now, $\pi = \pi'. e_1$ is the required linearization. 

Let $\F{lo}_1$ be the linearization relation for $\F{merge}(e_\top(l''),a''',e_\top(b'''))$ (i.e. from the RHS in Eq. \eqref{eq:1op9}, without the event $e_1$) and let $\F{lo}_2$ be the linearization relation for $\F{merge}(l',a'',b'')$ (i.e. the LHS in Eq. \eqref{eq:1op9}). Then $\pi'$ according to the IH extends $\F{lo}_1$. We will show that for any pair of events $e,e'$ in $\F{merge}(e_\top(l''),a''',e_\top(b'''))$ , $e \xrightarrow{\F{lo}_2} e' \implies e \xrightarrow{\F{lo}_1} e'$. This ensures that if $\pi$ extends $\F{lo}_2$. Now, the $\F{vis}$ and $\F{rc}$ relation between $e'$ and $e$ remains the same while determining both $\F{lo}_1$ and $\F{lo}_2$. If $e \xrightarrow[\F{rc}]{\F{lo}_2} e'$, then $e'$ cannot be visible to any non-commutative event while calculating $\F{lo}_2$, but then the same should be true for $\F{lo}_1$ as well. If $e \xrightarrow[\F{vis}]{\F{lo}_2} e'$, then clearly $e \xrightarrow[\F{vis}]{\F{lo}_1} e'$. This concludes the proof that $\pi = \pi'. e_1$ must extend $\F{lo}_2$.\\

\noindent\textrm{\sc{Case 1.1.2}}:$(M_1^a \neq \phi \wedge M_2^a \neq \phi)$\\
Consider $e_1 \in M_1^a, e_2 \in M_2^a$ such that there does not exist $e \in M_i^a$ and $e_i \xrightarrow{\F{lo_m}} e$ (for $i=1,2$), i.e. each of the $e_i$s are maximal events according to $\F{lo_m}$. Since $\F{lo}$ ordering between events remains the same in all versions, and since versions $v_1$ and $v_2$ (which are being merged) were already linearizable, there would exist sequences leading to the states $a''$ and $b''$ such that $e_1$ and $e_2$ would appear at the end resp. Hence, there exists $a'''$ and $b'''$ such that $a'' = e_1(a''')$ and $b'' = e_1(b''')$. Since $e_1 \mid\mid_C e_2$, they are related to each other by $\F{rc}$ relation or they commute with each other i.e., $e_1 \xrightarrow{\F{rc}} e_2 \vee e_2 \xrightarrow{\F{rc}} e_1 \vee e_1 \rightleftarrows e_2$.
We will consider the case when $e_2 \xrightarrow{\F{rc}} e_1 \vee e_1 \rightleftarrows e_2$. $e_1 \xrightarrow{\F{rc}} e_2$ is handled by $\textrm{\sc{MergeCommutativity}}$.
The equation becomes 
\begin{equation}\label{eq:2op9}
\F{merge}(l', e_1(a'''), e_2(b''')) = e_1(\F{merge}(l', a''', e_2(b''')))
\end{equation}
which is the $\textrm{\sc{BottomUp-2-OP}}$ rule.\\

From the induction assumption, we get that $\F{merge}(l', a''', e_2(b'''))$ is already obtained by 
the linearization of events applied on the initial state $\sigma_0$. If $\pi'$ is the linearization for this merge, then $\pi = \pi'. e_1$ is the required linearization.

For this, we prove that $e_1$ is not linearized before any of the events in $M_1^a \textbackslash \{e_1\} \cup M_2^a$.
Clearly, $e_1$ is not linearized before any event in $M_1^a \textbackslash \{e_1\}$ because it is the maximal event on that branch.
Since $e_2 \xrightarrow{\F{rc}} e_1, e_1 \xrightarrow{\F{vis}} e_2$ is not possible.
$e_1 \xrightarrow{\F{rc}} e_2$ is not possible as $\F{rc}^+$ is irreflexive.
So $e_1 \xrightarrow{\F{lo}} e_2$ is not possible.
Let's assume there is some event $e$ in $ M_2^a \textbackslash \{e_2\}$ that comes $\F{lo}$ after $e_1$. There are 2 possibilities.
\begin{itemize}
	\item $e_1 \xrightarrow{\F{rc}} e:$ Since $e_2 \xrightarrow{\F{rc}} e_1$, this case is not possible due to $\textrm{\sc{no-rc-chain}}$ restriction.
	\item $e_1 \xrightarrow{\F{vis}} e:$ This is not possible as events in $M_2^a \textbackslash \{e_2\}$ are concurrent with $e_1$. 
		This is because every version is causally closed.
\end{itemize}

\noindent\textrm{\sc{Case 2}}: $(\mid L_1^{a} \cup L_2^{a} \mid > 0)$\\
The proof here will be identical to the proof of Inductive Case 1.1, substituting $L_1^a$ and $L_2^a$ for $M_1^a$ and $M_2^a$, and using the rules $\textrm{\sc{BottomUp-1-OP}}$, $\textrm{\sc{MergeCommutativity}}$ and  $\textrm{\sc{BottomUp-2-OP}}$.

\subsubsection{Case $({\textrm{\sc{Query}})}$:}
Assume that a query operation is applied on a replica $r$. Let $C = \langle N, H, L, G, vis \rangle$ 
be the configuration of the replica before the operation. According to the semantics, the configuration of the 
replica remains same after the query operation. By the induction hypothesis, Def.~\ref{def:lin} holds for the configuration $C$.
\end{proof}

\noindent \textbf{Theorem~\ref{thm:2}}
If an MRDT $\M{D}$ satisfies  the VCs $\psi^*(\textrm{\sc{BottomUp-2-OP}})$,  $\psi^*(\textrm{\sc{BottomUp-1-OP}})$,  \\$\psi^*(\textrm{\sc{BottomUp-0-OP}})$, $\textrm{\sc{MergeIdempotence}}$ and $\textrm{\sc{MergeCommutativity}}$, then $\M{D}$ is linearizable.

\begin{proof}
To prove that $\M{D}$ is linearizable, we will prove that any execution $\tau \in \llbracket \M{S_D} \rrbracket$ is linearizable, for which we will show that all of its configurations are linearizable.
Let $\tau = C_0 \xrightarrow{t_1} C_1 \xrightarrow{t_2} C_2 \ldots \xrightarrow{t_n} C$
be an execution of $\M{S_D}$, where $\{t_1,\dots, t_n\}$ are labels of the transition system. 
We prove by induction on the length of $\tau$. Base case of $C_0$ which consists of only 
one replica $r_0$ is trivially satisfied, as no operations are applied on the head version $v_0$ at $r_0$. 
Assuming the required result holds in the execution $C_0 \rightarrow^{*} C$, 
and suppose there is a new transition $C \rightarrow C^{'}$, we need to prove that $C$ is linearizable.
There are four cases corresponding to the four transition rules given in Fig. 8.

\subsubsection{Case $({\textrm{\sc{CreateBranch}})}$:}
Assume that a new replica $r^{'}$ is forked off from the origin replica $r$. Let $C = \langle N, H, L, G, vis \rangle$ 
and $C^{'} = \langle N^{'}, H^{'}, L^{'}, G^{'}, vis \rangle$ be the configurations of the replica before and after the branch creation.
According to the semantics, we have $L(H(r)) = L^{'}(H^{'}(r^{'}))$ and $N(H(r)) = N^{'}(H^{'}(r^{'}))$. We need to prove that Def.~\ref{def:lin}
holds for $C^{'}$. This is obvious since Def.~\ref{def:lin} holds for C by the induction assumption.

\subsubsection{Case $({\textrm{\sc{Apply}}})$:}
Assume that an event $e$ is applied on a replica $r$. Let $C = \langle N, H, L, G, vis \rangle$ 
and $C^{'} = \langle N^{'}, H^{'}, L^{'}, G^{'}, vis^{'} \rangle$ be the configurations of the replica before and after the apply operation.
By semantics we have $L^{'}(H^{'}(r)) = L(H(r)) \cup \{e\}$. We need to prove that Def.~\ref{def:lin} holds for $C^{'}$. 
By induction assumption, $\exists \pi.\ \F{lo}(C)_{\mid L(H(r))} \subseteq \pi \wedge N(H(r)) = \pi(\sigma_0)$. 
Here $\F{lo}(C^{'})_{\mid L^{'}(H^{'}(r))}$ is the linearization order $\F{lo}(C)_{\mid L(H(r))}.e$ and $\pi^{'} = \pi.e$.
We need to show that $\pi^{'}$ extends  $\F{lo}(C^{'})_{\mid L^{'}(H^{'}(r))}$. 
We have $N^{'}(H^{'}(r)) = e(\pi(\sigma_0))$. Event $e$ is visible to all events in $\pi$ according to the semantics of $\F{apply}$. 
Since $\forall e^{'} \in \pi. e^{'} \xrightarrow[\F{vis}]{\F{lo}} e$,  $e \xrightarrow[\F{vis}]{\F{lo}} e^{'}$ is not possible due to
anti-symmetry of $\F{vis}$. $e \xrightarrow[\F{rc}]{\F{lo}} e^{'}$ is also
not possible as it would require $e$ and $e^{'}$ to be concurrent events. 
Hence, $\pi^{'}$ is a total order which extends $\F{lo}(C^{'})_{\mid L^{'}(H^{'}(r))}$. This proves the required result.

\subsubsection{Case $({\textrm{\sc{Merge}}})$:}
Consider there is a $\F{merge}(r_1,r_2)$ transition to $C^{'}$ where $r_2$ merges with $r_1$.
Let $C = \langle N, H, L, G, vis \rangle$, $C' = \langle N', H', L', G', vis \rangle$ , and let $H(r_1) = v_1, H(r_2) = v_2$. Let $v_\top$ be the 
LCA of $v_1$ and $v_2$ in $G$. Let $N(v_1) = a$, $N(v_2) = b$, $N(v_\top) = l$. 
The transition will install a version $v_m$ with state $m = \F{merge}(l, a, b)$ at 
the replica $r_1$, leaving the other replicas unchanged. Also, $L'(v_m) = L(v_1) \cup L(v_2)$. We need to show that 
there exists a sequence $\pi_m$ of events in $L'(v_m)$ such that $\pi_m$ extends $\F{lo}(C')_{\mid L'(v_m)}$ and $m = \pi(\sigma_0)$.
For ease of readability, we use $L_1$ for $L(v_1)$, $L_2$ for $L(v_2)$ and $L_\top$ for $L(v_\top)$, and $\F{lo_m}$ for $\F{lo}(C')_{\mid L'(v_m)}$. 

By induction assumption, we know that\\
$\exists \pi_a.\ \F{lo}(C)_{\mid L(v_1)} \subseteq \pi_a \wedge a = \pi_a(\sigma_0)$\\
$\exists \pi_b.\ \F{lo}(C)_{\mid L(v_2)} \subseteq \pi_b \wedge b = \pi_b(\sigma_0)$\\
$\exists \pi_\top.\ \F{lo}(C)_{\mid L(v_\top)} \subseteq \pi_\top \wedge l = \pi_\top(\sigma_0)$\\

To start off, let's consider the set $L_1^a \cup L_2^a$. These are all local events of $v_1$ and $v_2$, which are not linearized before events of the LCA. 
We consider different cases depending on the size of this set.\\

\noindent\textrm{\sc{Case 1}}: $(\mid L_1^{a} \cup L_2^{a} \mid = 0)$\\
We note that in this case, $a,b$ can be defined as follows:
$a= {\pi_a}_{\mid (L_\top^{b} \cup L_1^{b} \cup L_\top^{a})}(\sigma_0)$,
$b = {\pi_b}_{\mid (L_\top^{b} \cup L_2^{b} \cup L_\top^{a})}(\sigma_0)$.\\
We need to show that there exists a sequence $\pi_m$ that extends $\F{lo_m}$ such that $\F{merge}(l,a,b) = \pi_m(\sigma_0)$.
Here, we induct on the size of the set $L_\top^a$.\\

\noindent\textrm{\sc{Base Case 1}}: $(L_\top^a  = \phi)$\\
Then $L_1^b \cup L_2^b = \phi$. So $l = a = b$. $\F{merge}(l, l, l) = l$ is handled by $\textrm{\sc{MergeIdempotence}}$.
We know that $l$ is correctly linearized, hence the required result follows.\\

\noindent\textrm{\sc{Inductive Case 1}}: $(\mid L_\top^a  \mid > 0)$\\
Let $L_\top^a = \{e_1^{\top}, \dots, e_{m-1}^{\top}, e_m^{\top}\}$. Let $S = \{e_1^{\top}, \dots, e_{m-1}^{\top}\}$.
By IH, for the set $S$, we have the required result. We define $l',a',b'$ based on the above set $S$: $l'=\pi_{l_{\mid L_\top^b \cup S}}(\sigma_0)$, $a' = \pi_{a_{\mid L_\top^b \cup \bigcup_{e \in S} L_1^b(e) \cup S}}(\sigma_0)$, $b' = \pi_{b_{\mid L_\top^b \cup \bigcup_{e \in S} L_2^b(e) \cup S}(\sigma_0)}$. Note that in this case, all the LCA events which are linearized after local events are already taken as part of the states $l',a',b'$. Now, suppose we add one more LCA event $e_m^{\top}$ to all states. We define $a'',b''$ such that
$a'' = {\pi_a}_{\mid L_1^b(e_m^{\top})} (a')$,
$b'' = {\pi_b}_{\mid L_2^b(e_m^{\top})} (b')$. 

Then, $l = e_m^{\top}(l'), a = e_m^{\top}(a''), b = e_m^{\top}(b'')$.  $e_m^{\top}$ is not linearized before any of the events in $L_\top^{b} \cup L_1^{b} \cup L_2^b \cup S$  based on the definition of $L_\top^a$.

Now, by $\textrm{\sc{BottomUp-0-OP}}$ rule,

\begin{equation}\label{eq:0}
	\F{merge}(e_m^{\top}(l'), e_m^{\top}(a''), e_m^{\top}(b'')) = e_m^{\top}(\F{merge}(l',a'',b''))
\end{equation}

We will now show prove $\textrm{\sc{BottomUp-0-OP}}$ rule, i.e. Eqn. \eqref{eq:0}:

\noindent\textrm{\sc{Proof of } Eq. \eqref{eq:0}}: 

Let $l_b = \pi_{l_{\mid L_\top^b}}(\sigma_0)$.

We first induct on $\mid L_\top^b \mid$ to show that $\F{merge}(e_m^{\top}(l_b), e_m^{\top}(l_b), e_m^{\top}(l_b)) = e_m^{\top}(\F{merge}(l_b, l_b, l_b))$

For the base case, we use $\psi^{L_\top^b}_\F{base-0op}$. For the inductive case, we use $\psi^{L_\top^b}_\F{ind-0op}$, whose pre-condition will be satisfied by the IH. 

Next, we induct on $\mid L_\top^a \setminus \{e_m^{\top}\} \mid$ to show Eqn. \eqref{eq:0}.

For the base case, we have $\mid L_\top^a \setminus \{e_m^{\top}\} \mid = 0$. In this case, the set $S = \emptyset$. Also, $l' = a' = b' = l_b$. Hence, we need to show the following: 

\begin{equation}\label{eq:100}
	\F{merge}(e_m^{\top}(l_b), e_m^{\top}({\pi_a}_{\mid L_1^b(e_m^{\top})} (a')), e_m^{\top}({\pi_b}_{\mid L_2^b(e_m^{\top})} (b'))) = e_m^{\top}(\F{merge}(l',{\pi_a}_{\mid L_1^b(e_m^{\top})} (a'),{\pi_b}_{\mid L_2^b(e_m^{\top})} (b')))
\end{equation}

We will now induct on $\mid L_1^b(e_m^{\top}) \cup L_2^b(e_m^{\top}) \mid$ to show Eqn. \eqref{eq:100}.

For the base case where $\mid L_1^b(e_m^{\top}) \cup L_2^b(e_m^{\top}) \mid = 0$, it directly follows from the outcome of the induction on $\mid L_\top^b \mid$.

For the inductive case, we use one of $\psi^{L_1^b}_\F{ind1-0op}$, $\psi^{L_1^b}_\F{ind2-0op}$, $\psi^{L_2^b}_\F{ind1-0op}$ or $\psi^{L_2^b}_\F{ind2-0op}$ depending on the event $e_b$ or $e$ to be added to $L_1^b(e_m^{\top})$ or $L_2^b(e_m^{\top})$, with the pre-condition of these VCs being inferred from the IH.

This completes the proof of Eqn. \eqref{eq:100}.

Now, we consider the inductive case for $\mid L_\top^a \setminus \{e_m^{\top}\} \mid$ to show Eqn. \eqref{eq:101}.  By IH, we get the following:
\begin{equation}\label{eq:3}
	\F{merge}(e_m^{\top}(l'''), e_m^{\top}(a'''), e_m^{\top}(b''')) = e_m^{\top}(\F{merge}(l''',a''',b'''))
\end{equation}

where for the set $S' = S \setminus e_{m-1}^{\top}$, $l'''=\pi_{l_{\mid L_\top^b \cup S'}}(\sigma_0)$, $a''' = \pi_{a_{\mid L_\top^b \cup \bigcup_{e \in S'} L_1^b(e) \cup S'}}(\sigma_0)$, $b''' = \pi_{b_{\mid L_\top^b \cup \bigcup_{e \in S'} L_2^b(e) \cup S'}(\sigma_0)}$. That is, we consider the effects of all event in $S$ except $e_{m-1}^{\top}$. 

Now, we first use $\psi^{L_\top^a}_\F{ind-0op}$ to apply $e_{m-1}^{\top}$ to $l'''$, $a'''$ and $b'''$. Note that the pre-condition for  $\psi^{L_\top^a}_\F{ind-0op}$ is satisfied due to Eqn. \eqref{eq:3}. 

Next, we use induct on $\mid L_1^b(e_{m-1}^{\top}) \cup L_2^b(e_{m-1}^{\top}) \mid$ using the VCs $\psi^{L_1^b}_\F{ind1-0op}$, $\psi^{L_1^b}_\F{ind2-0op}$, $\psi^{L_2^b}_\F{ind1-0op}$ or $\psi^{L_2^b}_\F{ind2-0op}$ to add all events in these sets. Finally, we induct on $\mid L_1^b(e_{m}^{\top}) \cup L_2^b(e_{m}^{\top}) \mid$ to again add all these events, thereby proving Eqn. \eqref{eq:101}.

Now that we have linearized $e_m^{\top}$ using Eqn. \eqref{eq:101}, we need to linearize the events that led to $\F{merge}(l',a'',b'')$. 
Let's denote $L_1^b(e_m^{\top})$ as $M_1^a$ and $L_2^b(e_m^{\top})$ as $M_2^a$. Now we induct on the size of the set $M_1^{a} \cup M_2^{a}$.\\

\noindent\textrm{\sc{Base Case 1.1}}:$(\mid M_1^{a} \cup M_2^{a} \mid = 0)$\\
$a'' = a', b'' = b'$. By induction assumption, $\exists \pi.\ \F{lo}(C)_{\mid (L_\top^b \cup \bigcup_{e \in S} L_1^b(e) \cup \bigcup_{e \in S} L_2^b(e) \cup S)} \subseteq \pi$\\ 
and $\F{merge}(l',a',b') = \pi(\sigma_0)$. Hence, $\pi_m = \pi. e_m^{\top}$.\\

\noindent\textrm{\sc{Inductive Case 1.1}}:$(\mid M_1^{a} \cup M_2^{a} \mid > 0)$\\
We have 2 cases here: \\(1.1.1) Either of $M_1^{a}$ or $M_2^{a}$ is $\phi$ \\(1.1.2) Both $M_1^{a}$ and $M_2^{a}$ are not $\phi$.\\

\noindent\textrm{\sc{Case 1.1.1}}:$(M_1^a \neq \phi \wedge M_2^a = \phi)$\\
Consider $e_1 \in M_1^a$ such that there does not exist $e \in M_1^a$ and $e_1 \xrightarrow{\F{lo_m}} e$, i.e. $e_1$ is the maximal event according to $\F{lo_m}$. Since $\F{lo}$ ordering between events remains the same in all versions, and since versions $v_1$ and $v_2$ (which are being merged) were already linearizable, there would exist sequences leading to the states $a$ such that $e_1$ would appear at the end. Hence, there exists $a'''$ such that $a'' = e_1(a''')$. Since $M_2^a$ is empty, all local events in $L_2$ are linearized before the rest of the  LCA events.
Suppose $L_\top^a \setminus \{e_m^{\top}\} \neq \phi$ or $L_\top^b \neq \phi$, the last event which leads to the state $l',b''$ must be an LCA event.  Let's consider $e_\top$ to be the maximal event in $L_\top$ according to $\F{lo_m}$.
Hence there exists states $l'', b'''$ such that $l' = e_\top(l''), b'' = e_\top(b''')$.
By $\textrm{\sc{BottomUp-1-OP}}$ rule
\begin{equation}\label{eq:1op}
	\F{merge}(e_\top(l''), e_1(a'''), e_\top(b''')) = e_1(\F{merge}(e_\top(l''),a''',e_\top(b''')))
\end{equation}

Again, we prove $\textrm{\sc{BottomUp-1-OP}}$ rule using the same induction scheme that we showed for $\textrm{\sc{BottomUp-0-OP}}$. Briefly, we use $\psi^{L_\top^b}_\F{base-1op}$ and $\psi^{L_\top^b}_\F{ind-1op}$ for induction on $\mid L_\top^b \mid$. Then, we use $\psi^{L_\top^a}_\F{ind-1op}$, $\psi^{L_1^b}_\F{ind1-1op}$, $\psi^{L_1^b}_\F{ind2-1op}$, $\psi^{L_2^b}_\F{ind1-1op}$ and $\psi^{L_2^b}_\F{ind2-1op}$ to build the event sets $L_\top^a \setminus \{e_m^{\top}\}$ and $\sqcup_{e \in L_\top^a \setminus \{e_m^{\top}\}} L_1^b(e) \cup \sqcup_{e \in L_\top^a \setminus \{e_m^{\top}\}} L_2^b(e)$.

From the induction assumption, we get that $\F{merge}(e_\top(l''),a''',e_\top(b'''))$ is already obtained by 
the linearization of events applied on the initial state $\sigma_0$. That is, there exists a sequence $\pi'$ over events in $L_\top^b \cup \bigcup_{e \in S} L_1^b(e) \cup \bigcup_{e \in S} L_2^b(e) \cup S \cup M_1^a \setminus e_1$ which extends $\F{lo_m}$ relation such that $\F{merge}(e_\top(l''),a''',e_\top(b''')) = \pi'(\sigma_0)$. Now, $\pi = \pi' e_1$ is the required linearization. \\

\noindent\textrm{\sc{Case 1.1.2}}:$(M_1^a \neq \phi \wedge M_2^a \neq \phi)$\\
Consider $e_1 \in M_1^a, e_2 \in M_2^a$ such that there does not exist $e \in M_i^a$ and $e_i \xrightarrow{\F{lo_m}} e$ (for $i=1,2$), i.e. each of the $e_i$s are maximal events according to $\F{lo_m}$. Since $\F{lo}$ ordering between events remains the same in all versions, and since versions $v_1$ and $v_2$ (which are being merged) were already linearizable, there would exist sequences leading to the states $a''$ and $b''$ such that $e_1$ and $e_2$ would appear at the end resp. Hence, there exists $a'''$ and $b'''$ such that $a'' = e_1(a''')$ and $b'' = e_1(b''')$. Since $e_1 \mid\mid_C e_2$, they are related to each other by $\F{rc}$ relation or they commute with each other i.e., $e_1 \xrightarrow{\F{rc}} e_2 \vee e_2 \xrightarrow{\F{rc}} e_1 \vee e_1 \rightleftarrows e_2$.
We will consider the case when $e_2 \xrightarrow{\F{rc}} e_1 \vee e_1 \rightleftarrows e_2$. $e_1 \xrightarrow{\F{rc}} e_2$ is handled by $\textrm{\sc{MergeCommutativity}}$.
The equation becomes 
\begin{equation}\label{eq:2op10}
	\F{merge}(l', e_1(a'''), e_2(b''')) = e_1(\F{merge}(l', a''', e_2(b''')))
\end{equation}
which is the $\textrm{\sc{BottomUp-2-OP}}$ rule.\\

Again, we prove $\textrm{\sc{BottomUp-2-OP}}$ rule using the same induction scheme that we showed for $\textrm{\sc{BottomUp-1-OP}}$. Briefly, we use $\psi^{L_\top^b}_\F{base-2op}$ and $\psi^{L_\top^b}_\F{ind-2op}$ for induction on $\mid L_\top^b \mid$. Then, we use $\psi^{L_\top^a}_\F{ind-2op}$, $\psi^{L_1^b}_\F{ind1-2op}$, $\psi^{L_1^b}_\F{ind2-2op}$, $\psi^{L_2^b}_\F{ind1-2op}$ and $\psi^{L_2^b}_\F{ind2-2op}$ to build the event sets $L_\top^a \setminus \{e_m^{\top}\}$ and $\sqcup_{e \in L_\top^a \setminus \{e_m^{\top}\}} L_1^b(e) \cup \sqcup_{e \in L_\top^a \setminus \{e_m^{\top}\}} L_2^b(e)$.\\

From the induction assumption, we get that $\F{merge}(l', a''', e_2(b'''))$ is already obtained by 
the linearization of events applied on the initial state $\sigma_0$. If $\pi'$ is the linearization for this merge, then $\pi = \pi' e_1$ is the required linearization.\\

\noindent\textrm{\sc{Case 2}}: $(\mid L_1^{a} \cup L_2^{a} \mid > 0)$\\
The proof here will be identical to the proof of Inductive Case 1.1, substituting $L_1^a$ and $L_2^a$ for $M_1^a$ and $M_2^a$, and using the rules $\textrm{\sc{BottomUp-1-OP}}$, $\textrm{\sc{MergeCommutativity}}$ and  $\textrm{\sc{BottomUp-2-OP}}$.

\subsubsection{Case $({\textrm{\sc{Query}})}$:}
Assume that a query operation is applied on a replica $r$. Let $C = \langle N, H, L, G, vis \rangle$ 
be the configuration of the replica before the operation. According to the semantics, the configuration of the 
replica remains same after the query operation. By the induction hypothesis, Def.~\ref{def:lin} holds for the configuration $C$.

\end{proof}

\begin{table}[ht]
\scriptsize
\makebox[\textwidth][c]{%
\begin{tabular}{|>{\raggedright\arraybackslash}p{2.2cm}|>{\raggedright\arraybackslash}p{3cm}|>{\raggedright\arraybackslash}p{3.5cm}|>{\raggedright\arraybackslash}p{4cm}|}
		 
  		\hline
  		\textbf{VC Name} & \multicolumn{2}{c|}{\textbf{Pre-condition}} & \textbf{Post-condition} \\
		\hline

		$\textrm{\sc{MergeCommutativity}}$ & 
		$ $ & 
		$ $ &
		$\mu(l, a, b)= \mu(l, b, a)$\\
		\hline
		
		$\textrm{\sc{MergeIdempotence}}$ & 
		$ $ & 
		$ $ &
		$\mu(s, s, s)= s$\\
		\hline

		$\psi^{L_\top^b}_\F{base-2op}$ &
		$e_2 \xrightarrow{\F{rc}} e_1 \vee e_2 \rightleftarrows e_1$ & 
		$ $ &
		$\mu(\sigma_0, e_1(\sigma_0), e_2(\sigma_0)) = e_1(\mu(\sigma_0, \sigma_0, e_2(\sigma_0)))$\\
		\hline

		$\psi^{L_\top^b}_\F{ind-2op}$ &
		$e_2 \xrightarrow{\F{rc}} e_1 \vee e_2 \rightleftarrows e_1$ &
		$\mu(l, e_1(l), e_2(l)) = e_1(\mu(l, l, e_2(l)))$ & 
		$\mu(e_\top(l), e_1 \comp e_\top(l), e_2 \comp e_\top(l)) = e_1(\mu(e_\top(l), e_\top(l), e_2 \comp e_\top(l)))$\\
		\hline
		
		$\psi^{L_\top^a}_\F{ind-2op}$ &
		$(e_2 \xrightarrow{\F{rc}} e_1 \vee e_2 \rightleftarrows e_1) \wedge (\exists e. e \xrightarrow{\F{rc}} e_\top)$ &
		$\mu(l, e_1(a), e_2(b)) = e_1(\mu(l, a, e_2(b)))$ & 
		$\mu(e_\top(l), e_1 \comp e_\top(a), e_2 \comp e_\top(b)) = e_1(\mu(e_\top(l), e_\top(a), e_2 \comp e_\top(b)))$\\
		\hline

		$\psi^{L_1^b}_\F{ind1-2op}$ & 
		$(e_2 \xrightarrow{\F{rc}} e_1 \vee e_2 \rightleftarrows e_1) \wedge e_b \xrightarrow{\F{rc}} e_\top$ & 
		$\mu(e_\top(l), e_1 \comp e_\top(a), e_2 \comp e_\top(b)) = e_1(\mu(e_\top(l), e_\top(a), e_2 \comp e_\top(b)))$ & 
		$\mu(e_\top(l), e_1 \comp e_\top \comp e_b(a), e_2 \comp e_\top(b)) = e_1(\mu(e_\top(l), e_\top \comp e_b(a), e_2 \comp e_\top(b)))$\\
		\hline
		
		$\psi^{L_1^b}_\F{ind2-2op}$ & 
		$(e_2 \xrightarrow{\F{rc}} e_1 \vee e_2 \rightleftarrows e_1) \wedge e_b \xrightarrow{\F{rc}} e_\top \wedge (\neg e \rightleftarrows e_b \vee e \xrightarrow{\F{rc}} e_\top)$ &
		$\mu(e_\top(l), e_1 \comp e_\top \comp e_b(a), e_2 \comp e_\top(b)) = e_1(\mu(e_\top(l), e_\top \comp e_b(a), e_2 \comp e_\top(b)))$ & 
		$\mu(e_\top(l), e_1 \comp e_\top \comp e_b(e(a)), e_2 \comp e_\top(b)) = e_1(\mu(e_\top(l), e_\top \comp e_b \comp e(a), e_2 \comp e_\top(b)))$\\
		\hline

		$\psi^{L_2^b}_\F{ind1-2op}$ & 
		$(e_2 \xrightarrow{\F{rc}} e_1 \vee e_2 \rightleftarrows e_1) \wedge e_b \xrightarrow{\F{rc}} e_\top$ &
		$\mu(e_\top(l), e_1 \comp e_\top(a), e_2 \comp e_\top(b)) = e_1(\mu(e_\top(l), e_\top(a), e_2 \comp e_\top(b)))$ & 
		$\mu(e_\top(l), e_1 \comp e_\top(a), e_2 \comp e_\top \comp e_b(b)) = e_1(\mu(e_\top(l), e_\top(a), e_2 \comp e_\top \comp e_b(b)))$\\
		\hline

		$\psi^{L_2^b}_\F{ind2-2op}$ & 
		$(e_2 \xrightarrow{\F{rc}} e_1 \vee e_2 \rightleftarrows e_1) \wedge e_b \xrightarrow{\F{rc}} e_\top$ & 
		$\mu(e_\top(l), e_1 \comp e_\top(a), e_2 \comp e_\top(b)) = e_1(\mu(e_\top(l), e_\top(a), e_2 \comp e_\top(b)))$ & 
		$\mu(e_\top(l), e_1 \comp e_\top \comp e_b(a), e_2 \comp e_\top(b)) = e_1(\mu(e_\top(l), e_\top \comp e_b(a), e_2 \comp e_\top(b)))$\\
		\hline
				
		$\psi_\F{ind-2op}^{L_1^a}$ & 
		$e_2 \xrightarrow{\F{rc}} e_1 \vee e_2 \rightleftarrows e_1$ & 
		$\mu(l, e_1(a), e_2(b)) = e_1(\mu(l, a, e_2(b)))$ & 
		$\mu(l, e_1 \comp e_1'(a), e_2(b)) = e_1(\mu(l, e_1'(a), e_2(b)))$\\
		\hline
		
		$\psi_\F{ind-2op}^{L_2^a}$ & 
		$e_2 \xrightarrow{\F{rc}} e_1 \vee e_2 \rightleftarrows e_1$ &
		$\mu(l, e_1(a), e_2(b)) = e_1(\mu(l, a, e_2(b)))$ & 
		$\mu(l, e_1(a), e_2 \comp e_2'(b)) = e_1(\mu(l, a, e_2 \comp e_2'(b)))$\\
		\hline

		$\psi^{L_\top^b}_\F{base-1op}$ &
		$ $ & 
		$ $ &
		$\mu(\sigma_0, e_1(\sigma_0), \sigma_0) = e_1(\mu(\sigma_0, \sigma_0, \sigma_0))$\\
		\hline

		$\psi^{L_\top^b}_\F{ind-1op}$ &
		$ $ &
		$\mu(l, e_1(l), l) = e_1(\mu(l, l, l))$ & 
		$\mu(e_\top(l), e_1 \comp e_\top(l), e_\top(l)) = e_1(\mu(e_\top(l), e_\top(l), e_\top(l)))$\\
		\hline

		$\psi^{L_\top^a}_\F{ind-1op}$ &
		$\exists e. e \xrightarrow{\F{rc}} e_\top$ &
		$\mu(e_\top'(l), e_1(a), e_\top'(b)) = e_1(\mu(e_\top'(l), a, e_\top'(b)))$ & 
		$\mu(e_\top \comp e_\top'(l), e_1 \comp e_\top(a), e_\top \comp e_\top'(b)) = e_1(\mu(e_\top \comp e_\top'(l), e_\top(a), e_\top \comp e_\top'(b)))$\\
		\hline
		
		$\psi^{L_1^b}_\F{ind1-1op}$ &
		$e_b \xrightarrow{\F{rc}} e_\top$ & 
		$\mu(e_\top(l), e_1 \comp e_\top(a), e_\top(b))) = e_1(\mu(e_\top(l), e_\top(a), e_\top(b)))$ & 
		$\mu(e_\top(l), e_1 \comp e_\top \comp e_b(a), e_\top(b)) = e_1(\mu(e_\top(l), e_\top \comp e_b(a), e_\top(b)))$\\
		\hline

		$\psi^{L_1^b}_\F{ind2-1op}$ &
		$e_b \xrightarrow{\F{rc}} e_\top \wedge (\neg e \rightleftarrows e_b \vee e \xrightarrow{\F{rc}} e_\top)$ &
		$\mu(e_\top(l), e_1 \comp e_\top \comp e_b(a), e_\top(b)) = e_1(\mu(e_\top(l), e_\top \comp e_b(a), e_\top(b)))$ & 
		$\mu(e_\top(l), e_1 \comp e_\top \comp e_b \comp e(a), e_\top(b)) = e_1(\mu(e_\top(l), e_\top \comp e_b \comp e(a), e_\top(b)))$\\
		\hline
		
		$\psi^{L_2^b}_\F{ind1-1op}$ & 
		$e_b \xrightarrow{\F{rc}} e_\top$ &
		$\mu(e_\top(l), e_1 \comp e_\top(a), e_\top(b)) = e_1(\mu(e_\top(l), e_\top(a), e_\top(b)))$ & 
		$\mu(e_\top(l), e_1 \comp e_\top(a), e_\top \comp e_b(b)) = e_1(\mu(e_\top(l), e_\top(a), e_\top \comp e_b(b)))$\\
		\hline

		$\psi^{L_2^b}_\F{ind2-1op}$ &
		$e_b \xrightarrow{\F{rc}} e_\top \wedge (\neg e \rightleftarrows e_b \vee e \xrightarrow{\F{rc}} e_\top)$ & 
		$\mu(e_\top(l), e_1 \comp e_\top(a), e_\top \comp e_b(b)) = e_1(\mu(e_\top(l), e_\top(a), e_\top \comp e_b(b)))$ & 
		$\mu(e_\top(l), e_1 \comp e_\top(a), e_\top \comp e_b \comp e(b)) = e_1(\mu(e_\top(l), e_\top(a), e_\top \comp e_b \comp e(b)))$\\
		\hline	
		
		$\psi_\F{ind-1op}^{L_1^a}$ & 
		$ $ &
		$\mu(e_\top(l), e_1(a), e_\top(b)) = e_1(\mu(e_\top(l), a, e_\top(b)))$ & 
		$\mu(e_\top(l), e_1 \comp e_1'(a), e_\top(b)) = e_1(\mu(e_\top(l), e_1'(a), e_\top(b)))$\\
		\hline
		
		$\psi^{L_\top^b}_\F{base-0op}$ &
		$ $ & 
		$ $ &
		$\mu(e_1(\sigma_0), e_1(\sigma_0), e_1(\sigma_0)) = e_1(\mu(\sigma_0, \sigma_0, \sigma_0))$\\
		\hline
		
		$\psi^{L_\top^b}_\F{ind-0op}$ &
		$ $ &
		$\mu(e_1(l), e_1(l), e_1(l)) = e_1(\mu(l, l, l))$ & 
		$\mu(e_1\comp e_\top(l), e_1 \comp e_\top(l), e_1\comp e_\top(l)) = e_1(\mu(e_\top(l), e_\top(l), e_\top(l)))$\\
		\hline
		
		$\psi^{L_\top^a}_\F{ind-0op}$ &
		$\exists e. e \xrightarrow{\F{rc}} e_\top$ &
		$\mu(e_1(l), e_1(a), e_1(b)) = e_1(\mu(l, a, b))$ &  
		$\mu(e_1\comp e_\top(l), e_1 \comp e_\top(a), e_1\comp e_\top(b)) = e_1(\mu(e_\top(l), e_\top(a), e_\top(b)))$\\
		\hline
		
		$\psi^{L_1^b}_\F{ind1-0op}$ &
		$ $ & 
		$\mu(e_1(l), e_1(a), e_1(b)) = e_1(\mu(l, a, b))$ &  
		$\mu(e_1(l), e_1 \comp e_b(a)), e_1(b)) = e_1(\mu(l, e_b(a), b))$\\
		\hline
		
		$\psi^{L_1^b}_\F{ind2-0op}$ &
		$\neg e \rightleftarrows e_b \vee e \xrightarrow{\F{rc}} e_1$ &
		$\mu(e_1(l), e_1 \comp e_b(a)), e_1(b)) = e_1(\mu(l, e_b(a), b))$ & 
		$\mu(e_1(l), e_1 \comp e_b \comp e(a))), e_1(b)) = e_1(\mu(l, e_b \comp e(a)), b))$\\
		\hline
		
		$\psi^{L_2^b}_\F{ind1-0op}$ & 
		$ $ &
		$\mu(e_1(l), e_1(a), e_1(b)) = e_1(\mu(l, a, b))$ &  
		$\mu(e_1(l), e_1(a), e_1 \comp e_b(b))) = e_1(\mu(l, a, e_b(b)))$\\
		\hline
		
		$\psi^{L_2^b}_\F{ind2-0op}$ &
		$\neg e \rightleftarrows e_b \vee e \xrightarrow{\F{rc}} e_1$ & 
		$\mu(e_1(l), e_1(a), e_1 \comp e_b(b))) = e_1(\mu(l, a, e_b(b)))$ & 
		$\mu(e_1(l), e_1(a), e_1 \comp e_b \comp e(b)))) = e_1(\mu(l, a, e_b \comp e(b))))$\\
		\hline	
		
\end{tabular}%
}
\caption{Complete set of Verification Conditions for MRDTs}
\label{tab:full}
\end{table}

\begin{table}[ht]
\scriptsize
\makebox[\textwidth][c]{%
\begin{tabular}{|>{\raggedright\arraybackslash}p{2.2cm}|>{\raggedright\arraybackslash}p{3cm}|>{\raggedright\arraybackslash}p{3.5cm}|>{\raggedright\arraybackslash}p{4cm}|}
		 
  		\hline
  		\textbf{VC Name} & \multicolumn{2}{c|}{\textbf{Pre-condition}} & \textbf{Post-condition} \\
		\hline

		$\textrm{\sc{MergeCommutativity}}$ & 
		$ $ & 
		$ $ &
		$\mu(a, b)= \mu(b, a)$\\
		\hline
		
		$\textrm{\sc{MergeIdempotence}}$ & 
		$ $ & 
		$ $ &
		$\mu(s, s)= s$\\
		\hline

		$\psi^{L_\top^b}_\F{base-2op}$ &
		$e_2 \xrightarrow{\F{rc}} e_1 \vee e_2 \rightleftarrows e_1$ & 
		$ $ &
		$\mu(e_1(\sigma_0), e_2(\sigma_0)) = e_1(\mu(\sigma_0, e_2(\sigma_0)))$\\
		\hline

		$\psi^{L_\top^b}_\F{ind-2op}$ &
		$e_2 \xrightarrow{\F{rc}} e_1 \vee e_2 \rightleftarrows e_1$ &
		$\mu(e_1(l), e_2(l)) = e_1(\mu(l, e_2(l)))$ & 
		$\mu(e_1 \comp e_\top(l), e_2 \comp e_\top(l)) = e_1(\mu(e_\top(l), e_2 \comp e_\top(l)))$\\
		\hline
		
		$\psi^{L_\top^a}_\F{ind-2op}$ &
		$(e_2 \xrightarrow{\F{rc}} e_1 \vee e_2 \rightleftarrows e_1) \wedge (\exists e. e \xrightarrow{\F{rc}} e_\top)$ &
		$\mu(e_1(a), e_2(b)) = e_1(\mu(a, e_2(b)))$ & 
		$\mu(e_1 \comp e_\top(a), e_2 \comp e_\top(b)) = e_1(\mu(e_\top(a), e_2 \comp e_\top(b)))$\\
		\hline

		$\psi^{L_1^b}_\F{ind1-2op}$ & 
		$(e_2 \xrightarrow{\F{rc}} e_1 \vee e_2 \rightleftarrows e_1) \wedge e_b \xrightarrow{\F{rc}} e_\top$ & 
		$\mu(e_1 \comp e_\top(a), e_2 \comp e_\top(b)) = e_1(\mu(e_\top(a), e_2 \comp e_\top(b)))$ & 
		$\mu(e_1 \comp e_\top \comp e_b(a), e_2 \comp e_\top(b)) = e_1(\mu(e_\top \comp e_b(a), e_2 \comp e_\top(b)))$\\
		\hline
		
		$\psi^{L_1^b}_\F{ind2-2op}$ & 
		$(e_2 \xrightarrow{\F{rc}} e_1 \vee e_2 \rightleftarrows e_1) \wedge e_b \xrightarrow{\F{rc}} e_\top \wedge (\neg e \rightleftarrows e_b \vee e \xrightarrow{\F{rc}} e_\top)$ &
		$\mu(e_1 \comp e_\top \comp e_b(a), e_2 \comp e_\top(b)) = e_1(\mu(e_\top \comp e_b(a), e_2 \comp e_\top(b)))$ & 
		$\mu(e_1 \comp e_\top \comp e_b(e(a)), e_2 \comp e_\top(b)) = e_1(\mu(e_\top \comp e_b \comp e(a), e_2 \comp e_\top(b)))$\\
		\hline

		$\psi^{L_2^b}_\F{ind1-2op}$ & 
		$(e_2 \xrightarrow{\F{rc}} e_1 \vee e_2 \rightleftarrows e_1) \wedge e_b \xrightarrow{\F{rc}} e_\top$ &
		$\mu(e_1 \comp e_\top(a), e_2 \comp e_\top(b)) = e_1(\mu(e_\top(a), e_2 \comp e_\top(b)))$ & 
		$\mu(e_1 \comp e_\top(a), e_2 \comp e_\top \comp e_b(b)) = e_1(\mu(e_\top(a), e_2 \comp e_\top \comp e_b(b)))$\\
		\hline

		$\psi^{L_2^b}_\F{ind2-2op}$ & 
		$(e_2 \xrightarrow{\F{rc}} e_1 \vee e_2 \rightleftarrows e_1) \wedge e_b \xrightarrow{\F{rc}} e_\top$ & 
		$\mu(e_1 \comp e_\top(a), e_2 \comp e_\top(b)) = e_1(\mu(e_\top(a), e_2 \comp e_\top(b)))$ & 
		$\mu(e_1 \comp e_\top \comp e_b(a), e_2 \comp e_\top(b)) = e_1(\mu(e_\top \comp e_b(a), e_2 \comp e_\top(b)))$\\
		\hline
				
		$\psi_\F{ind-2op}^{L_1^a}$ & 
		$e_2 \xrightarrow{\F{rc}} e_1 \vee e_2 \rightleftarrows e_1$ & 
		$\mu(e_1(a), e_2(b)) = e_1(\mu(a, e_2(b)))$ & 
		$\mu(e_1 \comp e_1'(a), e_2(b)) = e_1(\mu(e_1'(a), e_2(b)))$\\
		\hline
		
		$\psi_\F{ind-2op}^{L_2^a}$ & 
		$e_2 \xrightarrow{\F{rc}} e_1 \vee e_2 \rightleftarrows e_1$ &
		$\mu(e_1(a), e_2(b)) = e_1(\mu(a, e_2(b)))$ & 
		$\mu(e_1(a), e_2 \comp e_2'(b)) = e_1(\mu(a, e_2 \comp e_2'(b)))$\\
		\hline

		$\psi^{L_\top^b}_\F{base-1op}$ &
		$ $ & 
		$ $ &
		$\mu(e_1(\sigma_0), \sigma_0) = e_1(\mu(\sigma_0, \sigma_0))$\\
		\hline

		$\psi^{L_\top^b}_\F{ind-1op}$ &
		$ $ &
		$\mu(e_1(l), l) = e_1(\mu(l, l))$ & 
		$\mu(e_1 \comp e_\top(l), e_\top(l)) = e_1(\mu(e_\top(l), e_\top(l)))$\\
		\hline

		$\psi^{L_\top^a}_\F{ind-1op}$ &
		$\exists e. e \xrightarrow{\F{rc}} e_\top$ &
		$\mu(e_1(a), e_\top'(b)) = e_1(\mu(a, e_\top'(b)))$ & 
		$\mu(e_1 \comp e_\top(a), e_\top \comp e_\top'(b)) = e_1(\mu(e_\top(a), e_\top \comp e_\top'(b)))$\\
		\hline
		
		$\psi^{L_1^b}_\F{ind1-1op}$ &
		$e_b \xrightarrow{\F{rc}} e_\top$ & 
		$\mu(e_1 \comp e_\top(a), e_\top(b))) = e_1(\mu(e_\top(a), e_\top(b)))$ & 
		$\mu(e_1 \comp e_\top \comp e_b(a), e_\top(b)) = e_1(\mu(e_\top \comp e_b(a), e_\top(b)))$\\
		\hline

		$\psi^{L_1^b}_\F{ind2-1op}$ &
		$e_b \xrightarrow{\F{rc}} e_\top \wedge (\neg e \rightleftarrows e_b \vee e \xrightarrow{\F{rc}} e_\top)$ &
		$\mu(e_1 \comp e_\top \comp e_b(a), e_\top(b)) = e_1(\mu(e_\top \comp e_b(a), e_\top(b)))$ & 
		$\mu(e_1 \comp e_\top \comp e_b \comp e(a), e_\top(b)) = e_1(\mu(e_\top \comp e_b \comp e(a), e_\top(b)))$\\
		\hline
		
		$\psi^{L_2^b}_\F{ind1-1op}$ & 
		$e_b \xrightarrow{\F{rc}} e_\top$ &
		$\mu(e_1 \comp e_\top(a), e_\top(b)) = e_1(\mu(e_\top(a), e_\top(b)))$ & 
		$\mu(e_1 \comp e_\top(a), e_\top \comp e_b(b)) = e_1(\mu(e_\top(a), e_\top \comp e_b(b)))$\\
		\hline

		$\psi^{L_2^b}_\F{ind2-1op}$ &
		$e_b \xrightarrow{\F{rc}} e_\top \wedge (\neg e \rightleftarrows e_b \vee e \xrightarrow{\F{rc}} e_\top)$ & 
		$\mu(e_1 \comp e_\top(a), e_\top \comp e_b(b)) = e_1(\mu(e_\top(a), e_\top \comp e_b(b)))$ & 
		$\mu(e_1 \comp e_\top(a), e_\top \comp e_b \comp e(b)) = e_1(\mu(e_\top(a), e_\top \comp e_b \comp e(b)))$\\
		\hline	
		
		$\psi_\F{ind-1op}^{L_1^a}$ & 
		$ $ &
		$\mu(e_1(a), e_\top(b)) = e_1(\mu(a, e_\top(b)))$ & 
		$\mu(e_1 \comp e_1'(a), e_\top(b)) = e_1(\mu(e_1'(a), e_\top(b)))$\\
		\hline
		
		$\psi^{L_\top^b}_\F{base-0op}$ &
		$ $ & 
		$ $ &
		$\mu(e_1(\sigma_0), e_1(\sigma_0)) = e_1(\mu(\sigma_0, \sigma_0))$\\
		\hline
		
		$\psi^{L_\top^b}_\F{ind-0op}$ &
		$ $ &
		$\mu(e_1(l), e_1(l)) = e_1(\mu(l, l))$ & 
		$\mu(e_1 \comp e_\top(l), e_1\comp e_\top(l)) = e_1(\mu(e_\top(l), e_\top(l)))$\\
		\hline
		
		$\psi^{L_\top^a}_\F{ind-0op}$ &
		$\exists e. e \xrightarrow{\F{rc}} e_\top$ &
		$\mu(e_1(a), e_1(b)) = e_1(\mu(a, b))$ &  
		$\mu(e_1 \comp e_\top(a), e_1\comp e_\top(b)) = e_1(\mu(e_\top(a), e_\top(b)))$\\
		\hline
		
		$\psi^{L_1^b}_\F{ind1-0op}$ &
		$ $ & 
		$\mu(e_1(a), e_1(b)) = e_1(\mu(a, b))$ &  
		$\mu(e_1 \comp e_b(a)), e_1(b)) = e_1(\mu(e_b(a), b))$\\
		\hline
		
		$\psi^{L_1^b}_\F{ind2-0op}$ &
		$\neg e \rightleftarrows e_b \vee e \xrightarrow{\F{rc}} e_1$ &
		$\mu(e_1 \comp e_b(a)), e_1(b)) = e_1(\mu(e_b(a), b))$ & 
		$\mu(e_1 \comp e_b \comp e(a))), e_1(b)) = e_1(\mu(e_b \comp e(a)), b))$\\
		\hline
		
		$\psi^{L_2^b}_\F{ind1-0op}$ & 
		$ $ &
		$\mu(e_1(a), e_1(b)) = e_1(\mu(a, b))$ &  
		$\mu(e_1(a), e_1 \comp e_b(b))) = e_1(\mu(a, e_b(b)))$\\
		\hline
		
		$\psi^{L_2^b}_\F{ind2-0op}$ &
		$\neg e \rightleftarrows e_b \vee e \xrightarrow{\F{rc}} e_1$ & 
		$\mu(e_1(a), e_1 \comp e_b(b))) = e_1(\mu(a, e_b(b)))$ & 
		$\mu(e_1(a), e_1 \comp e_b \comp e(b)))) = e_1(\mu(a, e_b \comp e(b))))$\\
		\hline	
		
\end{tabular}%
}
\caption{Complete set of Verification Conditions for CRDTs}
\label{tab:full_crdts}
\end{table}

\subsection{Buggy MRDT implementation in~\cite{Vimala}}
\label{subsec:bug_impl}
\begin{figure}[ht]
\small
\begin{algorithmic} [1]
	\State $\Sigma = (\mathbb{N} \times \F{bool})$
	\State $O = \{\F{enable}, \F{disable}\}$
	\State $Q = \{\F{rd} \}$
	\State $\sigma_0 = (0, \F{false})$
	\State ${\F{do}(\sigma,\_,\_,\F{enable}}) = (\F{fst}(\sigma) + 1, \F{true})$
	\State ${\F{do}(\sigma,\_,\_,\F{disable}}) = (\F{fst}(\sigma), \F{false})$
	\State $\F{merge\_flag}((lc,lf), (ac,af), (bc,bf)) =
   			 \begin{cases}
      				\F{true}, & \F{if}\ af = \text{true}~\&\&~bf = \F{true} \\
				\F{false}, & \F{else\ if}\ af = \text{false}~\&\&~bf = \F{false} \\
				ac > lc, & \F{else\ if}\ af = \text{true}\\
				bc > lc, & \F{otherwise}
   			 \end{cases}$

	\State $\F{merge}(\sigma_\top,\sigma_a,\sigma_b) = (\F{fst}(\sigma_a) + \F{fst}(\sigma_b) - \F{fst}(\sigma_\top), \F{merge\_flag} (\sigma_\top, \sigma_a, \sigma_b))$
	\State $\F{query}(\sigma,rd) = \F{snd}(\sigma)$
	\State $\F{rc} = \{(\F{disable}, \F{enable})\}$
\end{algorithmic}
\caption{Enable-wins flag MRDT implementation from \cite{Vimala}}
\label{fig:ewflag_impl}
\end{figure}

\end{document}